\documentclass[twocolumn]{article}

\usepackage[T1]{fontenc}
\usepackage[utf8]{inputenc}
\usepackage{authblk}
\usepackage{cite}
\usepackage{amsmath,amssymb,amsfonts}
\usepackage{algorithmic}
\usepackage{graphicx}
\usepackage{textcomp}
\usepackage{xcolor}
\usepackage{setspace}
\usepackage{hyperref}
\usepackage{comment}
\usepackage{array}
\def\BibTeX{{\rm B\kern-.05em{\sc i\kern-.025em b}\kern-.08em
    T\kern-.1667em\lower.7ex\hbox{E}\kern-.125emX}}
    
\usepackage[]{subcaption}

\usepackage{booktabs} 
\renewcommand{\paragraph}[1]{\vspace{0.3\baselineskip}\noindent\textbf{#1}}

\usepackage{pgfplots, tikz}
\usepackage{pgfplotstable}
\usetikzlibrary{pgfplots.groupplots}
\usetikzlibrary{shapes.geometric}
\usetikzlibrary{positioning,fit,shapes.geometric,backgrounds, calc}
\usetikzlibrary{arrows,decorations.markings}
\usetikzlibrary{shapes.arrows}
\usetikzlibrary{patterns}
\tikzset{textnode/.style={inner sep=0pt,outer sep=0,execute at begin node={\strut}}}
\tikzstyle{state} = [textnode,circle, draw, inner sep=0pt, outer sep=0]

\usepackage{helvet}
\pgfplotsset{
  tick label style = {font=\scriptsize\sffamily},
  every axis label = {font=\scriptsize\sffamily},
  legend style = {font=\footnotesize\sffamily},
  title style = {font=\footnotesize\sffamily},
  label style = {font=\footnotesize\sffamily}
}
\selectcolormodel{cmyk}

\usepackage{mathtools}
\DeclarePairedDelimiter\ceil{\lceil}{\rceil}
\DeclarePairedDelimiter\floor{\lfloor}{\rfloor}

\usepackage{multirow}
\usepackage{graphicx}
\usepackage{listings}
\usepackage{enumitem}

\def\rot#1{\rotatebox{90}{#1}}

\newenvironment{customlegend}[1][]{%
    \begingroup
    \csname pgfplots@init@cleared@structures\endcsname
    \pgfplotsset{#1}%
}{%
    \csname pgfplots@createlegend\endcsname
    \endgroup
}%

\def\addlegendimage{\csname pgfplots@addlegendimage\endcsname}

\pgfplotsset{
    jitter/.style={
        y filter/.code={\pgfmathparse{\pgfmathresult+rnd*#1}}
    },
    jitter/.default=0.05
}

\newcommand{\nop}[1]{}

\newcommand*{\eg}{{\em e.g.}}
\newcommand*{\etal}{{\em et al.~}}
\newcommand*{\ie}{{\em i.e.}}

\begin{document}

\title{Dynamics of Team Library Adoptions: \\ An Exploration of GitHub Commit Logs}

\author[1]{Pamela Bilo Thomas \thanks{pthomas4@nd.edu}}
\author[1]{Rachel Krohn \thanks{rkrohn@nd.edu}}
\author[1]{Tim Weninger \thanks{tweninge@nd.edu}}
\affil[1]{Department of Computer Science and Engineering, University of Notre Dame}

\date{\vspace{-5ex}}

\renewcommand\Authands{ and }

\maketitle
\begin{abstract}
When a group of people strives to understand new information, struggle ensues as various ideas compete for attention.  Steep learning curves are surmounted as teams learn together.  To understand how these team dynamics play out in software development, we explore Git logs, which provide a complete change history of software repositories.  In these repositories, we observe code additions, which represent successfully implemented ideas, and code deletions, which represent ideas that have failed or been superseded. By examining the patterns between these commit types, we can begin to understand how teams adopt new information. We specifically study what happens after a software library is adopted by a project, \ie, when a library is used for the first time in the project. We find that a variety of factors, including team size, library popularity, and prevalence on Stack Overflow are associated with how quickly teams learn and successfully adopt new software libraries.
\end{abstract}

\maketitle

\section{Introduction}

Online collaboration systems like GitHub provide a powerful setting in which to study the process of learning new information and adopting new technology in a group setting. By investigating how software developers adopt and use software libraries in this specific context, we may better understand how humans learn new technical information and incorporate concepts~\cite{krohn2019library}.

This work aims to answer the following research questions:
\begin{itemize}[noitemsep, topsep=0pt]
    \item[RQ1] What are the events that happen when a team adopts a library for the first time?
    \item[RQ2] Are commits containing new libraries more likely to have deletions than other types of commits?
    \item[RQ3] Do the answers to these questions vary by library type, team size, or the amount of information available on Stack Overflow?
    \item[RQ4] Do team members fight over the adoption and usage of a new library?
\end{itemize}

\section{Data Acquisition}

First, we issued a query to the GitHub search API for projects written primarily in Python. GitHub returned repository IDs of the 1,000 most popular Python projects on the site. We then found all GitHub users who made at least one commit to a repository in this set and retrieved all their Python projects. We did this breadth-first style crawling two additional times, culminating in 259,923 projects with 89,311 contributing GitHub users.

Of these, we were able to clone 259,613 Python repositories to disk. These repositories constitute about 13\% of all Python repositories on GitHub as of September 2018. The full dataset of cloned projects occupies about 8 TB of disk space. Source code and repository IDs are available at \url{https://github.com/tweninger/growingpains}.

Additionally, we downloaded all question posts from Stack Overflow from its inception until September 2018. Appropriate tagging tends to increase viewership of questions~\cite{saha2013discriminative}, so we filtered out any posts that were not tagged as Python posts, then extracted all libraries mentioned in each question.


An important complication that arises in Git commit histories is that stashes, reverts, branches, and other Git commands can result in a non-monotonic chronology of edits. Fortunately, each commit keeps a pointer to its parent(s). Because the order of actions is more important than exact times, we enforce a monotonic chronology according to the commit graph.

\begin{figure*}[t]
    \centering
    \begin{subfigure}{0.25\textwidth}
        \pgfplotstableread{
x   px
1	0.065364284
2	0.083105853
3	0.071027879
4	0.05928872
5	0.051107115
6	0.044126763
7	0.037812489
8	0.032861175
9	0.02951923
10	0.026639305
11	0.023820983
12	0.021199019
13	0.019389441
14	0.018307544
15	0.016451763
16	0.015115755
17	0.013764347
18	0.012882658
19	0.012278182
20	0.011743009
21	0.01061876
22	0.010110538
23	0.009097944
24	0.008978589
25	0.008112302
26	0.007758086
27	0.007430822
28	0.00714206
29	0.006737792
30	0.006591486
31	0.006141016
32	0.00554424
33	0.005428735
34	0.00507067
35	0.005132272
36	0.004905113
37	0.00438149
38	0.004427692
39	0.004339138
40	0.00415818
41	0.003807815
42	0.003765463
43	0.003484401
44	0.003488251
45	0.003418948
46	0.003203339
47	0.003033932
48	0.003110935
49	0.003053182
50	0.002710518
51	0.002645065
52	0.002683566
53	0.002521859
54	0.002610413
55	0.002410204
56	0.002159944
57	0.002171494
58	0.002337051
59	0.002152243
60	0.002140693
61	0.002029038
62	0.001948184
63	0.00207139
64	0.001894282
65	0.00183653
66	0.001767227
67	0.001821129
68	0.001540067
69	0.001528516
70	0.001663272
71	0.001501565
72	0.00138221
73	0.001524666
74	0.001439962
75	0.001424562
76	0.001413011
77	0.001359109
78	0.001239754
79	0.001305207
80	0.001224353
81	0.001309057
82	0.001197402
83	0.001266705
84	0.001235904
85	0.0011512
86	0.000985643
87	0.001178151
88	0.001024144
89	0.001016444
90	0.00092404
91	0.001116548
92	0.000958692
93	0.001024144
94	0.001062646
95	0.001001043
96	0.000974092
97	0.000908639
98	0.000827786
99	0.000839336
100	0.000881688
101	0.000816235
102	0.000750783
103	0.000754633
104	0.000839336
105	0.000727682
106	0.000727682
107	0.000731532
108	0.000639128
109	0.000816235
110	0.000673779
111	0.00069688
112	0.000669929
113	0.000754633
114	0.000677629
115	0.00068148
116	0.000650678
117	0.000604476
118	0.000635278
119	0.000639128
120	0.000627577
121	0.000592926
122	0.000619877
123	0.000523623
124	0.000546724
125	0.000562124
126	0.000616027
127	0.000542874
128	0.000531323
129	0.000477421
130	0.000546724
131	0.000492821
132	0.000488971
133	0.00046972
134	0.00045047
135	0.000392717
136	0.00045817
137	0.00045817
138	0.000423518
139	0.000500522
140	0.000504372
141	0.000512072
142	0.000477421
143	0.000492821
144	0.000427369
145	0.000473571
146	0.000435069
147	0.000419668
148	0.000400417
149	0.000385017
150	0.000354215
151	0.000377316
152	0.000350365
153	0.000350365
154	0.000388867
155	0.00045047
156	0.000400417
157	0.000358066
158	0.000411968
159	0.000446619
160	0.000327264
161	0.000350365
162	0.000319564
163	0.000334965
164	0.000265662
165	0.000408118
166	0.000411968
167	0.000319564
168	0.000288763
169	0.000257961
170	0.000308013
171	0.000323414
172	0.000296463
173	0.000296463
174	0.000257961
175	0.000261811
176	0.000292613
177	0.000281062
178	0.000315714
179	0.00023486
180	0.000296463
181	0.000296463
182	0.000319564
183	0.00023101
184	0.000327264
185	0.000265662
186	0.000257961
187	0.000261811
188	0.000265662
189	0.000250261
190	0.00023871
191	0.000261811
192	0.000250261
193	0.000288763
194	0.000242561
195	0.000254111
196	0.00023871
197	0.000215609
198	0.000200209
199	0.000250261
200	0.00022331
201	0.000269512
202	0.000184808
203	0.000204059
204	0.00023101
205	0.00023486
206	0.000254111
207	0.000254111
208	0.000200209
209	0.000180958
210	0.000184808
211	0.00023486
212	0.00023486
213	0.000180958
214	0.000204059
215	0.000142456
216	0.000180958
217	0.000188658
218	0.000130906
219	0.000188658
220	0.000177108
221	0.00023486
222	0.000184808
223	0.000192508
224	0.000196359
225	0.00022331
226	0.000180958
227	0.000115505
228	0.000157857
229	0.000211759
230	0.000196359
231	0.000173258
232	0.000154007
233	0.000157857
234	0.000192508
235	0.000173258
236	0.000138606
237	0.000134756
238	0.000142456
239	0.000192508
240	0.000161707
241	0.000157857
242	0.000150157
243	0.000192508
244	0.000154007
245	0.000173258
246	0.000123205
247	0.000157857
248	0.000169407
249	0.000138606
250	0.000165557
251	0.000134756
252	0.000138606
253	0.000107805
254	0.000169407
255	0.000150157
256	0.000200209
257	0.000103955
258	0.000142456
259	9.2404E-05
260	0.000115505
261	0.000134756
262	0.000142456
263	0.000100104
264	0.000107805
265	9.62542E-05
266	0.000142456
267	0.000127056
268	0.000111655
269	0.000103955
270	0.000100104
271	0.000115505
272	0.000146306
273	0.000107805
274	0.000103955
275	0.000115505
276	0.000130906
277	0.000130906
278	0.000119355
279	0.000142456
280	8.85538E-05
281	0.000103955
282	0.000100104
283	8.47037E-05
284	0.000111655
285	0.000119355
286	0.000111655
287	0.000107805
288	0.000100104
289	0.000119355
290	0.000142456
291	0.000115505
292	0.000107805
293	0.000150157
294	0.000107805
295	0.000103955
296	9.2404E-05
297	0.000127056
298	0.000119355
299	0.000103955
300	0.000100104
301	0.000111655
302	0.000103955
303	0.000127056
304	8.47037E-05
305	0.000107805
306	7.31532E-05
307	8.47037E-05
308	9.2404E-05
309	8.08535E-05
310	9.2404E-05
311	0.000127056
312	9.62542E-05
313	0.000138606
314	8.08535E-05
315	6.54528E-05
316	0.000115505
317	6.16027E-05
318	8.85538E-05
319	0.000100104
320	9.2404E-05
321	8.85538E-05
322	5.00522E-05
323	0.000100104
324	6.16027E-05
325	7.31532E-05
326	8.08535E-05
327	6.9303E-05
328	9.2404E-05
329	0.000103955
330	7.31532E-05
331	6.54528E-05
332	6.54528E-05
333	6.54528E-05
334	8.47037E-05
335	8.08535E-05
336	8.47037E-05
337	7.31532E-05
338	6.16027E-05
339	5.00522E-05
340	9.2404E-05
341	0.000103955
342	8.47037E-05
343	0.000100104
344	6.16027E-05
345	6.9303E-05
346	6.16027E-05
347	4.6202E-05
348	0.000107805
349	7.70033E-05
350	6.54528E-05
351	7.31532E-05
352	9.2404E-05
353	8.08535E-05
354	7.31532E-05
355	7.70033E-05
356	8.85538E-05
357	8.85538E-05
358	3.85017E-05
359	5.77525E-05
360	3.85017E-05
361	3.08013E-05
362	0.000103955
363	5.39023E-05
364	5.39023E-05
365	4.23518E-05
366	4.23518E-05
367	6.54528E-05
368	6.54528E-05
369	8.47037E-05
370	4.6202E-05
371	7.70033E-05
372	7.31532E-05
373	5.77525E-05
374	6.16027E-05
375	6.16027E-05
376	5.77525E-05
377	8.08535E-05
378	4.6202E-05
379	6.9303E-05
380	6.9303E-05
381	6.16027E-05
382	5.39023E-05
383	6.16027E-05
384	6.54528E-05
385	5.77525E-05
386	5.39023E-05
387	6.16027E-05
388	6.54528E-05
389	7.70033E-05
390	5.39023E-05
391	7.70033E-05
392	5.39023E-05
393	8.47037E-05
394	5.39023E-05
395	4.23518E-05
396	6.54528E-05
397	7.31532E-05
398	5.39023E-05
399	6.16027E-05
400	3.46515E-05
401	6.54528E-05
402	8.08535E-05
403	5.77525E-05
404	5.39023E-05
405	4.6202E-05
406	4.6202E-05
407	7.31532E-05
408	5.77525E-05
409	3.85017E-05
410	5.39023E-05
411	5.77525E-05
412	2.69512E-05
413	4.6202E-05
414	4.6202E-05
415	4.23518E-05
416	5.77525E-05
417	4.6202E-05
418	5.77525E-05
419	3.85017E-05
420	4.23518E-05
421	3.85017E-05
422	6.54528E-05
423	6.16027E-05
424	5.00522E-05
425	8.47037E-05
426	3.08013E-05
427	4.6202E-05
428	6.54528E-05
429	7.70033E-05
430	4.23518E-05
431	5.39023E-05
432	3.08013E-05
433	3.08013E-05
434	5.39023E-05
435	5.77525E-05
436	4.23518E-05
437	6.16027E-05
438	3.85017E-05
439	3.85017E-05
440	5.39023E-05
441	5.39023E-05
442	4.6202E-05
443	6.16027E-05
444	3.46515E-05
445	4.23518E-05
446	4.6202E-05
447	5.39023E-05
448	3.85017E-05
449	6.54528E-05
450	5.39023E-05
451	5.77525E-05
452	5.00522E-05
453	7.70033E-05
454	3.08013E-05
455	4.23518E-05
456	5.00522E-05
457	6.9303E-05
458	5.77525E-05
459	3.46515E-05
460	5.39023E-05
461	7.70033E-05
462	3.08013E-05
463	5.39023E-05
464	5.39023E-05
465	3.85017E-05
466	3.46515E-05
467	3.46515E-05
468	6.54528E-05
469	5.39023E-05
470	2.3101E-05
471	3.46515E-05
472	5.77525E-05
473	2.69512E-05
474	0.000100104
475	2.3101E-05
476	1.92508E-05
477	5.39023E-05
478	5.00522E-05
479	4.23518E-05
480	4.23518E-05
481	2.69512E-05
482	3.46515E-05
483	2.3101E-05
484	4.6202E-05
485	3.85017E-05
486	4.23518E-05
487	7.31532E-05
488	3.85017E-05
489	2.3101E-05
490	1.54007E-05
491	3.46515E-05
492	4.6202E-05
493	8.08535E-05
494	3.46515E-05
495	4.6202E-05
496	2.3101E-05
497	3.46515E-05
498	4.23518E-05
499	3.85017E-05
500	3.85017E-05
501	4.6202E-05
502	3.85017E-05
503	8.08535E-05
504	3.08013E-05
505	3.08013E-05
506	2.69512E-05
507	4.23518E-05
508	4.6202E-05
509	3.85017E-05
510	3.08013E-05
511	4.6202E-05
512	4.23518E-05
513	1.54007E-05
514	2.69512E-05
515	3.46515E-05
516	3.85017E-05
517	1.92508E-05
518	2.3101E-05
519	5.77525E-05
520	2.3101E-05
521	1.92508E-05
522	5.39023E-05
523	3.46515E-05
524	4.6202E-05
525	1.54007E-05
526	1.92508E-05
527	3.46515E-05
528	3.85017E-05
529	5.39023E-05
530	3.46515E-05
531	3.46515E-05
532	7.31532E-05
533	2.3101E-05
534	2.3101E-05
535	2.69512E-05
536	5.00522E-05
537	3.46515E-05
538	3.46515E-05
539	3.46515E-05
540	2.69512E-05
541	4.23518E-05
542	4.6202E-05
543	3.46515E-05
544	3.08013E-05
545	6.9303E-05
546	1.92508E-05
547	3.08013E-05
548	5.00522E-05
549	4.23518E-05
550	1.54007E-05
551	3.46515E-05
552	2.69512E-05
553	2.3101E-05
554	4.6202E-05
555	3.46515E-05
556	2.3101E-05
557	1.92508E-05
558	5.39023E-05
559	3.85017E-05
560	2.3101E-05
561	2.69512E-05
562	1.54007E-05
563	2.3101E-05
564	2.3101E-05
565	3.85017E-05
566	1.15505E-05
567	2.3101E-05
568	1.54007E-05
569	3.46515E-05
570	3.08013E-05
571	3.08013E-05
572	4.23518E-05
573	3.08013E-05
574	1.92508E-05
575	5.00522E-05
576	1.92508E-05
577	4.6202E-05
578	3.46515E-05
579	1.92508E-05
580	4.23518E-05
581	2.3101E-05
582	5.00522E-05
583	3.08013E-05
584	3.08013E-05
585	3.46515E-05
586	3.46515E-05
587	2.3101E-05
588	3.46515E-05
589	1.92508E-05
590	3.08013E-05
591	2.3101E-05
592	2.69512E-05
593	1.92508E-05
594	1.15505E-05
595	3.46515E-05
596	3.46515E-05
597	1.92508E-05
598	4.6202E-05
599	5.39023E-05
600	3.08013E-05
601	2.3101E-05
602	3.46515E-05
603	2.69512E-05
604	3.85017E-05
605	1.92508E-05
606	1.54007E-05
607	3.08013E-05
608	2.3101E-05
609	1.92508E-05
610	5.00522E-05
611	3.85017E-05
612	1.92508E-05
613	1.92508E-05
614	3.08013E-05
615	1.92508E-05
616	1.92508E-05
617	3.85017E-05
618	2.3101E-05
619	1.92508E-05
620	3.85017E-05
621	2.69512E-05
622	1.54007E-05
623	5.39023E-05
624	1.54007E-05
625	1.92508E-05
626	3.46515E-05
627	1.92508E-05
628	2.3101E-05
629	2.69512E-05
630	1.92508E-05
631	2.69512E-05
632	6.54528E-05
633	1.92508E-05
634	1.92508E-05
635	2.69512E-05
636	1.54007E-05
637	1.15505E-05
638	2.69512E-05
639	2.69512E-05
640	2.69512E-05
641	1.15505E-05
642	2.3101E-05
643	1.54007E-05
644	1.54007E-05
645	1.54007E-05
646	3.46515E-05
647	2.69512E-05
648	2.3101E-05
649	3.46515E-05
650	3.46515E-05
651	3.85017E-05
652	2.69512E-05
653	1.15505E-05
654	3.08013E-05
655	1.92508E-05
656	1.54007E-05
657	3.85017E-05
658	1.15505E-05
659	3.08013E-05
660	2.3101E-05
661	1.92508E-05
662	1.54007E-05
663	1.15505E-05
664	3.08013E-05
665	3.46515E-05
666	2.3101E-05
667	3.08013E-05
668	3.85017E-05
669	1.54007E-05
670	1.15505E-05
671	1.54007E-05
672	1.15505E-05
673	3.08013E-05
674	2.69512E-05
675	1.54007E-05
676	3.46515E-05
677	1.54007E-05
678	2.69512E-05
679	1.92508E-05
680	1.15505E-05
681	1.92508E-05
682	1.54007E-05
683	2.3101E-05
684	1.15505E-05
685	1.15505E-05
686	1.15505E-05
687	1.92508E-05
688	3.08013E-05
689	3.85017E-05
690	1.15505E-05
691	1.54007E-05
692	1.92508E-05
693	2.69512E-05
694	1.54007E-05
695	1.92508E-05
696	7.70033E-06
697	3.85017E-06
698	2.69512E-05
699	7.70033E-06
700	2.69512E-05
701	2.69512E-05
702	1.15505E-05
703	1.15505E-05
704	1.15505E-05
705	2.3101E-05
706	1.54007E-05
707	2.3101E-05
708	1.54007E-05
709	2.69512E-05
710	7.70033E-06
711	3.08013E-05
712	1.54007E-05
713	2.3101E-05
714	1.54007E-05
715	1.15505E-05
716	3.85017E-06
717	1.54007E-05
718	1.54007E-05
719	1.92508E-05
720	1.15505E-05
721	1.15505E-05
722	2.69512E-05
723	7.70033E-06
724	1.54007E-05
725	3.46515E-05
726	1.15505E-05
727	7.70033E-06
728	3.08013E-05
729	2.3101E-05
730	1.92508E-05
731	2.69512E-05
732	3.85017E-06
733	1.92508E-05
734	2.3101E-05
735	1.54007E-05
736	1.15505E-05
737	3.85017E-06
738	1.54007E-05
739	1.54007E-05
741	3.85017E-06
742	1.92508E-05
743	2.69512E-05
744	1.15505E-05
745	1.54007E-05
746	1.15505E-05
747	1.15505E-05
748	1.54007E-05
749	1.54007E-05
750	1.15505E-05
751	2.3101E-05
752	1.15505E-05
753	7.70033E-06
754	1.92508E-05
755	7.70033E-06
756	3.85017E-06
757	1.92508E-05
759	2.3101E-05
760	3.08013E-05
761	1.54007E-05
762	1.15505E-05
763	7.70033E-06
764	1.92508E-05
765	7.70033E-06
766	3.85017E-06
767	1.54007E-05
768	7.70033E-06
769	1.92508E-05
770	2.3101E-05
771	7.70033E-06
772	1.54007E-05
773	3.85017E-06
774	2.3101E-05
775	3.85017E-06
776	3.85017E-06
777	3.85017E-06
778	7.70033E-06
779	1.92508E-05
780	3.08013E-05
781	1.15505E-05
782	2.3101E-05
783	1.15505E-05
784	2.69512E-05
785	1.54007E-05
786	7.70033E-06
787	1.15505E-05
788	1.54007E-05
789	1.54007E-05
790	1.54007E-05
791	7.70033E-06
792	7.70033E-06
793	1.15505E-05
794	1.15505E-05
795	1.54007E-05
796	1.15505E-05
797	1.92508E-05
798	3.85017E-06
799	1.92508E-05
800	1.15505E-05
801	7.70033E-06
802	1.92508E-05
803	1.54007E-05
804	3.85017E-06
805	3.85017E-06
806	1.54007E-05
807	1.15505E-05
808	1.54007E-05
809	1.54007E-05
810	1.15505E-05
811	1.92508E-05
812	7.70033E-06
813	1.54007E-05
814	3.85017E-06
815	1.54007E-05
816	1.15505E-05
817	2.69512E-05
818	1.92508E-05
819	1.92508E-05
820	1.92508E-05
821	1.54007E-05
822	2.69512E-05
823	1.15505E-05
824	1.54007E-05
825	1.15505E-05
826	1.15505E-05
827	2.3101E-05
828	3.08013E-05
829	1.15505E-05
830	7.70033E-06
832	1.54007E-05
833	3.08013E-05
834	2.69512E-05
835	1.54007E-05
836	2.69512E-05
837	1.92508E-05
838	1.15505E-05
839	1.92508E-05
840	1.15505E-05
841	2.3101E-05
842	7.70033E-06
843	1.54007E-05
844	1.54007E-05
845	7.70033E-06
846	2.69512E-05
847	1.92508E-05
848	1.54007E-05
849	1.92508E-05
850	1.54007E-05
851	1.15505E-05
852	1.92508E-05
853	3.85017E-05
854	1.54007E-05
855	2.69512E-05
856	2.69512E-05
857	1.92508E-05
858	2.3101E-05
859	1.54007E-05
860	1.15505E-05
861	1.15505E-05
862	1.15505E-05
863	3.08013E-05
864	2.69512E-05
865	1.15505E-05
866	1.15505E-05
867	2.3101E-05
868	1.15505E-05
869	1.92508E-05
870	1.54007E-05
871	7.70033E-06
872	1.15505E-05
873	1.54007E-05
874	7.70033E-06
875	1.15505E-05
876	2.3101E-05
877	1.54007E-05
878	1.15505E-05
879	2.69512E-05
880	1.15505E-05
881	1.54007E-05
882	1.92508E-05
883	1.54007E-05
884	1.92508E-05
885	1.15505E-05
886	7.70033E-06
887	1.54007E-05
888	1.92508E-05
890	1.54007E-05
891	7.70033E-06
892	1.54007E-05
893	1.15505E-05
894	2.69512E-05
895	1.54007E-05
896	7.70033E-06
897	1.54007E-05
898	2.3101E-05
899	3.85017E-06
900	1.15505E-05
901	3.85017E-06
902	1.92508E-05
903	1.92508E-05
904	1.15505E-05
905	1.92508E-05
906	3.85017E-06
907	1.92508E-05
908	1.15505E-05
909	3.85017E-06
910	3.85017E-06
911	2.3101E-05
912	7.70033E-06
913	1.54007E-05
914	1.15505E-05
915	1.15505E-05
916	3.85017E-06
917	1.54007E-05
918	1.15505E-05
919	3.85017E-06
920	3.85017E-06
921	7.70033E-06
922	7.70033E-06
923	3.85017E-06
924	1.15505E-05
925	2.69512E-05
926	1.54007E-05
927	1.15505E-05
928	3.85017E-06
929	7.70033E-06
930	1.54007E-05
931	1.15505E-05
932	1.54007E-05
933	1.15505E-05
934	1.92508E-05
935	3.85017E-06
936	1.15505E-05
937	1.92508E-05
938	7.70033E-06
939	7.70033E-06
940	1.54007E-05
941	1.92508E-05
942	3.85017E-06
943	7.70033E-06
944	1.54007E-05
945	1.54007E-05
946	3.85017E-06
947	1.15505E-05
948	3.85017E-06
949	1.15505E-05
950	3.85017E-06
951	3.85017E-06
952	2.69512E-05
953	7.70033E-06
954	2.69512E-05
955	7.70033E-06
956	1.92508E-05
957	1.92508E-05
958	1.15505E-05
959	1.92508E-05
960	1.15505E-05
961	1.54007E-05
963	3.85017E-06
964	2.3101E-05
965	1.54007E-05
966	7.70033E-06
967	7.70033E-06
968	1.15505E-05
969	1.54007E-05
970	1.92508E-05
971	1.15505E-05
972	1.15505E-05
973	7.70033E-06
974	1.54007E-05
975	1.15505E-05
976	1.54007E-05
977	1.15505E-05
978	1.15505E-05
979	1.15505E-05
980	1.54007E-05
981	7.70033E-06
982	3.85017E-06
983	3.85017E-06
984	3.08013E-05
985	1.54007E-05
986	1.92508E-05
988	3.85017E-06
989	1.54007E-05
990	7.70033E-06
991	7.70033E-06
992	3.85017E-06
993	3.85017E-06
994	1.15505E-05
995	1.92508E-05
996	1.15505E-05
997	3.85017E-06
998	1.54007E-05
999	3.85017E-06
1000	1.15505E-05
1001	1.54007E-05
1002	7.70033E-06
1003	3.08013E-05
1004	1.54007E-05
1005	2.3101E-05
1006	1.54007E-05
1007	7.70033E-06
1008	1.92508E-05
1009	3.85017E-06
1010	1.92508E-05
1011	3.08013E-05
1012	1.54007E-05
1013	3.46515E-05
1014	1.15505E-05
1015	2.69512E-05
1016	2.3101E-05
1017	1.92508E-05
1018	3.85017E-06
1019	1.92508E-05
1020	3.46515E-05
1021	1.15505E-05
1022	9.62542E-05
1023	1.92508E-05
1024	1.92508E-05
1025	1.92508E-05
1026	7.70033E-06
1027	1.15505E-05
1028	1.92508E-05
1029	2.69512E-05
1030	1.54007E-05
1031	7.70033E-06
1032	1.15505E-05
1033	3.85017E-06
1034	1.54007E-05
1035	1.15505E-05
1036	7.70033E-06
1037	1.92508E-05
1038	2.69512E-05
1039	7.70033E-06
1040	2.3101E-05
1041	2.3101E-05
1042	3.85017E-06
1043	2.3101E-05
1044	1.15505E-05
1045	7.70033E-06
1046	7.70033E-06
1047	7.70033E-06
1048	7.70033E-06
1049	7.70033E-06
1050	7.70033E-06
1051	3.85017E-06
1052	1.15505E-05
1053	1.15505E-05
1055	1.54007E-05
1056	1.54007E-05
1057	3.85017E-06
1058	7.70033E-06
1059	3.85017E-06
1060	1.54007E-05
1061	3.85017E-06
1062	1.15505E-05
1063	1.54007E-05
1064	7.70033E-06
1065	7.70033E-06
1067	2.69512E-05
1068	7.70033E-06
1069	7.70033E-06
1070	7.70033E-06
1074	7.70033E-06
1075	7.70033E-06
1076	7.70033E-06
1077	1.15505E-05
1078	1.15505E-05
1079	3.85017E-06
1080	1.54007E-05
1081	1.15505E-05
1082	7.70033E-06
1083	1.15505E-05
1084	7.70033E-06
1085	7.70033E-06
1086	3.85017E-06
1087	1.54007E-05
1088	1.92508E-05
1089	3.85017E-06
1090	3.85017E-06
1091	1.92508E-05
1092	1.92508E-05
1093	7.70033E-06
1094	7.70033E-06
1095	7.70033E-06
1096	3.85017E-06
1097	3.85017E-06
1098	7.70033E-06
1099	3.85017E-06
1100	3.85017E-06
1101	1.54007E-05
1102	1.15505E-05
1103	3.85017E-06
1104	1.54007E-05
1105	3.85017E-06
1106	1.54007E-05
1107	7.70033E-06
1108	7.70033E-06
1109	1.15505E-05
1110	7.70033E-06
1111	1.15505E-05
1112	1.54007E-05
1113	3.85017E-06
1114	7.70033E-06
1115	1.92508E-05
1116	1.15505E-05
1117	1.54007E-05
1118	7.70033E-06
1119	2.3101E-05
1120	1.15505E-05
1121	1.15505E-05
1122	3.85017E-06
1123	1.54007E-05
1124	3.85017E-06
1125	1.54007E-05
1126	3.85017E-06
1127	1.92508E-05
1128	7.70033E-06
1129	1.92508E-05
1130	3.85017E-06
1131	1.92508E-05
1132	1.15505E-05
1133	1.54007E-05
1134	1.15505E-05
1136	3.85017E-06
1137	2.3101E-05
1138	7.70033E-06
1139	3.85017E-06
1140	7.70033E-06
1141	3.85017E-06
1142	7.70033E-06
1143	7.70033E-06
1144	3.85017E-06
1145	7.70033E-06
1146	3.85017E-06
1147	7.70033E-06
1148	1.15505E-05
1149	1.15505E-05
1150	7.70033E-06
1151	1.54007E-05
1152	1.54007E-05
1153	3.85017E-06
1154	1.54007E-05
1155	1.54007E-05
1156	3.85017E-06
1157	3.85017E-06
1158	7.70033E-06
1159	3.85017E-06
1160	1.15505E-05
1161	3.85017E-06
1162	3.85017E-06
1163	7.70033E-06
1164	1.54007E-05
1165	7.70033E-06
1166	7.70033E-06
1169	7.70033E-06
1171	7.70033E-06
1172	2.3101E-05
1174	3.85017E-06
1176	1.92508E-05
1180	7.70033E-06
1182	3.85017E-06
1185	7.70033E-06
1186	1.92508E-05
1188	7.70033E-06
1189	7.70033E-06
1190	3.85017E-06
1192	3.85017E-06
1193	1.15505E-05
1194	7.70033E-06
1195	7.70033E-06
1196	3.85017E-06
1198	7.70033E-06
1199	7.70033E-06
1200	1.92508E-05
1202	1.15505E-05
1203	7.70033E-06
1204	1.54007E-05
1205	1.54007E-05
1206	3.85017E-06
1208	1.15505E-05
1209	3.85017E-06
1210	7.70033E-06
1211	3.85017E-06
1212	7.70033E-06
1213	3.85017E-06
1214	3.85017E-06
1215	3.85017E-06
1216	7.70033E-06
1217	3.85017E-06
1218	3.85017E-06
1219	7.70033E-06
1220	7.70033E-06
1224	7.70033E-06
1226	7.70033E-06
1227	1.54007E-05
1228	3.85017E-06
1229	3.85017E-06
1230	3.85017E-06
1233	3.85017E-06
1234	1.15505E-05
1235	7.70033E-06
1236	3.85017E-06
1238	3.85017E-06
1239	1.54007E-05
1240	3.85017E-06
1241	1.15505E-05
1242	1.15505E-05
1244	7.70033E-06
1245	1.92508E-05
1246	3.85017E-06
1247	7.70033E-06
1248	3.85017E-06
1249	1.15505E-05
1250	7.70033E-06
1251	3.85017E-06
1252	7.70033E-06
1253	3.85017E-06
1254	1.15505E-05
1255	3.85017E-06
1258	1.15505E-05
1260	1.92508E-05
1261	7.70033E-06
1262	3.85017E-06
1263	3.85017E-06
1265	7.70033E-06
1266	7.70033E-06
1267	3.85017E-06
1269	7.70033E-06
1270	3.85017E-06
1271	3.85017E-06
1272	3.85017E-06
1273	1.15505E-05
1275	3.85017E-06
1276	7.70033E-06
1277	1.15505E-05
1278	3.85017E-06
1279	3.85017E-06
1280	7.70033E-06
1282	3.85017E-06
1285	3.85017E-06
1286	7.70033E-06
1287	1.15505E-05
1290	1.54007E-05
1291	7.70033E-06
1293	1.15505E-05
1294	3.85017E-06
1296	7.70033E-06
1297	1.15505E-05
1298	3.85017E-06
1299	3.85017E-06
1300	3.85017E-06
1302	3.85017E-06
1303	7.70033E-06
1304	7.70033E-06
1305	3.85017E-06
1307	7.70033E-06
1308	3.85017E-06
1309	3.85017E-06
1311	7.70033E-06
1312	7.70033E-06
1313	7.70033E-06
1315	1.15505E-05
1316	7.70033E-06
1319	1.15505E-05
1320	7.70033E-06
1321	1.54007E-05
1322	7.70033E-06
1323	1.54007E-05
1324	3.85017E-06
1325	1.15505E-05
1327	7.70033E-06
1328	3.85017E-06
1329	7.70033E-06
1331	1.15505E-05
1332	7.70033E-06
1333	7.70033E-06
1334	3.85017E-06
1335	3.85017E-06
1336	7.70033E-06
1338	2.3101E-05
1339	3.85017E-06
1340	7.70033E-06
1341	7.70033E-06
1342	3.85017E-06
1343	3.85017E-06
1344	7.70033E-06
1345	3.85017E-06
1349	7.70033E-06
1350	1.15505E-05
1352	7.70033E-06
1353	2.69512E-05
1354	3.85017E-06
1356	3.85017E-06
1357	3.85017E-06
1358	7.70033E-06
1360	7.70033E-06
1361	1.15505E-05
1363	1.54007E-05
1364	7.70033E-06
1365	7.70033E-06
1366	1.15505E-05
1367	7.70033E-06
1368	3.85017E-06
1369	1.54007E-05
1370	3.85017E-06
1372	3.85017E-06
1373	3.85017E-06
1374	3.85017E-06
1376	7.70033E-06
1378	3.85017E-06
1379	1.15505E-05
1380	7.70033E-06
1381	3.85017E-06
1382	3.85017E-06
1383	3.85017E-06
1384	3.85017E-06
1385	1.15505E-05
1386	7.70033E-06
1387	7.70033E-06
1388	3.85017E-06
1389	1.92508E-05
1390	7.70033E-06
1391	3.85017E-06
1392	3.85017E-06
1393	7.70033E-06
1394	3.85017E-06
1395	2.69512E-05
1396	7.70033E-06
1398	1.54007E-05
1399	3.85017E-06
1400	7.70033E-06
1401	3.85017E-06
1402	3.85017E-06
1403	3.85017E-06
1404	7.70033E-06
1406	3.85017E-06
1407	1.54007E-05
1408	7.70033E-06
1410	3.85017E-06
1412	7.70033E-06
1413	3.85017E-06
1414	1.15505E-05
1416	3.85017E-06
1417	3.85017E-06
1419	3.85017E-06
1420	1.54007E-05
1422	1.54007E-05
1423	3.85017E-06
1426	3.85017E-06
1427	7.70033E-06
1429	7.70033E-06
1431	1.15505E-05
1432	7.70033E-06
1433	3.85017E-06
1435	3.85017E-06
1436	3.85017E-06
1437	3.85017E-06
1438	7.70033E-06
1440	3.85017E-06
1442	1.15505E-05
1443	1.15505E-05
1444	7.70033E-06
1445	3.85017E-06
1446	7.70033E-06
1448	7.70033E-06
1449	7.70033E-06
1450	3.85017E-06
1452	1.15505E-05
1455	1.92508E-05
1456	7.70033E-06
1457	3.85017E-06
1460	3.85017E-06
1461	7.70033E-06
1462	7.70033E-06
1463	1.15505E-05
1464	3.85017E-06
1465	1.15505E-05
1467	7.70033E-06
1468	3.85017E-06
1471	7.70033E-06
1472	1.15505E-05
1474	1.15505E-05
1475	3.85017E-06
1479	3.85017E-06
1480	7.70033E-06
1482	3.85017E-06
1484	3.85017E-06
1486	7.70033E-06
1489	7.70033E-06
1490	3.85017E-06
1492	3.85017E-06
1493	7.70033E-06
1494	1.15505E-05
1496	7.70033E-06
1497	3.85017E-06
1499	3.85017E-06
1500	3.85017E-06
1501	3.85017E-06
1502	1.15505E-05
1503	7.70033E-06
1505	3.85017E-06
1506	3.85017E-06
1507	3.85017E-06
1509	3.85017E-06
1510	3.85017E-06
1513	7.70033E-06
1515	3.85017E-06
1516	3.85017E-06
1517	1.15505E-05
1520	7.70033E-06
1523	1.15505E-05
1524	7.70033E-06
1527	7.70033E-06
1528	7.70033E-06
1529	1.15505E-05
1530	3.85017E-06
1531	3.85017E-06
1533	3.85017E-06
1534	1.54007E-05
1535	3.85017E-06
1536	3.85017E-06
1537	3.85017E-06
1538	1.15505E-05
1539	7.70033E-06
1540	3.85017E-06
1541	3.85017E-06
1542	3.85017E-06
1543	3.85017E-06
1544	3.85017E-06
1545	3.85017E-06
1547	7.70033E-06
1548	3.85017E-06
1549	7.70033E-06
1551	3.85017E-06
1552	1.54007E-05
1553	3.85017E-06
1554	1.54007E-05
1555	7.70033E-06
1556	7.70033E-06
1557	3.85017E-06
1559	3.85017E-06
1560	1.15505E-05
1563	7.70033E-06
1564	3.85017E-06
1568	3.85017E-06
1570	1.15505E-05
1571	1.15505E-05
1572	7.70033E-06
1573	7.70033E-06
1574	1.15505E-05
1576	3.85017E-06
1579	7.70033E-06
1581	3.85017E-06
1582	3.85017E-06
1584	3.85017E-06
1585	7.70033E-06
1586	7.70033E-06
1587	7.70033E-06
1591	7.70033E-06
1592	3.85017E-06
1594	7.70033E-06
1595	3.85017E-06
1596	3.85017E-06
1597	3.85017E-06
1599	7.70033E-06
1600	3.85017E-06
1601	3.85017E-06
1603	7.70033E-06
1604	3.85017E-06
1605	7.70033E-06
1606	7.70033E-06
1608	7.70033E-06
1612	1.15505E-05
1613	7.70033E-06
1614	3.85017E-06
1616	3.85017E-06
1618	3.85017E-06
1623	3.85017E-06
1624	3.85017E-06
1626	7.70033E-06
1630	3.85017E-06
1631	3.85017E-06
1632	3.85017E-06
1635	1.15505E-05
1636	7.70033E-06
1639	3.85017E-06
1640	3.85017E-06
1642	3.85017E-06
1643	1.54007E-05
1644	7.70033E-06
1646	3.85017E-06
1650	3.85017E-06
1651	7.70033E-06
1652	7.70033E-06
1653	7.70033E-06
1654	3.85017E-06
1655	7.70033E-06
1656	3.85017E-06
1658	3.85017E-06
1662	3.85017E-06
1663	3.85017E-06
1670	3.85017E-06
1671	7.70033E-06
1672	1.15505E-05
1676	7.70033E-06
1678	3.85017E-06
1680	1.15505E-05
1681	3.85017E-06
1684	3.85017E-06
1685	3.85017E-06
1686	3.85017E-06
1687	7.70033E-06
1688	3.85017E-06
1691	3.85017E-06
1697	3.85017E-06
1698	7.70033E-06
1700	3.85017E-06
1701	7.70033E-06
1705	3.85017E-06
1706	3.85017E-06
1709	3.85017E-06
1712	7.70033E-06
1714	7.70033E-06
1715	7.70033E-06
1716	3.85017E-06
1720	3.85017E-06
1721	1.15505E-05
1722	7.70033E-06
1725	3.85017E-06
1726	3.85017E-06
1729	3.85017E-06
1731	3.85017E-06
1732	7.70033E-06
1733	7.70033E-06
1734	3.85017E-06
1735	3.85017E-06
1739	3.85017E-06
1742	3.85017E-06
1743	3.85017E-06
1747	7.70033E-06
1750	7.70033E-06
1751	3.85017E-06
1753	1.54007E-05
1755	3.85017E-06
1756	3.85017E-06
1757	3.85017E-06
1758	3.85017E-06
1759	3.85017E-06
1760	3.85017E-06
1761	1.15505E-05
1762	7.70033E-06
1764	3.85017E-06
1765	3.85017E-06
1767	1.54007E-05
1769	7.70033E-06
1770	7.70033E-06
1771	1.15505E-05
1772	3.85017E-06
1773	3.85017E-06
1775	3.85017E-06
1778	3.85017E-06
1779	3.85017E-06
1780	7.70033E-06
1782	3.85017E-06
1784	7.70033E-06
1785	3.85017E-06
1789	7.70033E-06
1791	7.70033E-06
1793	3.85017E-06
1794	3.85017E-06
1795	7.70033E-06
1796	7.70033E-06
1797	3.85017E-06
1799	7.70033E-06
1800	3.85017E-06
1801	3.85017E-06
1804	3.85017E-06
1807	3.85017E-06
1808	3.85017E-06
1809	1.54007E-05
1811	3.85017E-06
1812	3.85017E-06
1815	7.70033E-06
1818	3.85017E-06
1821	3.85017E-06
1823	1.15505E-05
1824	3.85017E-06
1827	3.85017E-06
1830	7.70033E-06
1833	3.85017E-06
1834	3.85017E-06
1837	3.85017E-06
1838	7.70033E-06
1839	7.70033E-06
1846	7.70033E-06
1847	3.85017E-06
1849	3.85017E-06
1851	3.85017E-06
1855	7.70033E-06
1856	3.85017E-06
1858	7.70033E-06
1859	3.85017E-06
1860	3.85017E-06
1861	3.85017E-06
1862	3.85017E-06
1866	3.85017E-06
1872	7.70033E-06
1873	3.85017E-06
1874	7.70033E-06
1875	3.85017E-06
1877	3.85017E-06
1878	7.70033E-06
1879	3.85017E-06
1882	3.85017E-06
1885	3.85017E-06
1886	3.85017E-06
1888	3.85017E-06
1889	3.85017E-06
1891	7.70033E-06
1892	7.70033E-06
1893	7.70033E-06
1896	3.85017E-06
1897	3.85017E-06
1898	7.70033E-06
1899	3.85017E-06
1901	7.70033E-06
1902	7.70033E-06
1903	7.70033E-06
1909	3.85017E-06
1910	3.85017E-06
1914	3.85017E-06
1916	3.85017E-06
1917	7.70033E-06
1918	3.85017E-06
1920	3.85017E-06
1923	7.70033E-06
1924	7.70033E-06
1927	7.70033E-06
1928	3.85017E-06
1933	3.85017E-06
1935	3.85017E-06
1936	7.70033E-06
1939	3.85017E-06
1942	3.85017E-06
1943	1.15505E-05
1944	3.85017E-06
1946	3.85017E-06
1949	7.70033E-06
1953	3.85017E-06
1954	3.85017E-06
1955	3.85017E-06
1957	3.85017E-06
1960	3.85017E-06
1967	3.85017E-06
1968	3.85017E-06
1971	1.15505E-05
1972	7.70033E-06
1973	3.85017E-06
1977	7.70033E-06
1978	3.85017E-06
1980	3.85017E-06
1984	3.85017E-06
1985	3.85017E-06
1986	3.85017E-06
1989	3.85017E-06
1990	7.70033E-06
1991	7.70033E-06
1995	3.85017E-06
2001	1.15505E-05
2004	7.70033E-06
2005	3.85017E-06
2006	7.70033E-06
2010	3.85017E-06
2015	3.85017E-06
2018	7.70033E-06
2020	3.85017E-06
2021	3.85017E-06
2023	7.70033E-06
2024	3.85017E-06
2025	7.70033E-06
2027	3.85017E-06
2028	3.85017E-06
2029	3.85017E-06
2032	3.85017E-06
2033	3.85017E-06
2034	7.70033E-06
2035	7.70033E-06
2036	3.85017E-06
2037	3.85017E-06
2042	3.85017E-06
2043	1.15505E-05
2048	3.85017E-06
2051	1.15505E-05
2052	3.85017E-06
2053	7.70033E-06
2054	3.85017E-06
2055	3.85017E-06
2056	3.85017E-06
2060	3.85017E-06
2061	3.85017E-06
2062	7.70033E-06
2063	3.85017E-06
2068	3.85017E-06
2071	3.85017E-06
2072	7.70033E-06
2073	7.70033E-06
2074	3.85017E-06
2078	7.70033E-06
2081	3.85017E-06
2082	7.70033E-06
2090	3.85017E-06
2092	3.85017E-06
2093	3.85017E-06
2096	3.85017E-06
2097	7.70033E-06
2098	7.70033E-06
2099	3.85017E-06
2100	1.92508E-05
2102	3.85017E-06
2111	3.85017E-06
2113	3.85017E-06
2116	3.85017E-06
2118	3.85017E-06
2119	3.85017E-06
2121	3.85017E-06
2122	3.85017E-06
2123	7.70033E-06
2129	3.85017E-06
2131	7.70033E-06
2133	7.70033E-06
2135	7.70033E-06
2138	3.85017E-06
2142	3.85017E-06
2147	7.70033E-06
2148	3.85017E-06
2151	7.70033E-06
2153	1.15505E-05
2158	7.70033E-06
2160	3.85017E-06
2163	3.85017E-06
2164	7.70033E-06
2167	7.70033E-06
2170	7.70033E-06
2176	3.85017E-06
2177	3.85017E-06
2182	3.85017E-06
2184	3.85017E-06
2188	3.85017E-06
2192	3.85017E-06
2193	3.85017E-06
2196	3.85017E-06
2199	3.85017E-06
2200	3.85017E-06
2204	3.85017E-06
2211	7.70033E-06
2212	3.85017E-06
2213	3.85017E-06
2216	7.70033E-06
2217	3.85017E-06
2220	3.85017E-06
2221	3.85017E-06
2222	3.85017E-06
2223	7.70033E-06
2226	3.85017E-06
2228	3.85017E-06
2240	3.85017E-06
2241	7.70033E-06
2242	7.70033E-06
2243	7.70033E-06
2244	3.85017E-06
2245	7.70033E-06
2248	3.85017E-06
2254	7.70033E-06
2257	3.85017E-06
2264	7.70033E-06
2265	3.85017E-06
2266	3.85017E-06
2274	3.85017E-06
2282	3.85017E-06
2284	7.70033E-06
2285	7.70033E-06
2289	3.85017E-06
2290	7.70033E-06
2291	3.85017E-06
2292	3.85017E-06
2293	3.85017E-06
2294	3.85017E-06
2297	3.85017E-06
2302	3.85017E-06
2307	3.85017E-06
2312	7.70033E-06
2320	3.85017E-06
2321	7.70033E-06
2323	3.85017E-06
2324	3.85017E-06
2325	7.70033E-06
2328	3.85017E-06
2330	3.85017E-06
2331	1.15505E-05
2334	7.70033E-06
2335	3.85017E-06
2340	3.85017E-06
2347	3.85017E-06
2348	3.85017E-06
2350	3.85017E-06
2351	3.85017E-06
2354	3.85017E-06
2355	3.85017E-06
2359	3.85017E-06
2360	3.85017E-06
2361	7.70033E-06
2362	3.85017E-06
2364	3.85017E-06
2365	3.85017E-06
2366	3.85017E-06
2367	3.85017E-06
2373	3.85017E-06
2376	3.85017E-06
2378	3.85017E-06
2380	3.85017E-06
2382	3.85017E-06
2383	3.85017E-06
2386	7.70033E-06
2388	3.85017E-06
2389	3.85017E-06
2393	3.85017E-06
2396	3.85017E-06
2397	3.85017E-06
2398	3.85017E-06
2401	1.15505E-05
2405	3.85017E-06
2406	1.15505E-05
2410	7.70033E-06
2411	7.70033E-06
2415	3.85017E-06
2418	3.85017E-06
2426	3.85017E-06
2433	3.85017E-06
2435	3.85017E-06
2437	3.85017E-06
2438	7.70033E-06
2439	3.85017E-06
2443	1.54007E-05
2446	3.85017E-06
2447	7.70033E-06
2449	3.85017E-06
2450	3.85017E-06
2452	3.85017E-06
2454	3.85017E-06
2458	7.70033E-06
2461	3.85017E-06
2463	3.85017E-06
2464	3.85017E-06
2470	7.70033E-06
2472	3.85017E-06
2473	3.85017E-06
2482	3.85017E-06
2483	3.85017E-06
2487	3.85017E-06
2490	7.70033E-06
2491	7.70033E-06
2492	3.85017E-06
2501	3.85017E-06
2505	3.85017E-06
2511	7.70033E-06
2512	3.85017E-06
2513	3.85017E-06
2521	3.85017E-06
2522	3.85017E-06
2523	3.85017E-06
2525	3.85017E-06
2532	3.85017E-06
2533	3.85017E-06
2534	3.85017E-06
2535	3.85017E-06
2536	3.85017E-06
2538	3.85017E-06
2545	7.70033E-06
2546	3.85017E-06
2547	3.85017E-06
2549	3.85017E-06
2550	3.85017E-06
2551	3.85017E-06
2552	3.85017E-06
2553	3.85017E-06
2554	7.70033E-06
2555	3.85017E-06
2557	3.85017E-06
2558	3.85017E-06
2563	7.70033E-06
2566	3.85017E-06
2569	3.85017E-06
2570	3.85017E-06
2571	3.85017E-06
2578	3.85017E-06
2581	1.15505E-05
2582	3.85017E-06
2587	3.85017E-06
2589	3.85017E-06
2591	7.70033E-06
2595	3.85017E-06
2596	3.85017E-06
2597	7.70033E-06
2599	3.85017E-06
2601	7.70033E-06
2613	3.85017E-06
2617	3.85017E-06
2618	3.85017E-06
2625	3.85017E-06
2626	3.85017E-06
2630	3.85017E-06
2633	3.85017E-06
2634	7.70033E-06
2638	1.15505E-05
2639	3.85017E-06
2643	7.70033E-06
2644	3.85017E-06
2659	3.85017E-06
2667	3.85017E-06
2669	3.85017E-06
2670	3.85017E-06
2675	3.85017E-06
2676	7.70033E-06
2683	3.85017E-06
2687	3.85017E-06
2693	3.85017E-06
2694	3.85017E-06
2699	3.85017E-06
2700	3.85017E-06
2703	3.85017E-06
2707	3.85017E-06
2710	3.85017E-06
2713	3.85017E-06
2716	3.85017E-06
2719	3.85017E-06
2721	3.85017E-06
2725	3.85017E-06
2729	7.70033E-06
2731	3.85017E-06
2733	3.85017E-06
2734	3.85017E-06
2740	3.85017E-06
2743	3.85017E-06
2744	3.85017E-06
2746	3.85017E-06
2749	3.85017E-06
2752	3.85017E-06
2753	3.85017E-06
2754	3.85017E-06
2761	3.85017E-06
2766	3.85017E-06
2771	3.85017E-06
2777	3.85017E-06
2778	3.85017E-06
2779	3.85017E-06
2783	3.85017E-06
2788	3.85017E-06
2791	3.85017E-06
2793	3.85017E-06
2798	3.85017E-06
2810	3.85017E-06
2811	3.85017E-06
2816	3.85017E-06
2817	3.85017E-06
2821	3.85017E-06
2822	3.85017E-06
2826	3.85017E-06
2828	3.85017E-06
2829	3.85017E-06
2832	3.85017E-06
2836	3.85017E-06
2838	3.85017E-06
2840	3.85017E-06
2842	1.15505E-05
2843	3.85017E-06
2844	7.70033E-06
2850	3.85017E-06
2852	3.85017E-06
2857	7.70033E-06
2864	3.85017E-06
2865	3.85017E-06
2869	7.70033E-06
2874	3.85017E-06
2876	3.85017E-06
2877	3.85017E-06
2879	3.85017E-06
2883	3.85017E-06
2885	3.85017E-06
2895	7.70033E-06
2896	3.85017E-06
2898	3.85017E-06
2909	3.85017E-06
2910	3.85017E-06
2914	3.85017E-06
2916	3.85017E-06
2918	7.70033E-06
2919	3.85017E-06
2933	7.70033E-06
2935	3.85017E-06
2937	3.85017E-06
2944	7.70033E-06
2953	3.85017E-06
2956	7.70033E-06
2957	1.15505E-05
2973	3.85017E-06
2974	3.85017E-06
2977	3.85017E-06
2980	3.85017E-06
2985	3.85017E-06
2994	3.85017E-06
3000	3.85017E-06
3003	3.85017E-06
3014	7.70033E-06
3015	3.85017E-06
3016	3.85017E-06
3020	3.85017E-06
3021	3.85017E-06
3023	3.85017E-06
3028	3.85017E-06
3029	3.85017E-06
3033	3.85017E-06
3037	7.70033E-06
3053	3.85017E-06
3055	3.85017E-06
3056	3.85017E-06
3057	3.85017E-06
3058	3.85017E-06
3061	7.70033E-06
3062	3.85017E-06
3067	3.85017E-06
3069	7.70033E-06
3078	7.70033E-06
3081	3.85017E-06
3083	3.85017E-06
3085	3.85017E-06
3092	3.85017E-06
3099	3.85017E-06
3100	3.85017E-06
3110	3.85017E-06
3111	7.70033E-06
3112	3.85017E-06
3113	7.70033E-06
3114	3.85017E-06
3121	3.85017E-06
3130	3.85017E-06
3131	3.85017E-06
3134	3.85017E-06
3140	3.85017E-06
3148	3.85017E-06
3163	3.85017E-06
3165	3.85017E-06
3168	3.85017E-06
3169	3.85017E-06
3170	7.70033E-06
3174	3.85017E-06
3177	3.85017E-06
3182	3.85017E-06
3184	3.85017E-06
3185	3.85017E-06
3194	3.85017E-06
3202	3.85017E-06
3205	3.85017E-06
3206	7.70033E-06
3207	1.15505E-05
3209	7.70033E-06
3217	3.85017E-06
3222	7.70033E-06
3223	3.85017E-06
3224	3.85017E-06
3233	3.85017E-06
3235	3.85017E-06
3243	3.85017E-06
3244	3.85017E-06
3245	3.85017E-06
3251	7.70033E-06
3257	3.85017E-06
3262	7.70033E-06
3272	7.70033E-06
3274	3.85017E-06
3281	3.85017E-06
3284	3.85017E-06
3285	3.85017E-06
3288	3.85017E-06
3293	3.85017E-06
3294	3.85017E-06
3295	3.85017E-06
3301	3.85017E-06
3307	3.85017E-06
3313	3.85017E-06
3316	3.85017E-06
3321	3.85017E-06
3323	3.85017E-06
3340	3.85017E-06
3348	3.85017E-06
3355	7.70033E-06
3371	3.85017E-06
3381	3.85017E-06
3382	7.70033E-06
3389	3.85017E-06
3391	3.85017E-06
3402	3.85017E-06
3404	3.85017E-06
3407	3.85017E-06
3411	3.85017E-06
3414	3.85017E-06
3415	3.85017E-06
3422	3.85017E-06
3440	3.85017E-06
3446	3.85017E-06
3451	3.85017E-06
3453	3.85017E-06
3463	3.85017E-06
3467	3.85017E-06
3468	3.85017E-06
3469	3.85017E-06
3473	3.85017E-06
3474	3.85017E-06
3476	3.85017E-06
3480	3.85017E-06
3481	3.85017E-06
3483	3.85017E-06
3486	3.85017E-06
3487	3.85017E-06
3491	7.70033E-06
3495	3.85017E-06
3498	3.85017E-06
3502	7.70033E-06
3505	3.85017E-06
3507	3.85017E-06
3511	7.70033E-06
3515	3.85017E-06
3518	3.85017E-06
3521	7.70033E-06
3523	7.70033E-06
3524	3.85017E-06
3533	3.85017E-06
3536	3.85017E-06
3555	7.70033E-06
3556	3.85017E-06
3559	3.85017E-06
3561	3.85017E-06
3563	3.85017E-06
3566	3.85017E-06
3584	3.85017E-06
3586	3.85017E-06
3588	3.85017E-06
3594	3.85017E-06
3595	3.85017E-06
3596	3.85017E-06
3597	7.70033E-06
3600	3.85017E-06
3617	3.85017E-06
3620	3.85017E-06
3625	3.85017E-06
3626	3.85017E-06
3642	3.85017E-06
3646	3.85017E-06
3648	3.85017E-06
3655	3.85017E-06
3671	3.85017E-06
3672	3.85017E-06
3674	3.85017E-06
3677	3.85017E-06
3680	3.85017E-06
3683	3.85017E-06
3701	3.85017E-06
3708	3.85017E-06
3713	3.85017E-06
3716	3.85017E-06
3719	3.85017E-06
3720	3.85017E-06
3724	3.85017E-06
3740	3.85017E-06
3750	3.85017E-06
3756	3.85017E-06
3759	3.85017E-06
3762	3.85017E-06
3771	3.85017E-06
3786	3.85017E-06
3794	3.85017E-06
3796	3.85017E-06
3797	3.85017E-06
3816	3.85017E-06
3820	7.70033E-06
3827	3.85017E-06
3835	3.85017E-06
3840	7.70033E-06
3851	3.85017E-06
3852	3.85017E-06
3858	3.85017E-06
3864	3.85017E-06
3865	3.85017E-06
3878	3.85017E-06
3884	3.85017E-06
3892	3.85017E-06
3895	3.85017E-06
3896	3.85017E-06
3902	3.85017E-06
3908	7.70033E-06
3927	3.85017E-06
3932	3.85017E-06
3941	3.85017E-06
3950	3.85017E-06
3955	3.85017E-06
3962	3.85017E-06
3964	3.85017E-06
3970	3.85017E-06
3975	3.85017E-06
3977	3.85017E-06
3978	3.85017E-06
3984	3.85017E-06
3987	3.85017E-06
3989	3.85017E-06
3995	3.85017E-06
4000	3.85017E-06
4003	3.85017E-06
4005	3.85017E-06
4006	3.85017E-06
4008	3.85017E-06
4018	3.85017E-06
4022	3.85017E-06
4024	7.70033E-06
4027	3.85017E-06
4058	3.85017E-06
4059	3.85017E-06
4076	3.85017E-06
4094	3.85017E-06
4099	3.85017E-06
4105	3.85017E-06
4106	3.85017E-06
4109	3.85017E-06
4127	3.85017E-06
4130	3.85017E-06
4138	3.85017E-06
4158	3.85017E-06
4169	3.85017E-06
4170	3.85017E-06
4175	7.70033E-06
4181	3.85017E-06
4183	3.85017E-06
4184	3.85017E-06
4218	3.85017E-06
4224	3.85017E-06
4238	3.85017E-06
4246	3.85017E-06
4251	3.85017E-06
4260	3.85017E-06
4274	3.85017E-06
4275	3.85017E-06
4282	3.85017E-06
4286	3.85017E-06
4293	3.85017E-06
4294	3.85017E-06
4305	3.85017E-06
4318	3.85017E-06
4328	3.85017E-06
4333	3.85017E-06
4363	3.85017E-06
4369	3.85017E-06
4390	3.85017E-06
4395	7.70033E-06
4397	3.85017E-06
4411	3.85017E-06
4414	3.85017E-06
4421	3.85017E-06
4423	3.85017E-06
4424	3.85017E-06
4425	3.85017E-06
4467	7.70033E-06
4476	3.85017E-06
4486	7.70033E-06
4502	3.85017E-06
4516	3.85017E-06
4529	3.85017E-06
4536	1.54007E-05
4539	3.85017E-06
4558	3.85017E-06
4568	3.85017E-06
4585	3.85017E-06
4593	3.85017E-06
4598	3.85017E-06
4607	3.85017E-06
4614	3.85017E-06
4619	3.85017E-06
4623	3.85017E-06
4640	3.85017E-06
4655	3.85017E-06
4657	3.85017E-06
4658	3.85017E-06
4666	3.85017E-06
4682	3.85017E-06
4711	3.85017E-06
4721	3.85017E-06
4723	3.85017E-06
4730	3.85017E-06
4746	3.85017E-06
4780	3.85017E-06
4781	3.85017E-06
4796	3.85017E-06
4797	3.85017E-06
4798	3.85017E-06
4814	3.85017E-06
4819	3.85017E-06
4820	3.85017E-06
4822	3.85017E-06
4834	3.85017E-06
4852	3.85017E-06
4856	3.85017E-06
4858	3.85017E-06
4871	3.85017E-06
4887	3.85017E-06
4897	3.85017E-06
4901	7.70033E-06
4912	3.85017E-06
4926	3.85017E-06
4933	3.85017E-06
4943	3.85017E-06
4955	3.85017E-06
4966	3.85017E-06
4980	3.85017E-06
4993	3.85017E-06
5000	3.85017E-06
5003	7.70033E-06
5005	3.85017E-06
5029	3.85017E-06
5036	3.85017E-06
5037	3.85017E-06
5043	3.85017E-06
5044	3.85017E-06
5045	3.85017E-06
5073	7.70033E-06
5085	3.85017E-06
5102	3.85017E-06
5104	3.85017E-06
5118	7.70033E-06
5129	3.85017E-06
5142	3.85017E-06
5145	3.85017E-06
5166	3.85017E-06
5169	3.85017E-06
5172	3.85017E-06
5178	3.85017E-06
5183	3.85017E-06
5194	3.85017E-06
5209	3.85017E-06
5221	3.85017E-06
5253	3.85017E-06
5270	3.85017E-06
5277	3.85017E-06
5285	3.85017E-06
5292	3.85017E-06
5303	3.85017E-06
5308	3.85017E-06
5315	3.85017E-06
5317	3.85017E-06
5328	3.85017E-06
5331	3.85017E-06
5336	3.85017E-06
5351	3.85017E-06
5352	3.85017E-06
5388	3.85017E-06
5394	3.85017E-06
5405	3.85017E-06
5415	3.85017E-06
5416	3.85017E-06
5447	3.85017E-06
5448	3.85017E-06
5449	3.85017E-06
5465	3.85017E-06
5472	3.85017E-06
5487	3.85017E-06
5492	3.85017E-06
5567	3.85017E-06
5607	1.15505E-05
5610	7.70033E-06
5611	3.85017E-06
5618	3.85017E-06
5626	7.70033E-06
5629	3.85017E-06
5632	3.85017E-06
5634	3.85017E-06
5637	3.85017E-06
5651	3.85017E-06
5654	3.85017E-06
5690	3.85017E-06
5696	3.85017E-06
5702	3.85017E-06
5703	3.85017E-06
5706	3.85017E-06
5708	3.85017E-06
5712	7.70033E-06
5718	3.85017E-06
5747	3.85017E-06
5750	3.85017E-06
5759	3.85017E-06
5773	3.85017E-06
5776	3.85017E-06
5792	3.85017E-06
5797	3.85017E-06
5803	3.85017E-06
5817	3.85017E-06
5824	3.85017E-06
5829	3.85017E-06
5839	3.85017E-06
5849	3.85017E-06
5850	3.85017E-06
5864	3.85017E-06
5894	3.85017E-06
5896	3.85017E-06
5899	3.85017E-06
5920	3.85017E-06
5922	3.85017E-06
5924	3.85017E-06
5963	3.85017E-06
5968	3.85017E-06
5978	3.85017E-06
5988	3.85017E-06
6006	3.85017E-06
6014	3.85017E-06
6028	3.85017E-06
6039	3.85017E-06
6055	3.85017E-06
6116	3.85017E-06
6126	3.85017E-06
6132	3.85017E-06
6145	3.85017E-06
6146	3.85017E-06
6167	3.85017E-06
6175	3.85017E-06
6204	3.85017E-06
6216	3.85017E-06
6222	3.85017E-06
6234	3.85017E-06
6237	3.85017E-06
6260	3.85017E-06
6265	3.85017E-06
6271	3.85017E-06
6279	3.85017E-06
6284	3.85017E-06
6288	3.85017E-06
6299	3.85017E-06
6302	3.85017E-06
6330	3.85017E-06
6334	3.85017E-06
6335	3.85017E-06
6344	3.85017E-06
6349	7.70033E-06
6354	3.85017E-06
6430	3.85017E-06
6436	3.85017E-06
6447	3.85017E-06
6451	3.85017E-06
6458	3.85017E-06
6468	3.85017E-06
6475	3.85017E-06
6491	3.85017E-06
6496	3.85017E-06
6502	3.85017E-06
6521	3.85017E-06
6539	3.85017E-06
6557	3.85017E-06
6562	7.70033E-06
6564	3.85017E-06
6577	3.85017E-06
6592	3.85017E-06
6603	3.85017E-06
6607	3.85017E-06
6615	7.70033E-06
6616	3.85017E-06
6645	3.85017E-06
6648	3.85017E-06
6653	3.85017E-06
6661	3.85017E-06
6706	3.85017E-06
6707	3.85017E-06
6708	3.85017E-06
6715	3.85017E-06
6727	3.85017E-06
6741	3.85017E-06
6751	3.85017E-06
6755	3.85017E-06
6773	3.85017E-06
6785	3.85017E-06
6799	3.85017E-06
6800	3.85017E-06
6821	3.85017E-06
6835	3.85017E-06
6844	3.85017E-06
6845	3.85017E-06
6864	3.85017E-06
6881	3.85017E-06
6882	3.85017E-06
6895	3.85017E-06
6907	3.85017E-06
6937	3.85017E-06
6939	3.85017E-06
6975	3.85017E-06
6977	3.85017E-06
6979	3.85017E-06
7007	3.85017E-06
7020	3.85017E-06
7025	3.85017E-06
7027	3.85017E-06
7028	3.85017E-06
7041	3.85017E-06
7057	3.85017E-06
7069	3.85017E-06
7075	3.85017E-06
7081	3.85017E-06
7111	3.85017E-06
7116	3.85017E-06
7119	3.85017E-06
7123	3.85017E-06
7135	3.85017E-06
7136	3.85017E-06
7150	3.85017E-06
7171	3.85017E-06
7176	3.85017E-06
7195	3.85017E-06
7196	7.70033E-06
7205	3.85017E-06
7209	3.85017E-06
7220	3.85017E-06
7224	3.85017E-06
7261	3.85017E-06
7266	3.85017E-06
7278	3.85017E-06
7279	3.85017E-06
7293	3.85017E-06
7305	3.85017E-06
7313	3.85017E-06
7321	3.85017E-06
7325	3.85017E-06
7329	3.85017E-06
7330	3.85017E-06
7358	3.85017E-06
7364	3.85017E-06
7367	3.85017E-06
7419	3.85017E-06
7449	3.85017E-06
7470	3.85017E-06
7475	3.85017E-06
7490	3.85017E-06
7492	3.85017E-06
7496	3.85017E-06
7503	3.85017E-06
7544	3.85017E-06
7548	3.85017E-06
7559	3.85017E-06
7561	3.85017E-06
7565	3.85017E-06
7587	3.85017E-06
7589	3.85017E-06
7600	3.85017E-06
7615	3.85017E-06
7645	3.85017E-06
7685	3.85017E-06
7711	3.85017E-06
7726	3.85017E-06
7745	3.85017E-06
7765	3.85017E-06
7767	7.70033E-06
7801	3.85017E-06
7810	3.85017E-06
7833	3.85017E-06
7837	3.85017E-06
7844	3.85017E-06
7864	3.85017E-06
7881	3.85017E-06
7895	3.85017E-06
7919	3.85017E-06
7946	3.85017E-06
7956	3.85017E-06
7978	3.85017E-06
7981	3.85017E-06
7984	3.85017E-06
8002	3.85017E-06
8011	3.85017E-06
8014	3.85017E-06
8028	3.85017E-06
8031	3.85017E-06
8032	3.85017E-06
8035	3.85017E-06
8036	7.70033E-06
8037	3.85017E-06
8038	7.70033E-06
8039	3.85017E-06
8040	7.70033E-06
8049	3.85017E-06
8060	3.85017E-06
8069	3.85017E-06
8075	3.85017E-06
8090	1.54007E-05
8092	3.85017E-06
8096	3.85017E-06
8103	3.85017E-06
8107	3.85017E-06
8116	3.85017E-06
8118	3.85017E-06
8119	3.85017E-06
8183	7.70033E-06
8194	3.85017E-06
8196	3.85017E-06
8212	3.85017E-06
8218	7.70033E-06
8251	3.85017E-06
8254	3.85017E-06
8259	3.85017E-06
8270	1.54007E-05
8271	7.70033E-06
8273	3.85017E-06
8315	7.70033E-06
8324	3.85017E-06
8325	3.85017E-06
8329	3.85017E-06
8330	3.85017E-06
8355	3.85017E-06
8362	3.85017E-06
8390	3.85017E-06
8401	3.85017E-06
8418	3.85017E-06
8430	3.85017E-06
8451	3.85017E-06
8453	3.85017E-06
8468	3.85017E-06
8492	3.85017E-06
8497	3.85017E-06
8524	3.85017E-06
8535	3.85017E-06
8536	3.85017E-06
8553	3.85017E-06
8566	3.85017E-06
8578	3.85017E-06
8608	3.85017E-06
8612	3.85017E-06
8699	3.85017E-06
8740	3.85017E-06
8765	3.85017E-06
8766	3.85017E-06
8852	3.85017E-06
8856	3.85017E-06
8896	3.85017E-06
8902	3.85017E-06
8903	3.85017E-06
8950	3.85017E-06
8973	3.85017E-06
9023	3.85017E-06
9057	3.85017E-06
9066	3.85017E-06
9095	3.85017E-06
9175	3.85017E-06
9224	3.85017E-06
9253	3.85017E-06
9254	3.85017E-06
9278	3.85017E-06
9289	3.85017E-06
9296	3.85017E-06
9311	3.85017E-06
9327	3.85017E-06
9355	3.85017E-06
9368	3.85017E-06
9376	3.85017E-06
9384	3.85017E-06
9388	3.85017E-06
9454	3.85017E-06
9491	3.85017E-06
9515	3.85017E-06
9528	3.85017E-06
9532	3.85017E-06
9580	3.85017E-06
9610	3.85017E-06
9627	3.85017E-06
9635	3.85017E-06
9637	3.85017E-06
9693	3.85017E-06
9699	3.85017E-06
9715	3.85017E-06
9723	3.85017E-06
9742	7.70033E-06
9784	3.85017E-06
9793	3.85017E-06
9794	3.85017E-06
9809	3.85017E-06
9811	3.85017E-06
9827	3.85017E-06
9838	3.85017E-06
9848	3.85017E-06
9858	3.85017E-06
9928	3.85017E-06
9936	3.85017E-06
9962	3.85017E-06
9991	7.70033E-06
9992	3.85017E-06
9993	3.85017E-06
10000	3.85017E-06
10023	3.85017E-06
10031	3.85017E-06
10032	3.85017E-06
10037	3.85017E-06
10046	3.85017E-06
10050	3.85017E-06
10066	3.85017E-06
10094	3.85017E-06
10115	3.85017E-06
10119	3.85017E-06
10132	3.85017E-06
10133	3.85017E-06
10166	3.85017E-06
10167	3.85017E-06
10196	3.85017E-06
10227	3.85017E-06
10241	3.85017E-06
10242	3.85017E-06
10285	3.85017E-06
10307	3.85017E-06
10329	3.85017E-06
10330	3.85017E-06
10388	3.85017E-06
10392	3.85017E-06
10428	3.85017E-06
10481	3.85017E-06
10491	3.85017E-06
10513	3.85017E-06
10517	3.85017E-06
10523	7.70033E-06
10532	3.85017E-06
10534	3.85017E-06
10560	3.85017E-06
10611	3.85017E-06
10618	3.85017E-06
10638	3.85017E-06
10652	3.85017E-06
10676	3.85017E-06
10689	3.85017E-06
10786	3.85017E-06
10814	3.85017E-06
10862	3.85017E-06
10954	3.85017E-06
10977	3.85017E-06
10980	3.85017E-06
10999	3.85017E-06
11021	3.85017E-06
11051	3.85017E-06
11087	3.85017E-06
11267	3.85017E-06
11306	7.70033E-06
11311	3.85017E-06
11332	3.85017E-06
11338	3.85017E-06
11382	3.85017E-06
11427	3.85017E-06
11431	3.85017E-06
11433	3.85017E-06
11446	3.85017E-06
11486	3.85017E-06
11515	3.85017E-06
11534	7.70033E-06
11559	3.85017E-06
11563	3.85017E-06
11591	3.85017E-06
11632	3.85017E-06
11638	3.85017E-06
11673	3.85017E-06
11688	3.85017E-06
11723	3.85017E-06
11778	3.85017E-06
11791	3.85017E-06
11793	3.85017E-06
11805	3.85017E-06
11834	3.85017E-06
11958	3.85017E-06
11983	3.85017E-06
12038	3.85017E-06
12102	3.85017E-06
12131	3.85017E-06
12151	3.85017E-06
12160	3.85017E-06
12179	3.85017E-06
12195	3.85017E-06
12199	3.85017E-06
12293	3.85017E-06
12335	3.85017E-06
12341	3.85017E-06
12358	3.85017E-06
12376	3.85017E-06
12405	3.85017E-06
12444	3.85017E-06
12457	3.85017E-06
12490	3.85017E-06
12538	3.85017E-06
12602	3.85017E-06
12642	3.85017E-06
12653	3.85017E-06
12671	3.85017E-06
12687	7.70033E-06
12720	3.85017E-06
12721	3.85017E-06
12796	3.85017E-06
12837	3.85017E-06
12846	3.85017E-06
12961	3.85017E-06
12974	3.85017E-06
12981	3.85017E-06
12990	3.85017E-06
13007	3.85017E-06
13156	3.85017E-06
13233	3.85017E-06
13252	3.85017E-06
13294	3.85017E-06
13296	7.70033E-06
13318	3.85017E-06
13322	3.85017E-06
13385	3.85017E-06
13501	3.85017E-06
13509	3.85017E-06
13524	3.85017E-06
13551	3.85017E-06
13571	3.85017E-06
13661	3.85017E-06
13792	3.85017E-06
13807	3.85017E-06
13838	3.85017E-06
13955	3.85017E-06
13974	3.85017E-06
14071	3.85017E-06
14210	3.85017E-06
14288	3.85017E-06
14337	3.85017E-06
14359	3.85017E-06
14450	3.85017E-06
14618	3.85017E-06
14648	7.70033E-06
14713	3.85017E-06
14724	3.85017E-06
14740	3.85017E-06
14856	3.85017E-06
15008	3.85017E-06
15097	3.85017E-06
15281	3.85017E-06
15446	3.85017E-06
15573	3.85017E-06
15608	3.85017E-06
15644	3.85017E-06
15719	3.85017E-06
15728	3.85017E-06
15826	3.85017E-06
16013	3.85017E-06
16016	3.85017E-06
16208	3.85017E-06
16284	3.85017E-06
16308	3.85017E-06
16343	3.85017E-06
16472	3.85017E-06
16485	3.85017E-06
16578	3.85017E-06
16618	3.85017E-06
16663	3.85017E-06
16669	3.85017E-06
16725	3.85017E-06
16803	3.85017E-06
16890	3.85017E-06
16905	3.85017E-06
17115	3.85017E-06
17279	3.85017E-06
17340	3.85017E-06
17355	3.85017E-06
17994	3.85017E-06
18057	3.85017E-06
18121	3.85017E-06
18182	3.85017E-06
18196	3.85017E-06
18282	3.85017E-06
18365	3.85017E-06
18621	3.85017E-06
18638	3.85017E-06
18715	3.85017E-06
18759	3.85017E-06
18883	3.85017E-06
19142	3.85017E-06
19466	3.85017E-06
19480	3.85017E-06
19514	3.85017E-06
19520	3.85017E-06
19640	3.85017E-06
19743	3.85017E-06
19820	3.85017E-06
19914	3.85017E-06
19944	3.85017E-06
19947	3.85017E-06
19957	3.85017E-06
19983	3.85017E-06
20324	3.85017E-06
20414	3.85017E-06
20479	3.85017E-06
20635	3.85017E-06
20665	3.85017E-06
20696	3.85017E-06
20888	3.85017E-06
20907	3.85017E-06
21080	3.85017E-06
21300	3.85017E-06
21607	3.85017E-06
21723	3.85017E-06
21851	3.85017E-06
21919	3.85017E-06
21921	3.85017E-06
22180	3.85017E-06
22190	3.85017E-06
22229	3.85017E-06
22550	3.85017E-06
22650	3.85017E-06
22984	3.85017E-06
23019	3.85017E-06
23069	3.85017E-06
23099	3.85017E-06
23173	3.85017E-06
23322	3.85017E-06
23559	3.85017E-06
23781	3.85017E-06
23844	3.85017E-06
24009	3.85017E-06
24219	3.85017E-06
24741	3.85017E-06
24936	3.85017E-06
25143	3.85017E-06
25179	3.85017E-06
25507	3.85017E-06
25552	3.85017E-06
26588	3.85017E-06
26615	3.85017E-06
26635	3.85017E-06
26672	3.85017E-06
26771	3.85017E-06
27064	3.85017E-06
27165	3.85017E-06
27224	3.85017E-06
27286	3.85017E-06
27549	3.85017E-06
27770	3.85017E-06
27944	3.85017E-06
28076	3.85017E-06
28087	3.85017E-06
28115	3.85017E-06
28477	3.85017E-06
29256	3.85017E-06
29502	3.85017E-06
29558	3.85017E-06
29698	3.85017E-06
29735	3.85017E-06
30192	3.85017E-06
30226	3.85017E-06
30244	3.85017E-06
30930	3.85017E-06
31218	3.85017E-06
31394	3.85017E-06
31522	3.85017E-06
31587	3.85017E-06
31693	3.85017E-06
31735	3.85017E-06
32544	3.85017E-06
33065	3.85017E-06
34022	3.85017E-06
34075	3.85017E-06
34185	3.85017E-06
34206	3.85017E-06
34351	3.85017E-06
34415	3.85017E-06
34970	3.85017E-06
35114	3.85017E-06
35171	3.85017E-06
35320	3.85017E-06
36252	3.85017E-06
36364	3.85017E-06
36827	3.85017E-06
36848	3.85017E-06
37451	3.85017E-06
37674	3.85017E-06
37824	3.85017E-06
37938	3.85017E-06
38545	3.85017E-06
38575	3.85017E-06
38806	3.85017E-06
39076	3.85017E-06
39080	3.85017E-06
39529	3.85017E-06
39638	3.85017E-06
39834	3.85017E-06
39855	3.85017E-06
40270	3.85017E-06
40366	3.85017E-06
40568	3.85017E-06
41150	3.85017E-06
42287	3.85017E-06
42414	3.85017E-06
44039	3.85017E-06
44291	3.85017E-06
45117	3.85017E-06
45243	3.85017E-06
45245	3.85017E-06
45399	3.85017E-06
45668	3.85017E-06
45677	3.85017E-06
45750	3.85017E-06
45770	3.85017E-06
45861	3.85017E-06
48393	3.85017E-06
48479	3.85017E-06
48822	3.85017E-06
49024	3.85017E-06
49697	3.85017E-06
49879	3.85017E-06
49941	3.85017E-06
49942	3.85017E-06
50015	3.85017E-06
50038	3.85017E-06
50160	3.85017E-06
50441	3.85017E-06
53088	3.85017E-06
58930	3.85017E-06
59173	3.85017E-06
59356	3.85017E-06
60425	3.85017E-06
61620	3.85017E-06
62194	3.85017E-06
63382	3.85017E-06
63395	3.85017E-06
63651	3.85017E-06
64172	3.85017E-06
72632	3.85017E-06
78486	3.85017E-06
81052	3.85017E-06
82947	3.85017E-06
87948	3.85017E-06
88145	3.85017E-06
88736	3.85017E-06
89226	3.85017E-06
89747	3.85017E-06
89923	3.85017E-06
91111	3.85017E-06
91366	3.85017E-06
91578	3.85017E-06
91916	3.85017E-06
92063	3.85017E-06
93576	3.85017E-06
94141	3.85017E-06
97211	3.85017E-06
97618	3.85017E-06
98644	3.85017E-06
98683	3.85017E-06
99192	3.85017E-06
99324	3.85017E-06
101158	3.85017E-06
105516	3.85017E-06
115366	3.85017E-06
125468	3.85017E-06
}{\table}
\begin{tikzpicture}

    \begin{axis}[
    title=Commit Distribution,
    width=1.80in,
    clip=false,
    y label style={at={(axis description cs:0.1,.5)}},
    height=1.75in,
    ymode=log,
    xmode=log,
    xlabel = {Commits},
    ylabel = {$p(x)$},
    ]
    \addplot [only marks, mark=*, blue, mark size=0.5pt, fill opacity=0.2, draw opacity=0.2]  table [x=x, y=px]   {\table};
    \node[anchor=north east] at (rel axis cs:-0.05,-0.01) {\textbf{(a)}};
    \end{axis}
\end{tikzpicture}
        \caption{\label{fig:commit_dist}}
    \end{subfigure}
    \begin{subfigure}{0.25\textwidth}
\pgfplotstableread{
x   px
1	0.032710153
2	0.046165393
3	0.060562227
4	0.065938865
5	0.065215611
6	0.063905568
7	0.060930677
8	0.053752729
9	0.047857533
10	0.04470524
11	0.039737991
12	0.035848799
13	0.031782205
14	0.027497271
15	0.025423035
16	0.022134279
17	0.021219978
18	0.018736354
19	0.016007096
20	0.014901747
21	0.013509825
22	0.012172489
23	0.011203603
24	0.010480349
25	0.009634279
26	0.008201419
27	0.007218886
28	0.006700328
29	0.006686681
30	0.006345524
31	0.005594978
32	0.005076419
33	0.004271288
34	0.004257642
35	0.004107533
36	0.003588974
37	0.003397926
38	0.003056769
39	0.003316048
40	0.002442686
41	0.002906659
42	0.002415393
43	0.002101528
44	0.002374454
45	0.002046943
46	0.001883188
47	0.001582969
48	0.002033297
49	0.001869541
50	0.001419214
51	0.001378275
52	0.001664847
53	0.001514738
54	0.001405568
55	0.00143286
56	0.001296397
57	0.001337336
58	0.000927948
59	0.001118996
60	0.001050764
61	0.000914301
62	0.000887009
63	0.000805131
64	0.00095524
65	0.000614083
66	0.000559498
67	0.000436681
68	0.00058679
69	0.000518559
70	0.000436681
71	0.00058679
72	0.000559498
73	0.000409389
74	0.000463974
75	0.000641376
76	0.000559498
77	0.000450328
78	0.000545852
79	0.000504913
80	0.000450328
81	0.000300218
82	0.000436681
83	0.000354803
84	0.000409389
85	0.00036845
86	0.000286572
87	0.000245633
88	0.000272926
89	0.000300218
90	0.000259279
91	0.000272926
92	0.00047762
93	0.000204694
94	0.000231987
95	0.000204694
96	0.000245633
97	0.000259279
98	0.000259279
99	0.000272926
100	0.000313865
101	0.000272926
102	0.000245633
103	0.000245633
104	0.000245633
105	0.000327511
106	0.000231987
107	0.000150109
108	0.000191048
109	0.000231987
110	0.000259279
111	0.000150109
112	0.000231987
113	0.000218341
114	0.000191048
115	0.000191048
116	0.000177402
117	0.000259279
118	0.00010917
119	8.18777E-05
120	0.000218341
121	0.00010917
122	0.000204694
123	0.000163755
124	0.000191048
125	0.000204694
126	0.000150109
127	0.000300218
128	0.00010917
129	0.000286572
130	0.00010917
131	0.000204694
132	9.5524E-05
133	0.000150109
134	5.45852E-05
135	0.000150109
136	0.000163755
137	0.000163755
138	0.000136463
139	0.000245633
140	0.000122817
141	0.000150109
142	0.000191048
143	0.000177402
144	0.000122817
145	0.000163755
146	0.00010917
147	0.000204694
148	0.000177402
149	9.5524E-05
150	6.82314E-05
151	4.09389E-05
152	0.000150109
153	4.09389E-05
154	9.5524E-05
155	0.000122817
156	2.72926E-05
157	5.45852E-05
158	5.45852E-05
159	0.00010917
160	0.00010917
161	0.000122817
162	2.72926E-05
163	8.18777E-05
164	6.82314E-05
165	0.000122817
166	9.5524E-05
167	2.72926E-05
168	8.18777E-05
169	9.5524E-05
170	9.5524E-05
171	8.18777E-05
172	4.09389E-05
173	5.45852E-05
174	4.09389E-05
175	9.5524E-05
176	4.09389E-05
177	0.00010917
178	9.5524E-05
179	4.09389E-05
180	4.09389E-05
181	0.000122817
182	9.5524E-05
183	5.45852E-05
184	5.45852E-05
185	6.82314E-05
186	0.000122817
187	8.18777E-05
188	4.09389E-05
190	6.82314E-05
191	0.000122817
192	8.18777E-05
193	4.09389E-05
194	0.000122817
195	4.09389E-05
196	1.36463E-05
197	5.45852E-05
198	5.45852E-05
200	8.18777E-05
201	5.45852E-05
202	4.09389E-05
203	9.5524E-05
204	4.09389E-05
205	5.45852E-05
206	6.82314E-05
207	4.09389E-05
208	4.09389E-05
209	6.82314E-05
210	5.45852E-05
211	9.5524E-05
212	5.45852E-05
213	1.36463E-05
214	2.72926E-05
215	2.72926E-05
216	4.09389E-05
217	4.09389E-05
218	2.72926E-05
219	2.72926E-05
220	9.5524E-05
221	1.36463E-05
222	4.09389E-05
223	9.5524E-05
224	1.36463E-05
225	1.36463E-05
227	0.00010917
228	4.09389E-05
229	4.09389E-05
230	4.09389E-05
231	1.36463E-05
232	4.09389E-05
233	2.72926E-05
234	2.72926E-05
235	8.18777E-05
236	4.09389E-05
237	4.09389E-05
238	8.18777E-05
239	2.72926E-05
240	4.09389E-05
241	4.09389E-05
242	2.72926E-05
244	2.72926E-05
246	6.82314E-05
247	2.72926E-05
248	4.09389E-05
249	1.36463E-05
250	4.09389E-05
251	1.36463E-05
252	1.36463E-05
253	1.36463E-05
254	4.09389E-05
255	4.09389E-05
256	4.09389E-05
257	4.09389E-05
259	1.36463E-05
260	1.36463E-05
261	2.72926E-05
262	1.36463E-05
263	4.09389E-05
264	4.09389E-05
266	2.72926E-05
267	1.36463E-05
268	2.72926E-05
269	4.09389E-05
270	1.36463E-05
271	1.36463E-05
272	2.72926E-05
273	2.72926E-05
274	4.09389E-05
275	2.72926E-05
276	6.82314E-05
277	2.72926E-05
278	1.36463E-05
279	2.72926E-05
280	4.09389E-05
281	4.09389E-05
283	6.82314E-05
284	2.72926E-05
285	1.36463E-05
286	4.09389E-05
287	1.36463E-05
288	2.72926E-05
289	2.72926E-05
290	4.09389E-05
291	1.36463E-05
292	2.72926E-05
293	2.72926E-05
294	1.36463E-05
295	1.36463E-05
296	2.72926E-05
297	5.45852E-05
298	4.09389E-05
299	5.45852E-05
300	2.72926E-05
301	1.36463E-05
302	2.72926E-05
303	1.36463E-05
304	4.09389E-05
305	4.09389E-05
306	5.45852E-05
307	6.82314E-05
308	1.36463E-05
309	8.18777E-05
311	4.09389E-05
312	1.36463E-05
313	5.45852E-05
314	1.36463E-05
315	1.36463E-05
316	8.18777E-05
317	5.45852E-05
318	5.45852E-05
319	4.09389E-05
320	4.09389E-05
321	2.72926E-05
322	4.09389E-05
323	1.36463E-05
324	5.45852E-05
326	8.18777E-05
327	1.36463E-05
328	4.09389E-05
329	2.72926E-05
330	2.72926E-05
331	1.36463E-05
332	4.09389E-05
333	4.09389E-05
336	5.45852E-05
337	2.72926E-05
338	1.36463E-05
339	2.72926E-05
340	4.09389E-05
341	5.45852E-05
343	1.36463E-05
344	4.09389E-05
345	2.72926E-05
347	2.72926E-05
349	1.36463E-05
350	1.36463E-05
351	5.45852E-05
352	2.72926E-05
353	1.36463E-05
356	2.72926E-05
357	1.36463E-05
358	4.09389E-05
359	5.45852E-05
360	6.82314E-05
361	1.36463E-05
362	1.36463E-05
363	4.09389E-05
364	2.72926E-05
365	4.09389E-05
366	1.36463E-05
367	1.36463E-05
368	1.36463E-05
369	2.72926E-05
370	1.36463E-05
372	4.09389E-05
374	2.72926E-05
375	1.36463E-05
376	1.36463E-05
377	4.09389E-05
378	1.36463E-05
380	1.36463E-05
381	5.45852E-05
382	6.82314E-05
383	2.72926E-05
384	4.09389E-05
385	1.36463E-05
386	2.72926E-05
387	1.36463E-05
389	5.45852E-05
390	2.72926E-05
391	2.72926E-05
392	1.36463E-05
393	1.36463E-05
394	2.72926E-05
395	2.72926E-05
396	2.72926E-05
397	2.72926E-05
398	4.09389E-05
399	2.72926E-05
400	1.36463E-05
405	2.72926E-05
406	4.09389E-05
407	1.36463E-05
408	4.09389E-05
410	6.82314E-05
412	2.72926E-05
413	5.45852E-05
414	1.36463E-05
415	1.36463E-05
416	2.72926E-05
417	2.72926E-05
418	1.36463E-05
420	1.36463E-05
424	1.36463E-05
425	2.72926E-05
426	1.36463E-05
427	1.36463E-05
428	2.72926E-05
430	1.36463E-05
433	5.45852E-05
435	1.36463E-05
436	4.09389E-05
437	2.72926E-05
438	2.72926E-05
439	2.72926E-05
440	2.72926E-05
441	2.72926E-05
442	1.36463E-05
445	4.09389E-05
446	2.72926E-05
447	2.72926E-05
449	1.36463E-05
450	2.72926E-05
452	1.36463E-05
453	1.36463E-05
454	2.72926E-05
456	1.36463E-05
457	1.36463E-05
460	1.36463E-05
463	1.36463E-05
464	1.36463E-05
465	1.36463E-05
468	1.36463E-05
469	4.09389E-05
470	4.09389E-05
473	1.36463E-05
476	1.36463E-05
478	4.09389E-05
479	5.45852E-05
480	1.36463E-05
483	1.36463E-05
488	1.36463E-05
489	2.72926E-05
492	2.72926E-05
493	1.36463E-05
494	1.36463E-05
497	1.36463E-05
500	2.72926E-05
501	1.36463E-05
504	2.72926E-05
505	1.36463E-05
508	1.36463E-05
510	1.36463E-05
511	2.72926E-05
513	1.36463E-05
516	1.36463E-05
518	2.72926E-05
519	1.36463E-05
521	1.36463E-05
524	2.72926E-05
525	2.72926E-05
527	1.36463E-05
530	1.36463E-05
533	1.36463E-05
535	2.72926E-05
536	5.45852E-05
537	2.72926E-05
538	1.36463E-05
539	1.36463E-05
540	1.36463E-05
541	1.36463E-05
543	1.36463E-05
544	1.36463E-05
545	2.72926E-05
554	1.36463E-05
555	1.36463E-05
556	1.36463E-05
558	1.36463E-05
559	1.36463E-05
560	1.36463E-05
565	2.72926E-05
566	1.36463E-05
568	1.36463E-05
569	1.36463E-05
570	1.36463E-05
577	1.36463E-05
578	1.36463E-05
584	1.36463E-05
589	1.36463E-05
591	2.72926E-05
597	1.36463E-05
598	1.36463E-05
601	1.36463E-05
604	1.36463E-05
605	1.36463E-05
607	1.36463E-05
609	1.36463E-05
614	1.36463E-05
618	1.36463E-05
619	1.36463E-05
624	1.36463E-05
626	5.45852E-05
628	1.36463E-05
629	1.36463E-05
630	1.36463E-05
633	1.36463E-05
637	1.36463E-05
641	1.36463E-05
647	1.36463E-05
649	1.36463E-05
657	1.36463E-05
658	1.36463E-05
662	2.72926E-05
663	1.36463E-05
664	8.18777E-05
666	1.36463E-05
668	1.36463E-05
670	5.45852E-05
680	1.36463E-05
685	1.36463E-05
695	1.36463E-05
699	1.36463E-05
701	2.72926E-05
728	1.36463E-05
759	1.36463E-05
774	1.36463E-05
782	1.36463E-05
785	1.36463E-05
796	1.36463E-05
812	1.36463E-05
818	1.36463E-05
820	1.36463E-05
828	1.36463E-05
836	1.36463E-05
837	1.36463E-05
855	1.36463E-05
856	1.36463E-05
859	1.36463E-05
878	1.36463E-05
895	1.36463E-05
906	1.36463E-05
909	2.72926E-05
942	1.36463E-05
956	1.36463E-05
995	1.36463E-05
1103	1.36463E-05
1170	1.36463E-05
1193	1.36463E-05
1349	1.36463E-05
1474	1.36463E-05
}{\adoptions}
\begin{tikzpicture}
    \begin{axis}[
    title=Adoption Distribution,
    clip=false,
    width=1.80in,
    height=1.75in,
    ymode=log,
    xmode=log,
    xlabel = {Libraries},
    ylabel = {$p(x)$},
    y label style={at={(axis description cs:0.1,.5)}},
    ]
    \addplot [only marks, mark=*, blue, mark size=0.75pt, fill opacity=0.5, draw opacity=0.2]  table [x=x, y=px]   {\adoptions};
    \node[anchor=north east] at (rel axis cs:-0.05,-0.01) {\textbf{(b)}};
    \end{axis}
\end{tikzpicture}
        \caption{\label{fig:adopt_dist}}
    \end{subfigure}
    \begin{subfigure}{0.48\textwidth}
\pgfplotstableread{
x  mean  median  sum  stdev  q1  q3  cnt  inf sup
0	6.605275739	2	1713250	29.53497286	0	6	259376	6.491610478	6.718941
1	3.318329405	0	805332	17.19877578	0	3	242692	3.249902661	3.386756149
2	1.136170136	0	251414	9.233688399	0	1	221282	1.097696942	1.17464333
3	0.821645815	0	166684	10.19798266	0	0	202866	0.777268022	0.866023609
4	0.658405966	0	123434	7.325564945	0	0	187474	0.625245038	0.691566893
5	0.558639336	0	97320	6.759705349	0	0	174209	0.526896294	0.590382377
6	0.491375231	0	79961	5.527671934	0	0	162729	0.464517715	0.518232748
7	0.471273165	0	72044	5.3867295	0	0	152871	0.44426975	0.49827658
8	0.418197568	0	60348	5.037670789	0	0	144305	0.39220524	0.444189895
9	0.420811034	0	57479	5.507455476	0	0	136591	0.391603424	0.450018645
10	0.384562594	0	49872	4.745755664	0	0	129685	0.35873306	0.410392128
11	0.337856767	0	41727	3.767128541	0	0	123505	0.316846855	0.358866679
12	0.360991946	0	42579	4.902560797	0	0	117950	0.333013081	0.38897081
13	0.330652447	0	37330	4.395977501	0	0	112898	0.305009472	0.356295422
14	0.319436093	0	34532	4.195063877	0	0	108103	0.294428277	0.34444391
15	0.312781651	0	32483	3.602763282	0	0	103852	0.290869532	0.334693769
16	0.289635153	0	28936	3.196735731	0	0	99905	0.269812162	0.309458145
17	0.272603607	0	26252	2.629678632	0	0	96301	0.255994613	0.289212602
18	0.290545337	0	27012	3.590406153	0	0	92970	0.267465739	0.313624935
19	0.273420965	0	24545	3.363360898	0	0	89770	0.251418875	0.295423054
20	0.271911822	0	23584	3.536072314	0	0	86734	0.248378539	0.295445106
21	0.263552136	0	22131	3.37352611	0	0	83972	0.240734382	0.286369891
22	0.241249585	0	19623	3.110640375	0	0	81339	0.219872107	0.262627063
23	0.267887677	0	21150	4.600731017	0	0	78951	0.235795137	0.299980218
24	0.250117481	0	19161	3.763756616	0	0	76608	0.223464822	0.27677014
25	0.251926226	0	18768	3.602672027	0	0	74498	0.226055508	0.277796945
26	0.235817805	0	17085	4.010727597	0	0	72450	0.206612615	0.265022996
27	0.244785981	0	17265	3.812491939	0	0	70531	0.216649162	0.272922799
28	0.232965465	0	16001	3.508044865	0	0	68684	0.206729719	0.259201211
29	0.214549312	0	14360	3.008673203	0	0	66931	0.191755475	0.237343149
30	0.226198877	0	14745	3.009209974	0	0	65186	0.203097844	0.24929991
31	0.236097791	0	15017	4.24811332	0	0	63605	0.203083153	0.269112429
32	0.21187313	0	13173	2.69106621	0	0	62174	0.190719931	0.233026329
33	0.242957283	0	14765	4.611932631	0	0	60772	0.20628928	0.279625286
34	0.20688785	0	12297	3.023498101	0	0	59438	0.18258072	0.231194979
35	0.20466397	0	11892	3.231141325	0	0	58105	0.178391236	0.230936705
36	0.192855886	0	10960	2.763428174	0	0	56830	0.170135515	0.215576257
37	0.208254048	0	11601	3.338987236	0	0	55706	0.180525965	0.235982131
38	0.199105719	0	10865	3.05576134	0	0	54569	0.173466635	0.224744803
39	0.19606669	0	10478	3.139763793	0	0	53441	0.169446218	0.222687163
40	0.203624835	0	10662	2.833685436	0	0	52361	0.179352937	0.227896734
41	0.18609634	0	9554	2.793224026	0	0	51339	0.161934047	0.210258633
42	0.167129924	0	8418	1.847318429	0	0	50368	0.150996725	0.183263122
43	0.17015728	0	8417	2.106908858	0	0	49466	0.151589994	0.188724566
44	0.192287092	0	9334	2.618275449	0	0	48542	0.168994781	0.215579402
45	0.201623561	0	9612	4.26298649	0	0	47673	0.163355735	0.239891386
46	0.159187943	0	7457	1.264518921	0	0	46844	0.147736653	0.170639233
47	0.167958656	0	7735	2.210640437	0	0	46053	0.147768243	0.18814907
48	0.182268406	0	8244	3.245171516	0	0	45230	0.152360884	0.212175928
49	0.18701363	0	8315	6.030886983	0	0	44462	0.130954963	0.243072296
50	0.151960336	0	6651	1.048847099	0	0	43768	0.142134039	0.161786634
51	0.1681987	0	7246	2.12316463	0	0	43080	0.148149276	0.188248124
52	0.158836161	0	6731	2.816898036	0	0	42377	0.132015956	0.185656366
53	0.159097449	0	6635	2.370112255	0	0	41704	0.13634982	0.181845077
54	0.160984525	0	6606	1.926570488	0	0	41035	0.142343759	0.179625291
55	0.161014012	0	6504	2.697277766	0	0	40394	0.13470992	0.187318104
56	0.151122495	0	6018	2.35801967	0	0	39822	0.127962313	0.174282677
57	0.141142595	0	5539	1.145571203	0	0	39244	0.129808378	0.152476812
58	0.186837067	0	7222	3.060786449	0	0	38654	0.156323578	0.217350557
59	0.159364746	0	6071	2.711258975	0	0	38095	0.132138166	0.186591326
60	0.181585746	0	6818	3.638108901	0	0	37547	0.144786057	0.218385435
61	0.141274687	0	5229	0.95403046	0	0	37013	0.131555249	0.150994126
62	0.174931507	0	6385	3.60975111	0	0	36500	0.137898677	0.211964337
63	0.142841257	0	5138	2.190484652	0	0	35970	0.120203882	0.165478632
64	0.176097437	0	6246	4.298794341	0	0	35469	0.131359262	0.220835613
65	0.152913939	0	5350	1.796471665	0	0	34987	0.134089455	0.171738424
66	0.147344953	0	5089	1.880400798	0	0	34538	0.127513346	0.16717656
67	0.135570825	0	4617	1.086588722	0	0	34056	0.124030329	0.14711132
68	0.134609099	0	4530	1.165402013	0	0	33653	0.122157648	0.147060549
69	0.153922453	0	5121	3.474847416	0	0	33270	0.116583218	0.191261687
70	0.152974504	0	5022	2.611232914	0	0	32829	0.12472747	0.181221538
71	0.158321514	0	5135	3.359133351	0	0	32434	0.121763461	0.194879568
72	0.131320378	0	4210	1.214086268	0	0	32059	0.118030205	0.144610551
73	0.137053755	0	4342	2.148747987	0	0	31681	0.113392263	0.160715246
74	0.120246693	0	3763	1.185009365	0	0	31294	0.107117219	0.133376166
75	0.132645804	0	4103	1.920698898	0	0	30932	0.111240995	0.154050613
76	0.17775597	0	5434	4.830208596	0	0	30570	0.12360899	0.23190295
77	0.124871684	0	3771	0.916568409	0	0	30199	0.114533967	0.135209402
78	0.120726567	0	3609	0.913157281	0	0	29894	0.110374916	0.131078219
79	0.149690471	0	4425	2.864428937	0	0	29561	0.117036608	0.182344333
80	0.145289793	0	4249	2.894979847	0	0	29245	0.112109837	0.178469749
81	0.196398327	0	5682	6.243534043	0	0	28931	0.124452629	0.268344025
82	0.123846477	0	3543	1.446233101	0	0	28608	0.107087379	0.140605574
83	0.132996037	0	3759	1.181412498	0	0	28264	0.119222648	0.146769427
84	0.106278637	0	2969	0.698280153	0	0	27936	0.09809015	0.114467124
85	0.151872625	0	4197	1.476694172	0	0	27635	0.134461899	0.169283352
86	0.150498375	0	4122	3.868364356	0	0	27389	0.104684676	0.196312074
87	0.136697553	0	3703	3.718076289	0	0	27089	0.092420584	0.180974521
88	0.127797672	0	3426	1.102041004	0	0	26808	0.114605339	0.140990006
89	0.158785184	0	4214	4.083609763	0	0	26539	0.10965391	0.207916458
90	0.124515246	0	3275	2.383257929	0	0	26302	0.095712577	0.153317915
91	0.158655139	0	4129	4.628425363	0	0	26025	0.102421788	0.214888491
92	0.131175535	0	3380	3.228125629	0	0	25767	0.091759351	0.170591718
93	0.124700898	0	3179	0.946009998	0	0	25493	0.113087981	0.136313815
94	0.149787942	0	3779	4.662236819	0	0	25229	0.092257146	0.207318738
95	0.120241764	0	3004	1.175964335	0	0	24983	0.105659401	0.134824128
96	0.145133889	0	3588	2.277429765	0	0	24722	0.11674429	0.173523488
97	0.140866936	0	3448	1.804044731	0	0	24477	0.118266112	0.163467761
98	0.154223523	0	3741	4.795380423	0	0	24257	0.093875841	0.214571205
99	0.126023866	0	3031	1.398619241	0	0	24051	0.108347662	0.14370007
100	0.131880589	0	3141	2.833949581	0	0	23817	0.095888699	0.16787248
}{\table}
\begin{tikzpicture}
 \begin{groupplot}[
    group style={%
        group size=1 by 2,%
        x descriptions at=edge bottom,%
        vertical sep=0pt,%
    },
    clip=true,
    clip mode=individual,
    width=3.3in,
    xmin=0,
    xmax=100,
    ymode=log,
    legend cell align=left,
    legend pos=outer north east,
    legend style={draw=none}
    ]
    \nextgroupplot[
        title=Library Adoptions per commit,
        xticklabels={,,}, 
        height=1.5in, 
        ymin=0.05,
        y label style={at={(axis description cs:0.05,.5)}},
        ylabel = {avg adopts},
        ]
    \addplot [stack plots=y, fill=none, draw=none, forget plot]   table [x=x, y=inf]   {\table} \closedcycle;
    \addplot [stack plots=y, fill=gray!70, opacity=0.6, draw opacity=1, thin, smooth, area legend]   table [x=x, y expr=\thisrow{sup}-\thisrow{inf}]   {\table} \closedcycle;
    \addplot [stack plots=false, blue, thick, smooth]  table [x=x, y=mean]   {\table};
    \nextgroupplot[
        ybar, 
        bar width=0.8pt, 
        ymin=100,
        ymode=log,
        height=0.85in, 
        y label style={at={(axis description cs:0.05,.5)}},
        xlabel = {Commit \#},
        ylabel = {vol},
    ]
    \addplot []  table [x=x, y=sum]   {\table};
    \node[anchor=north east] at (rel axis cs:-0.05,-0.05) {\textbf{(c)}};
    \end{groupplot}
\end{tikzpicture}        
        \caption{\label{fig:adopt_per_commit}}
    \end{subfigure}
    
    \caption{(a) Number of commits per project follows a shifted power law distribution. (b) Number of libraries used per project also follows a shifted power law distribution. (c) When libraries are adopted into repositories, it tends to occur early in the repository history.}
    \label{fig:dists}
\end{figure*}
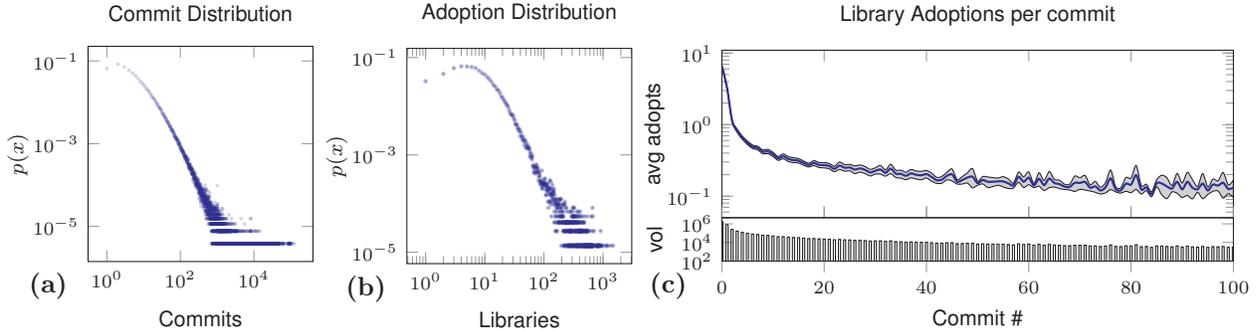

We combed through each Git commit log to find which libraries had been imported by using Regex pattern matching to identify import statements, such as `from $\ell$ import $f$" and `import $\ell$ as $f$", where $\ell$ is a library and $f$ is a function. We searched the log to find lines which referenced the functions contained in a library import using pattern matching to search for the libraries $\ell$ and functions $f$, along with indicator characters such as \texttt{.} and \texttt{(}. We gathered the author name, library used, and commit type.

Formally, for a project $p$ and a time $t$, we define a \textit{library adoption} to be an event ($p$, $\ell$, $t$) representing the first time $t$ that $\ell$ is found in $p$. 
The distribution of commit-activity per project, illustrated in Fig.~\ref{fig:commit_dist}, resembles a shifted power law distribution. Because of this dynamic, 50\% of projects were found to have 10 or fewer commits (\ie, median of 10) and 90\% of projects have 100 or fewer commits.

The distribution of the number of libraries adopted per project, illustrated in Fig.~\ref{fig:adopt_dist}, also resembles a shifted power law distribution, albeit with a larger offset than the commit-activity distribution of Fig.~\ref{fig:commit_dist}. However, the number of adoptions is less evenly distributed: 54\% of projects adopted 10 or fewer distinct libraries and 98\% of projects adopted 100 or fewer libraries.

Across all commits of all projects, we find that library adoptions occur more frequently within the first few commits. Figure~\ref{fig:adopt_per_commit} shows that a project's first commit adopts 6.4 libraries on average (with a median of 2 not illustrated). A project's second through fifth commits adopt 3.3, 1.1, 0.8, and 0.65 libraries on average (with median values of 0 not illustrated). In general, the average number of adoptions per commit appears to follow a Zipfian decay.


\section{Activity after Adoption}

%
%

\begin{figure*}[t]
    \centering
    \begin{minipage}{0.23\textwidth}
        \pgfplotstableread{
x  mean  ci std len
1	1.651243319	0.268344826	124.3277306	824607
2	1.153490169	0.210896099	92.00373456	731091
3	0.859202326	0.203605688	84.67123565	664339
4	0.845469499	0.357550317	142.2404817	607953
5	0.592620728	0.324996396	124.4607879	563386
6	0.545660992	0.133491075	49.28309535	523588
7	0.958971715	0.53884848	192.418113	489843
8	0.612923462	0.604793682	209.0440725	458944
9	0.29140521	0.086978782	29.19234654	432723
10	0.329555317	0.193305189	63.14599522	409924
11	0.24744261	0.068060012	21.67601108	389649
12	0.257003862	0.15565089	48.30966897	370053
13	0.22087338	0.100502264	30.43465659	352277
14	0.160015077	0.042064247	12.44785979	336405
15	0.192968426	0.067738068	19.59923281	321597
16	0.193829651	0.107885226	30.54580517	307949
17	0.156632105	0.082862458	22.96657436	295105
18	0.151560542	0.038463885	10.4474618	283410
19	0.168961931	0.081880358	21.81353701	272642
20	0.170350674	0.149971232	39.17684697	262146

}{\pos}

\pgfplotstableread{
x  mean  ci std len
1	-3.019223195	1.037067792	480.4873152	824607
2	-1.220897439	0.185654728	80.99214888	731091
3	-0.768173507	0.171583405	71.35448453	664339
4	-0.636285262	0.163460274	65.02768156	607953
5	-0.666687904	0.447195715	171.2583023	563386
6	-0.399717351	0.100645653	37.15701069	523588
7	-0.725532819	0.402519277	143.7361383	489843
8	-0.502135465	0.432424618	149.4655218	458944
9	-0.258146222	0.124241371	41.69864257	432723
10	-0.189622433	0.064941517	21.21410588	409924
11	-0.18394583	    0.060704007	19.33324281	389649
12	-0.23826764	    0.146337754	45.41913276	370053
13	-0.161422978	0.072489645	21.95171907	352277
14	-0.138586022	0.067127551	19.86471635	336405
15	-0.142471116	0.056043959	16.21567653	321597
16	-0.120249894	0.051523314	14.58792056	307949
17	-0.143031603	0.064510151	17.87995693	295105
18	-0.120478455	0.141609999	38.46374422	283410
19	-0.123048765	0.075244055	20.04557661	272642
20	-0.169176235	0.215951944	56.41292785	262146

}{\neg}

\pgfplotstableread{
x  mean
1	-1.367979876
2	-0.06740727
3	0.091028819
4	0.209184237
5	-0.074067176
6	0.145943641
7	0.233438896
8	0.110787997
9	0.033258988
10	0.139932884
11	0.063496781
12	0.018736222
13	0.059450402
14	0.021429055
15	0.05049731
16	0.073579758
17	0.013600502
18	0.031082088
19	0.045913166
20	0.001174439

}{\mid}

\begin{tikzpicture}

\pgfplotsset{
every axis legend/.append style={
at={(0.5,-0.650)},
anchor=north
},
}
 
 \begin{groupplot}[
    group style={%
        group size=1 by 2,%
        x descriptions at=edge bottom,%
        vertical sep=0pt,%
    },
    clip=true,
    clip mode=individual,
    width=1.75in,
    xmin=0,
    xmax=20,
    legend columns=3,
    legend style={draw=none}
    ]
    \nextgroupplot[
        title = Team Size 1,
        xticklabels={,,}, 
        height=1.5in, 
        ymax=2.2,
        ymin=-4,
        y label style={at={(axis description cs:0.08,.5)}},
        ylabel = {avg use (LOC$_\ell$)},
        yticklabels={,,-2,0,2}
        ]
    \addplot [stack plots=y, fill=none, draw=none, forget plot]   table [x=x, y expr=(\thisrow{mean} + \thisrow{ci})]   {\pos} \closedcycle;
    \addplot [stack plots=y, fill=gray!70, opacity=0.6, draw opacity=1, thin, smooth, area legend]   table [x=x, y expr=(\thisrow{mean} - \thisrow{ci}) - (\thisrow{mean} + \thisrow{ci}) ]   {\pos} \closedcycle;
    \addplot [stack plots=y, stack dir=minus, forget plot, draw=none] table [x=x, y expr = (\thisrow{mean} - \thisrow{ci})] {\pos};

    \addplot [stack plots=y, fill=none, draw=none, forget plot]   table [x=x, y expr=(\thisrow{mean} + \thisrow{ci})]   {\neg} \closedcycle;
    \addplot [stack plots=y, fill=gray!70, opacity=0.6, draw opacity=1, thin, smooth, area legend]   table [x=x,y expr=(\thisrow{mean} - \thisrow{ci}) - (\thisrow{mean} + \thisrow{ci}) ]   {\neg} \closedcycle;
    \addplot [stack plots=y, stack dir=minus, forget plot, draw=none] table [x=x, y expr = (\thisrow{mean} - \thisrow{ci})] {\neg};
    
    \addplot [stack plots=false, green, thick, smooth]  table [x=x, y=mean]
    {\pos};
    \addplot [stack plots=false, red, thick, smooth]  table [x=x, y=mean]   
    {\neg};
    
    \draw[ultra thin] (axis cs:\pgfkeysvalueof{/pgfplots/xmin},0) -- (axis cs:\pgfkeysvalueof{/pgfplots/xmax},0);
    
    \addplot [stack plots=false, black, thick, smooth]  table [x=x, y=mean]   
    {\mid};\legend{}
    
    \nextgroupplot[
        ybar, 
        bar width=2pt, 
        ymode=log,
        height=0.95in, 
        y label style={at={(axis description cs:0.08,.5)}},
        ylabel = {vol},
        ymax=1000000,
        ymin=100000,
        xticklabels={,0,5,10,15},
    ]
    \addplot []  table [x=x, y=len]   {\pos};
    \end{groupplot}
\end{tikzpicture}
    \end{minipage}
    \begin{minipage}{0.17\textwidth}
        \pgfplotstableread{
x  mean  ci std len
1	2.032319155	0.390489148	144.0994078	523124
2	1.405734328	0.211513512	74.65391497	478550
3	1.19091508	0.303706364	102.91952	441150
4	0.843524292	0.164701502	54.01549606	413181
5	1.049010875	0.438143063	139.1126715	387258
6	0.729374049	0.313571174	96.78039564	365933
7	0.735853468	0.321403713	96.76260802	348187
8	0.627459823	0.194976046	57.23694008	331047
9	0.822568942	0.430662961	123.2008865	314378
10	0.396675389	0.199320836	55.68490723	299826
11	0.501847419	0.261122479	71.44219734	287555
12	0.370474706	0.116694921	31.30837487	276514
13	0.433413131	0.275542234	72.36399275	264953
14	0.424682663	0.21634018	55.77621183	255343
15	0.263675166	0.09076918	22.98185734	246260
16	0.31461233	0.192380559	47.87100008	237861
17	0.289943585	0.094040461	22.99555631	229699
18	0.396114018	0.286053622	68.77245987	222042
19	0.339614186	0.115063511	27.2254449	215068
20	0.161675574	0.029095274	6.777355727	208438

}{\pos}

\pgfplotstableread{
x  mean  ci std len
1	-2.28571243	0.41616334	153.5737704	523124
2	-1.52653101	0.217528745	76.77699748	478550
3	-1.30500965	0.259847234	88.05660908	441150
4	-0.808520766	0.268092976	87.92375827	413181
5	-0.626034592	0.104346962	33.13069606	387258
6	-0.696400735	0.237177896	73.20242586	365933
7	-0.531818097	0.23377678	70.38142363	348187
8	-0.444286085	0.154829675	45.45161833	331047
9	-0.56727072	0.253591731	72.54565383	314378
10	-0.48255391	0.257368661	71.90191606	299826
11	-0.188362469	0.028945017	7.919255562	287555
12	-0.261334113	0.113762551	30.52164181	276514
13	-0.295431095	0.14529487	38.15791415	264953
14	-0.239308911	0.19490955	50.25102761	255343
15	-0.307661545	0.185976598	47.08743263	246260
16	-0.288803214	0.224145662	55.77526697	237861
17	-0.21094069	0.183189545	44.79503245	229699
18	-0.234517314	0.173795418	41.78355911	222042
19	-0.181266934	0.084952093	20.1007123	215068
20	-0.180976506	0.293452808	68.35591561	208438

}{\neg}

\pgfplotstableread{
x  mean
1	-0.253393275
2	-0.120796682
3	-0.11409457
4	0.035003526
5	0.422976283
6	0.032973314
7	0.204035371
8	0.183173737
9	0.255298222
10	-0.085878521
11	0.31348495
12	0.109140593
13	0.137982036
14	0.185373752
15	-0.043986379
16	0.025809116
17	0.079002895
18	0.161596704
19	0.158347252
20	-0.019300932

}{\mid}

\begin{tikzpicture}

\pgfplotsset{
every axis legend/.append style={
at={(0.5,-0.650)},
anchor=north
},
}
 
 \begin{groupplot}[
    group style={%
        group size=1 by 2,%
        x descriptions at=edge bottom,%
        vertical sep=0pt,%
    },
    clip=true,
    clip mode=individual,
    width=1.75in,
    xmin=0,
    xmax=20,
    legend columns=3,
    legend style={draw=none}
    ]
    \nextgroupplot[
        title = Team Size 2,
        xticklabels={,,}, 
        height=1.5in, 
        ymax=2.2,
        ymin=-4,
        y label style={at={(axis description cs:0.08,.5)}},
        yticklabels={,,},
        ]
    \addplot [stack plots=y, fill=none, draw=none, forget plot]   table [x=x, y expr=(\thisrow{mean} + \thisrow{ci})]   {\pos} \closedcycle;
    \addplot [stack plots=y, fill=gray!70, opacity=0.6, draw opacity=1, thin, smooth, area legend]   table [x=x, y expr=(\thisrow{mean} - \thisrow{ci}) - (\thisrow{mean} + \thisrow{ci}) ]   {\pos} \closedcycle;
    \addplot [stack plots=y, stack dir=minus, forget plot, draw=none] table [x=x, y expr = (\thisrow{mean} - \thisrow{ci})] {\pos};

    \addplot [stack plots=y, fill=none, draw=none, forget plot]   table [x=x, y expr=(\thisrow{mean} + \thisrow{ci})]   {\neg} \closedcycle;
    \addplot [stack plots=y, fill=gray!70, opacity=0.6, draw opacity=1, thin, smooth, area legend]   table [x=x,y expr=(\thisrow{mean} - \thisrow{ci}) - (\thisrow{mean} + \thisrow{ci}) ]   {\neg} \closedcycle;
    \addplot [stack plots=y, stack dir=minus, forget plot, draw=none] table [x=x, y expr = (\thisrow{mean} - \thisrow{ci})] {\neg};
    
    \addplot [stack plots=false, green, thick, smooth]  table [x=x, y=mean]
    {\pos};
    \addplot [stack plots=false, red, thick, smooth]  table [x=x, y=mean]   
    {\neg};
    
    \draw[ultra thin] (axis cs:\pgfkeysvalueof{/pgfplots/xmin},0) -- (axis cs:\pgfkeysvalueof{/pgfplots/xmax},0);
    
    \addplot [stack plots=false, black, thick, smooth]  table [x=x, y=mean]   
    {\mid};\legend{}
    
    \nextgroupplot[
        ybar, 
        bar width=2pt, 
        ymode=log,
        height=0.95in, 
        ymax=1000000,
        ymin=100000,
        yticklabels={,,},
        xticklabels={,0,5,10,15},
    ]
    \addplot []  table [x=x, y=len]   {\pos};
    \end{groupplot}
\end{tikzpicture}
    \end{minipage}
    \begin{minipage}{0.17\textwidth}
        \pgfplotstableread{
x  mean  ci std len
1	1.929151	0.332327057	119.6493369	497952
2	1.402172379	0.216374955	75.75029227	470818
3	1.407359181	0.38379334	131.188097	448841
4	0.956451968	0.18120385	60.76246089	431951
5	1.052755592	0.389524024	128.2198413	416237
6	0.781025128	0.277145313	89.87104709	403946
7	0.641322247	0.140513969	44.82142703	390869
8	1.339085144	0.616543453	193.7219026	379253
9	0.724237334	0.251009178	77.81794458	369215
10	0.859390478	0.307410666	93.92918508	358643
11	0.667341455	0.174888026	52.67807978	348529
12	0.589870031	0.182939249	54.44278929	340225
13	0.393521774	0.119062476	35.02846136	332500
14	0.582539333	0.1006909	29.25498801	324279
15	0.360437629	0.132081504	37.86210685	315664
16	0.553248332	0.2436437	68.96718771	307804
17	0.58736119	0.314332604	87.91637578	300511
18	0.426870409	0.143010747	39.46750026	292578
19	0.283243411	0.064960025	17.7278448	286101
20	0.261339057	0.048572007	13.10275791	279546

}{\pos}

\pgfplotstableread{
x  mean  ci std len
1	-2.473264252	0.450608028	162.2346134	497952
2	-2.642127161	0.651327469	228.0219819	470818
3	-1.357298082	0.319545701	109.2269931	448841
4	-1.056686067	0.253839509	85.11912541	431951
5	-1.018589327	0.36548087	120.3055431	416237
6	-1.024301295	0.362039931	117.400173	403946
7	-1.268517165	0.46158093	147.2360086	390869
8	-0.926111907	0.429499491	134.9514917	379253
9	-0.345378227	0.074466527	23.08613627	369215
10	-0.562055342	0.197698727	60.40675345	358643
11	-0.88808892	0.33603274	101.2165324	348529
12	-0.325775445	0.143899462	42.82453389	340225
13	-0.42164673	0.204894026	60.28030618	332500
14	-0.424273829	0.15590397	45.2967324	324279
15	-0.54853456	0.239534194	68.66418824	315664
16	-0.301133365	0.075040559	21.24141239	307804
17	-0.266106236	0.063168846	17.66783311	300511
18	-0.478598559	0.142411706	39.30217942	292578
19	-0.342197989	0.176669872	48.21389913	286101
20	-0.347000772	0.161187188	43.48176743	279546

}{\neg}

\pgfplotstableread{
x  mean
1	-0.544113253
2	-1.239954782
3	0.0500611
4	-0.100234099
5	0.034166265
6	-0.243276167
7	-0.627194918
8	0.412973237
9	0.378859107
10	0.297335136
11	-0.220747464
12	0.264094586
13	-0.028124956
14	0.158265505
15	-0.188096931
16	0.252114967
17	0.321254954
18	-0.05172815
19	-0.058954578
20	-0.085661715

}{\mid}

\begin{tikzpicture}

\pgfplotsset{
every axis legend/.append style={
at={(0.5,-0.650)},
anchor=north
},
}
 
 \begin{groupplot}[
    group style={%
        group size=1 by 2,%
        x descriptions at=edge bottom,%
        vertical sep=0pt,%
    },
    clip=true,
    clip mode=individual,
    width=1.75in,
    xmin=0,
    xmax=20,
    legend columns=3,
    legend style={draw=none}
    ]
    \nextgroupplot[
        title = Team Size 3-5,
        xticklabels={,,}, 
        height=1.5in, 
        ymax=2.2,
        ymin=-4,
        y label style={at={(axis description cs:0.08,.5)}},
        yticklabels={,,},
        ]
    \addplot [stack plots=y, fill=none, draw=none, forget plot]   table [x=x, y expr=(\thisrow{mean} + \thisrow{ci})]   {\pos} \closedcycle;
    \addplot [stack plots=y, fill=gray!70, opacity=0.6, draw opacity=1, thin, smooth, area legend]   table [x=x, y expr=(\thisrow{mean} - \thisrow{ci}) - (\thisrow{mean} + \thisrow{ci}) ]   {\pos} \closedcycle;
    \addplot [stack plots=y, stack dir=minus, forget plot, draw=none] table [x=x, y expr = (\thisrow{mean} - \thisrow{ci})] {\pos};

    \addplot [stack plots=y, fill=none, draw=none, forget plot]   table [x=x, y expr=(\thisrow{mean} + \thisrow{ci})]   {\neg} \closedcycle;
    \addplot [stack plots=y, fill=gray!70, opacity=0.6, draw opacity=1, thin, smooth, area legend]   table [x=x,y expr=(\thisrow{mean} - \thisrow{ci}) - (\thisrow{mean} + \thisrow{ci}) ]   {\neg} \closedcycle;
    \addplot [stack plots=y, stack dir=minus, forget plot, draw=none] table [x=x, y expr = (\thisrow{mean} - \thisrow{ci})] {\neg};
    
    \addplot [stack plots=false, green, thick, smooth]  table [x=x, y=mean]
    {\pos};
    \addplot [stack plots=false, red, thick, smooth]  table [x=x, y=mean]   
    {\neg};
    
    \draw[ultra thin] (axis cs:\pgfkeysvalueof{/pgfplots/xmin},0) -- (axis cs:\pgfkeysvalueof{/pgfplots/xmax},0);
    
    \addplot [stack plots=false, black, thick, smooth]  table [x=x, y=mean]   
    {\mid};
    \nextgroupplot[
        ybar, 
        bar width=2pt, 
        ymode=log,
        height=0.95in, 
        ymax=1000000,
        ymin=100000,
        yticklabels={,,},
        xticklabels={,0,5,10,15},
    ]
    \addplot []  table [x=x, y=len]   {\pos};
        
    \end{groupplot}
\end{tikzpicture}
    \end{minipage}
    \begin{minipage}{0.17\textwidth}
        \pgfplotstableread{
x  mean  ci std len
1	1.050850567	0.282333165	57.54468068	159584
2	0.750941434	0.098050972	19.7043593	155140
3	0.649013132	0.204355948	40.45898851	150577
4	0.553873642	0.0950642	18.60978549	147215
5	0.636505815	0.124343764	24.04350965	143632
6	0.64472753	0.16657971	31.89329309	140818
7	0.583553948	0.186334703	35.36437309	138372
8	0.730924484	0.304116402	57.20694271	135932
9	0.976807074	0.299351178	55.79806588	133469
10	0.99260832	1.360276141	251.8535629	131688
11	0.820900282	0.990993443	182.3625256	130087
12	0.799996505	0.377335269	68.86906641	127967
13	0.302849204	0.094573285	17.16167896	126499
14	0.561322115	0.192976047	34.78718491	124835
15	0.650074703	0.272893515	48.85595475	123127
16	0.331428621	0.073441384	13.05466031	121382
17	1.102919976	0.585756189	103.1548223	119138
18	0.42503473	0.239233942	41.82960838	117443
19	0.448454824	0.207425178	35.96648037	115499
20	0.494984841	0.233972036	40.32965296	114137

}{\pos}
\pgfplotstableread{
x  mean  ci std len
1	-2.241859104	0.456181256	92.97811233	159584
2	-1.61134390	    0.699801245	1171.150839	155140 
3	-1.216608564	0.364864707	72.23698227	150577
4	-0.838829039	0.19524087	38.22038917	147215
5	-0.926458896	0.236827624	45.79375028	143632
6	-0.589897163	0.125978828	24.11986233	140818
7	-0.650568356	0.359070333	68.14778484	138372
8	-0.74024272	0.590623431	111.1014092	135932
9	-0.363682035	0.188532949	35.14191581	133469
10	-0.349304955	0.123288772	22.8267743	131688
11	-0.545289333	0.245796252	45.23140452	130087
12	-0.532926179	0.339213063	61.91122022	127967
13	-0.549146811	1.00027142	181.5135957	126499
14	-0.398141595	0.152384289	27.46983651	124835
15	-0.431373577	0.442970462	79.30472375	123127
16	-1.165552672	1.482325016	263.492441	121382
17	-0.611259644	0.248352408	43.73619784	119138
18	-0.403864468	0.44033666	76.99204319	117443
19	-0.488043127	0.378021395	65.54700458	115499
20	-0.511196736	0.501939266	86.51904188	114137

}{\neg}
\pgfplotstableread{
x  mean
1	-1.191008536
2	-0.8604 
3	-0.567595432
4	-0.284955397
5	-0.289953081
6	0.054830367
7	-0.067014408
8	-0.009318235
9	0.613125038
10	0.643303365
11	0.275610949
12	0.267070326
13	-0.246297607
14	0.16318052
15	0.218701126
16	-0.834124051
17	0.491660332
18	0.021170262
19	-0.039588303
20	-0.016211895

}{\mid}

\begin{tikzpicture}

\pgfplotsset{
every axis legend/.append style={
at={(0.5,-0.650)},
anchor=north
},
}
 
 \begin{groupplot}[
    group style={%
        group size=1 by 2,%
        x descriptions at=edge bottom,%
        vertical sep=0pt,%
    },
    clip=true,
    clip mode=individual,
    width=1.75in,
    xmin=0,
    xmax=20,
    legend columns=3,
    legend style={draw=none}
    ]
    \nextgroupplot[
        title = Team Size 6-9,
        xticklabels={,,}, 
        height=1.5in, 
        ymax=2.2,
        ymin=-4,
        y label style={at={(axis description cs:0.08,.5)}},
        yticklabels={,,},
        ]
    \addplot [stack plots=y, fill=none, draw=none, forget plot]   table [x=x, y expr=(\thisrow{mean} + \thisrow{ci})]   {\pos} \closedcycle;
    \addplot [stack plots=y, fill=gray!70, opacity=0.6, draw opacity=1, thin, smooth, area legend]   table [x=x, y expr=(\thisrow{mean} - \thisrow{ci}) - (\thisrow{mean} + \thisrow{ci}) ]   {\pos} \closedcycle;
    \addplot [stack plots=y, stack dir=minus, forget plot, draw=none] table [x=x, y expr = (\thisrow{mean} - \thisrow{ci})] {\pos};

    \addplot [stack plots=y, fill=none, draw=none, forget plot]   table [x=x, y expr=(\thisrow{mean} + \thisrow{ci})]   {\neg} \closedcycle;
    \addplot [stack plots=y, fill=gray!70, opacity=0.6, draw opacity=1, thin, smooth, area legend]   table [x=x,y expr=(\thisrow{mean} - \thisrow{ci}) - (\thisrow{mean} + \thisrow{ci}) ]   {\neg} \closedcycle;
    \addplot [stack plots=y, stack dir=minus, forget plot, draw=none] table [x=x, y expr = (\thisrow{mean} - \thisrow{ci})] {\neg};
    
    \addplot [stack plots=false, green, thick, smooth]  table [x=x, y=mean]
    {\pos};
    \addplot [stack plots=false, red, thick, smooth]  table [x=x, y=mean]   
    {\neg};
    
    \draw[ultra thin] (axis cs:\pgfkeysvalueof{/pgfplots/xmin},0) -- (axis cs:\pgfkeysvalueof{/pgfplots/xmax},0);
    
    \addplot [stack plots=false, black, thick, smooth]  table [x=x, y=mean]   
    {\mid};\legend{}
    
    \nextgroupplot[
        ybar, 
        bar width=2pt, 
        ymode=log,
        height=0.95in, 
        ymax=1000000,
        ymin=100000,
        yticklabels={,,},
        xticklabels={,0,5,10,15},
    ]
    \addplot []  table [x=x, y=len]   {\pos};
    \end{groupplot}
\end{tikzpicture}
    \end{minipage}
    \begin{minipage}{0.17\textwidth}
        \pgfplotstableread{
x  mean  ci std len
1	1.637586069	0.351649057	111.6758934	387435
2	0.65854964	0.163973665	51.04189465	372224
3	0.341643807	0.029071153	8.93317118	362732
4	0.373245521	0.047925554	14.5680412	354950
5	0.473596635	0.05171692	15.57389312	348360
6	0.327974506	0.0617564	18.44141882	342550
7	0.426191882	0.21921352	65.01175089	337869
8	0.542986011	0.659301851	194.1221087	333028
9	0.292023865	0.082856886	24.22818173	328461
10	0.22722777	0.042008573	12.2200786	325066
11	0.258817848	0.064472764	18.64374935	321228
12	0.236628142	0.038664269	11.11580227	317513
13	0.195349703	0.028153691	8.053347104	314327
14	0.160452114	0.0272595	7.754491512	310864
15	0.210005149	0.043776243	12.39117273	307785
16	0.227892472	0.081097213	22.82593061	304330
17	0.298182178	0.037582866	10.51430997	300664
18	0.331939447	0.040715706	11.27226628	294441
19	0.228346273	0.141564887	38.95480948	290878
20	0.258880391	0.065428701	17.90339807	287630

}{\pos}

\pgfplotstableread{
x  mean  ci std len
1	-1.292118407	0.260338105	82.67757248	387435
2	-0.888578759	0.164608251	51.23942915	372224
3	-0.393580752	0.052277554	16.06418334	362732
4	-0.505307436	0.169555814	51.5402719	354950
5	-0.455338416	0.069717101	20.99441887	348360
6	-0.518664021	0.156759659	46.8108653	342550
7	-0.23761865	0.080257004	23.8016723	337869
8	-0.403392744	0.654759317	192.7846238	333028
9	-0.128770403	0.016772387	4.904413624	328461
10	-0.130079444	0.020998326	6.108305346	325066
11	-0.242047573	0.086758298	25.08811268	321228
12	-0.165817729	0.03820459	10.98364665	317513
13	-0.113360077	0.017988988	5.145739632	314327
14	-0.201265116	0.084573872	24.05867191	310864
15	-0.086854897	0.013227282	3.744074978	307785
16	-0.308135256	0.122407248	34.45321068	304330
17	-0.209026277	0.033888981	9.480896203	300664
18	-0.294902302	0.038758017	10.73027412	294441
19	-0.119439847	0.044500969	12.2454572	290878
20	-0.220033861	0.074300914	20.33112113	287630

}{\neg}

\pgfplotstableread{
x  mean
1	0.345467662
2	-0.230029119
3	-0.051936945
4	-0.132061915
5	0.018258219
6	-0.190689516
7	0.188573232
8	0.139593266
9	0.163253462
10	0.097148326
11	0.016770275
12	0.070810413
13	0.081989627
14	-0.040813002
15	0.123150252
16	-0.080242784
17	0.089155901
18	0.037037145
19	0.108906427
20	0.038846531

}{\mid}

\begin{tikzpicture}

\pgfplotsset{
every axis legend/.append style={
at={(0.5,-0.650)},
anchor=north
},
}
 
 \begin{groupplot}[
    group style={%
        group size=1 by 2,%
        x descriptions at=edge bottom,%
        vertical sep=0pt,%
    },
    clip=true,
    clip mode=individual,
    width=1.75in,
    xmin=0,
    xmax=20,
    legend columns=3,
    legend style={draw=none}
    ]
    \nextgroupplot[
        title = Team Size $\ge$10,
        xticklabels={,,}, 
        height=1.5in, 
        ymax=2.2,
        ymin=-4,
        y label style={at={(axis description cs:0.08,.5)}},
        yticklabels={,,},
        ]
    \addplot [stack plots=y, fill=none, draw=none, forget plot]   table [x=x, y expr=(\thisrow{mean} + \thisrow{ci})]   {\pos} \closedcycle;
    \addplot [stack plots=y, fill=gray!70, opacity=0.6, draw opacity=1, thin, smooth, area legend]   table [x=x, y expr=(\thisrow{mean} - \thisrow{ci}) - (\thisrow{mean} + \thisrow{ci}) ]   {\pos} \closedcycle;
    \addplot [stack plots=y, stack dir=minus, forget plot, draw=none] table [x=x, y expr = (\thisrow{mean} - \thisrow{ci})] {\pos};

    \addplot [stack plots=y, fill=none, draw=none, forget plot]   table [x=x, y expr=(\thisrow{mean} + \thisrow{ci})]   {\neg} \closedcycle;
    \addplot [stack plots=y, fill=gray!70, opacity=0.6, draw opacity=1, thin, smooth, area legend]   table [x=x,y expr=(\thisrow{mean} - \thisrow{ci}) - (\thisrow{mean} + \thisrow{ci}) ]   {\neg} \closedcycle;
    \addplot [stack plots=y, stack dir=minus, forget plot, draw=none] table [x=x, y expr = (\thisrow{mean} - \thisrow{ci})] {\neg};
    
    \addplot [stack plots=false, green, thick, smooth]  table [x=x, y=mean]
    {\pos};
    \addplot [stack plots=false, red, thick, smooth]  table [x=x, y=mean]   
    {\neg};
    
    \draw[ultra thin] (axis cs:\pgfkeysvalueof{/pgfplots/xmin},0) -- (axis cs:\pgfkeysvalueof{/pgfplots/xmax},0);
    
    \addplot [stack plots=false, black, thick, smooth]  table [x=x, y=mean]   
    {\mid};\legend{}
    
    \nextgroupplot[
        ybar, 
        bar width=2pt, 
        ymode=log,
        height=0.95in, 
        ymax=1000000,
        ymin=100000,
        yticklabels={,,},
        xticklabels={,0,5,10,15,20},
    ]
    \addplot []  table [x=x, y=len]   {\pos};
    \end{groupplot}
\end{tikzpicture}
    \end{minipage}

\begin{tikzpicture}
    \begin{customlegend}[legend columns=-1,
      legend style={
        draw=none,
        column sep=1ex,
      },
    legend entries={Additions, Deletions, Net Use}]
    \addlegendimage{green, mark=*}
    \addlegendimage{red, mark=*}
    \addlegendimage{black, mark=*}
    \end{customlegend}
\end{tikzpicture}

    \caption{Library additions, deletions and net-usage (in LOC) after the adoption event. }
    \label{fig:afteradoptionteamsize}
\end{figure*}
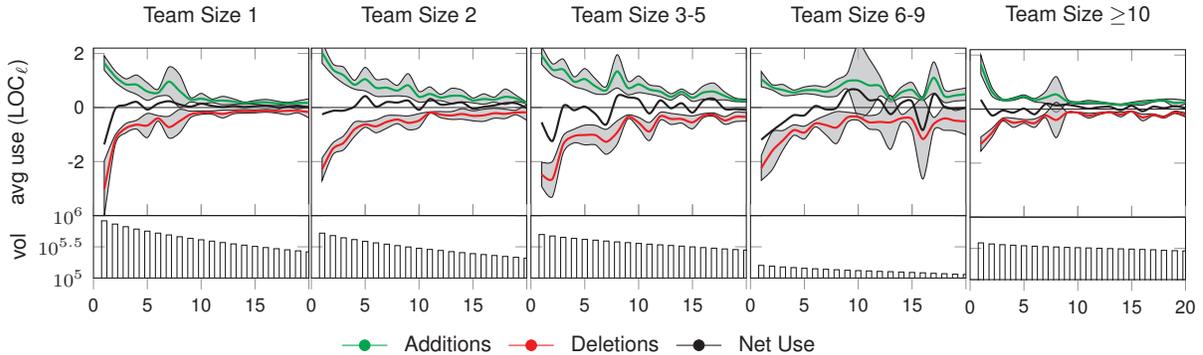

\begin{table}[]
    \centering
    \caption{Statistics surrounding newly adopted library $\ell$}
    \small{
    \begin{tabular}{@{}l|r@{}}
         Avg LOC that reference $\ell$ & 31.34 \\
         Median LOC that reference $\ell$ & 4 \\ \hline
         Avg inserted LOC that reference $\ell$ & 2.09 \\
         Avg deleted LOC that reference $\ell$ & 1.62 \\
    \end{tabular}
    }
    \label{tab:adoption_stats}
    
\end{table}

A simple (albeit poor) indicator of productivity in software projects is the number of lines of code (LOC) that are added and/or removed in a commit. Table \ref{tab:adoption_stats} shows there is a wide gap between the average and median LOC that reference a library $\ell$. Additionally, average LOC drops quickly after the first commit, as shown in Fig.~\ref{fig:afteradoptionteamsize}.

After the initial commit, we find that most of the following commits have only a small positive net productivity, and that the volume of activity of lines of code referencing $\ell$ in Fig.~\ref{fig:adopt_per_commit} tends towards zero rather quickly after the adoption.

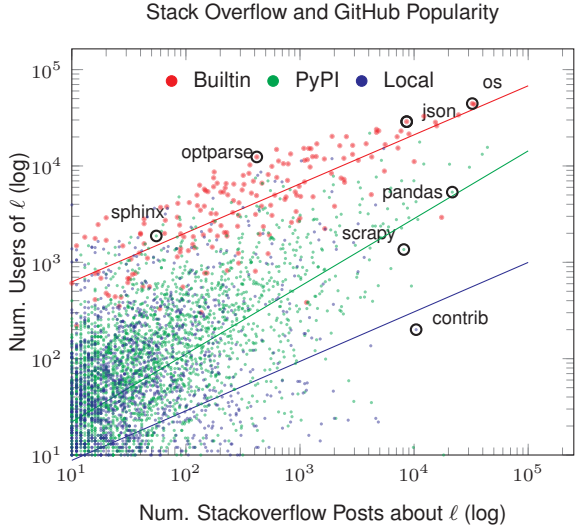
\begin{figure}[t]
    \centering
        \pgfplotstableread{
x	y	type lib
10	10	0	traci
10	10	0	mechanics
10	10	0	kaggleword2vecutility
12	10	0	mainsite
12	10	0	venue
13	10	0	dtype
13	10	0	gamelogic
13	10	0	celeryapp
14	10	0	third
15	10	0	rnn_cell
17	10	0	module_b
18	10	0	son
18	10	0	afm
18	10	0	mturk
18	10	0	pairwise_distances
20	10	0	_view
21	10	0	softwareproperties
21	10	0	submodules
22	10	0	executable
27	10	0	dynamodb
29	10	0	sub1
31	10	0	serving
33	10	0	methodtype
36	10	0	beta
49	10	0	delayed
50	10	0	firebase_admin
59	10	0	cred
60	10	0	collocations
105	10	0	dropdown
245	10	0	optionparser
283	10	0	figurecanvasqtagg
296	10	0	groupby
10	11	0	startfile
10	11	0	some_package
11	11	0	polymodel
11	11	0	median
11	11	0	etsconfig
11	11	0	twistd
11	11	0	bipartite
14	11	0	webclient
15	11	0	intent
16	11	0	skfuzzy
17	11	0	matplotlibwidget
19	11	0	stackoverflow
21	11	0	geolite2
25	11	0	_pywrap_tensorflow
31	11	0	col
32	11	0	httpconnection
34	11	0	chem
34	11	0	rcsetup
39	11	0	selectkbest
41	11	0	discriminant_analysis
44	11	0	second
51	11	0	gimpfu
58	11	0	doc2vec
59	11	0	textfield
67	11	0	errorcode
91	11	0	locators
93	11	0	global
154	11	0	_vendor
168	11	0	httperror
191	11	0	service_account
233	11	0	bytesio
10	12	0	ex1
10	12	0	quit
11	12	0	objecttype
11	12	0	userinterface
11	12	0	other_module
11	12	0	bidi
12	12	0	tuning
13	12	0	testresult
13	12	0	rtlsdr
15	12	0	mp4
15	12	0	mfcc
17	12	0	derivative
18	12	0	basecontroller
18	12	0	learn_io
19	12	0	cutils_ext
19	12	0	indexview
26	12	0	queryparser
27	12	0	clientsession
27	12	0	firefox_profile
28	12	0	googlecloudsdk
30	12	0	convexhull
37	12	0	sd
39	12	0	gencache
39	12	0	dbscan
41	12	0	coregraphics
46	12	0	ols
82	12	0	conv2d
85	12	0	load_model
116	12	0	reduce
120	12	0	offline
149	12	0	bigquery
415	12	0	render_template
430	12	0	ndimage
535	12	0	sqrt
10	13	0	verb
11	13	0	listings
11	13	0	basedoctemplate
11	13	0	_api
11	13	0	final
11	13	0	mplwidget
12	13	0	somelib
12	13	0	iplugin
13	13	0	contains
13	13	0	servicebase
13	13	0	phase
13	13	0	families
16	13	0	jvm
16	13	0	designer
17	13	0	rest_framework_mongoengine
17	13	0	euclidean_distances
18	13	0	rmsprop
22	13	0	sub2
22	13	0	modeller
23	13	0	exposure
23	13	0	gcs_oauth2_boto_plugin
23	13	0	endpoints_proto_datastore
24	13	0	pyqt
25	13	0	more
29	13	0	hadoop
30	13	0	assistant
39	13	0	aggregates
44	13	0	topo
104	13	0	useradmin
145	13	0	confusion_matrix
200	13	0	linearregression
249	13	0	logisticregression
253	13	0	apache
10	14	0	adaptor
11	14	0	get_terminal_size
12	14	0	adc
12	14	0	mfrc522
12	14	0	calculators
14	14	0	marrow
14	14	0	flac
15	14	0	candidate
16	14	0	booking
16	14	0	django_app
20	14	0	office
21	14	0	ij
21	14	0	htmlformatter
22	14	0	tkagg
23	14	0	richtext
23	14	0	a1
28	14	0	_split
30	14	0	octave
32	14	0	lazylinker_c
38	14	0	delaunay
41	14	0	occ
45	14	0	appium
79	14	0	someclass
81	14	0	hastraits
103	14	0	rcparams
133	14	0	apache_beam
170	14	0	combinations
220	14	0	gridspec
1504	14	0	httpresponse
10	15	0	chartit
10	15	0	phrases
10	15	0	shows
11	15	0	xively
11	15	0	libstdcxx
12	15	0	xor
14	15	0	directive
15	15	0	plaintext
15	15	0	patient
15	15	0	multiply
16	15	0	abstract_models
18	15	0	sklearn_pandas
18	15	0	mystuff
18	15	0	pyo
19	15	0	vector2
19	15	0	read_file
19	15	0	pointer
20	15	0	androidhelper
20	15	0	gauss
30	15	0	blogpost
32	15	0	mysqlconnection
33	15	0	priorityqueue
35	15	0	app_name
35	15	0	visuals
36	15	0	stdio
38	15	0	urandom
39	15	0	googledrive
115	15	0	filename
151	15	0	urlencode
11	16	0	webcam
12	16	0	sudo
12	16	0	nfc
12	16	0	baseplugin
14	16	0	reactors
15	16	0	insert
17	16	0	myflaskapp
18	16	0	app_config
19	16	0	accuracy
20	16	0	svd
20	16	0	inception_v3
22	16	0	_tkagg
30	16	0	chi2
30	16	0	notebookapp
30	16	0	gdalconst
32	16	0	m1
34	16	0	libc
37	16	0	business
40	16	0	stringtype
45	16	0	unpack
63	16	0	paraview
82	16	0	datareader
118	16	0	dumps
151	16	0	font_manager
304	16	0	tutorials
1195	16	0	division
10	17	0	xmlstream
10	17	0	nonlinearities
10	17	0	rlmain
11	17	0	initial
11	17	0	credit
11	17	0	chord
12	17	0	oauth2_plugin
14	17	0	confluent_kafka
21	17	0	caffe_pb2
22	17	0	staff
27	17	0	simple_salesforce
27	17	0	lxmlhtml
33	17	0	edu
39	17	0	ttypes
46	17	0	tcpserver
53	17	0	scrapy_splash
10	18	0	openface
10	18	0	__m2crypto
10	18	0	satchmo_store
14	18	0	evp
18	18	0	override_settings
19	18	0	skvideo
31	18	0	togglebutton
32	18	0	module_utils
32	18	0	wcs
35	18	0	hashers
10	19	0	skbio
10	19	0	bool
10	19	0	gateways
11	19	0	pools
12	19	0	osrng
18	19	0	stocks
20	19	0	basedir
28	19	0	socialaccount
41	19	0	ex48
63	19	0	enterprise
146	19	0	quad
169	19	0	permutations
205	19	0	build_ext
702	19	0	model_selection
10	20	0	minibatchkmeans
11	20	0	pygpu
11	20	0	logo
13	20	0	notestore
13	20	0	ssladapter
22	20	0	_lib
25	20	0	_mssql
27	20	0	naoqi
28	20	0	svmutil
28	20	0	uploads
30	20	0	requesthandler
32	20	0	setup_posix
32	20	0	txt
36	20	0	letters
56	20	0	sqs
58	20	0	child
158	20	0	module2
172	20	0	np_utils
397	20	0	spatial
413	20	0	namedtuple
11	21	0	sympify
12	21	0	information
12	21	0	cars
18	21	0	resizeimage
26	21	0	my_lib
30	21	0	googleplaces
71	21	0	some_module
1539	21	0	locals
10	22	0	txjsonrpc
16	22	0	sknn
17	22	0	filehandler
18	22	0	readwrite
20	22	0	sequtils
21	22	0	ex25
21	22	0	init_db
22	22	0	fast_rcnn
27	22	0	userinfo
29	22	0	pdfmetrics
63	22	0	mixture
78	22	0	pymel
84	22	0	agw
89	22	0	palettes
134	22	0	basecommand
173	22	0	get_user_model
300	22	0	naive_bayes
827	22	0	urlresolvers
4861	22	0	qtcore
12	23	0	networkmanager
12	23	0	layer1
13	23	0	smart_selects
13	23	0	odbc
13	23	0	twist
15	23	0	logreg
15	23	0	freecad
16	23	0	kernelmanager
16	23	0	global_vars
17	23	0	virtualbox
18	23	0	pymclevel
18	23	0	matplotlib_venn
19	23	0	dtypes
20	23	0	zinnia
54	23	0	interactiveshell
99	23	0	observers
130	23	0	geocoders
248	23	0	mimebase
2645	23	0	appengine
10	24	0	equipment
12	24	0	adafruit_ads1x15
12	24	0	subscribe
12	24	0	mymodels
13	24	0	sql_db
13	24	0	datetimewidget
15	24	0	rotate
15	24	0	dbn
16	24	0	complex
17	24	0	signing
21	24	0	subscriber
30	24	0	connectionerror
36	24	0	bge
39	24	0	primitive
61	24	0	logout
103	24	0	fits
109	24	0	quote
492	24	0	tk
2001	24	0	internet
4617	24	0	uix
10	25	0	change
11	25	0	xcb
12	25	0	ber
33	25	0	baseclass
87	25	0	indexes
119	25	0	cors
11	26	0	least_angle
12	26	0	secret_key
12	26	0	flask_uploads
14	26	0	dsa
26	26	0	catalogue
27	26	0	samples_generator
44	26	0	numbapro
63	26	0	employee
67	26	0	dropout
244	26	0	send_mail
276	26	0	listview
13	27	0	io_utils
14	27	0	textarea
15	27	0	winrm
15	27	0	rfe
16	27	0	printers
17	27	0	condition
17	27	0	campaign
18	27	0	protos
18	27	0	mathml
23	27	0	__check_build
33	27	0	myproj
57	27	0	basehandler
64	27	0	spyderlib
84	27	0	somemodule
276	27	0	deque
10	28	0	includes
10	28	0	wheezy
10	28	0	class_weight
10	28	0	_image
11	28	0	hbase
11	28	0	tuple
11	28	0	generation
13	28	0	hashmap
16	28	0	dictconfig
18	28	0	resnet50
20	28	0	_impl
21	28	0	check_random_state
30	28	0	projections
39	28	0	dll
43	28	0	sqla
65	28	0	httprequest
98	28	0	shared_task
117	28	0	mp3
10	29	0	archetypes
11	29	0	blogs
12	29	0	jnpr
13	29	0	initializers
13	29	0	vec3
16	29	0	vec
17	29	0	celery_app
18	29	0	voronoi
19	29	0	simplewebsocketserver
23	29	0	imagenet_utils
24	29	0	context_managers
28	29	0	_pylab_helpers
37	29	0	mul
72	29	0	trainers
138	29	0	pptx
149	29	0	grid_search
292	29	0	any
10	30	0	django_facebook
11	30	0	win32crypt
12	30	0	rlock
14	30	0	defines
17	30	0	monster
18	30	0	pdfdevice
23	30	0	misc_util
23	30	0	lldb
28	30	0	adam
38	30	0	global_settings
57	30	0	presentation
63	30	0	_path
66	30	0	behaviors
109	30	0	generics
291	30	0	authenticate
16	31	0	actstream
20	31	0	assembly
22	31	0	ttfonts
29	31	0	antlr4
43	31	0	drive
44	31	0	win32print
11	32	0	subject
11	32	0	_errors
15	32	0	multikernelmanager
16	32	0	shortener
16	32	0	amazonproduct
27	32	0	netsnmp
29	32	0	customers
31	32	0	ftpserver
44	32	0	id3
69	32	0	cuda
73	32	0	gdk
100	32	0	ffi
239	32	0	relationship
371	32	0	sgml
10	33	0	parseerror
10	33	0	mlabwrap
12	33	0	myparser
13	33	0	development
15	33	0	vec2d
15	33	0	ex47
16	33	0	libtcodpy
18	33	0	dsskey
22	33	0	decimal_precision
27	33	0	students
28	33	0	alexnet
36	33	0	basemodel
89	33	0	flaskapp
102	33	0	visa
159	33	0	count
185	33	0	create_app
190	33	0	apport
278	33	0	decomposition
12	34	0	clutter
12	34	0	mainmenu
12	34	0	m2
12	34	0	dot_parser
14	34	0	building
14	34	0	picture
16	34	0	combobox
16	34	0	klass
18	34	0	ui_main
18	34	0	led
24	34	0	exiftags
28	34	0	smartcard
29	34	0	ttransport
40	34	0	treeview
43	34	0	glade
129	34	0	graphlab
10	35	0	mainframe
10	35	0	tserver
11	35	0	sdk
12	35	0	espeak
15	35	0	imaging
15	35	0	scores
18	35	0	vehicle
20	35	0	answer
25	35	0	v3
36	35	0	school
48	35	0	imageqt
49	35	0	sa
73	35	0	supervised
235	35	0	ftp
10	36	0	_elementtree
11	36	0	flask_session
15	36	0	ultratb
16	36	0	dash_core_components
16	36	0	dash_html_components
22	36	0	xmlparser
26	36	0	itself
26	36	0	filestorage
32	36	0	gaussian_process
35	36	0	shaders
51	36	0	embedding
254	36	0	neighbors
10	37	0	momentjs
14	37	0	seq2seq
14	37	0	backend_qt4
17	37	0	telnet
20	37	0	market
22	37	0	asyncresult
33	37	0	videocapture
83	37	0	object_detection
153	37	0	contenttype
12	38	0	flask_restless
14	38	0	__doc__
16	38	0	rl_config
17	38	0	_distributor_init
19	38	0	topic
20	38	0	ho
62	38	0	signup
10	39	0	entries
12	39	0	djangocms_text_ckeditor
15	39	0	uinput
18	39	0	make
20	39	0	city
22	39	0	rooms
27	39	0	sentence
30	39	0	flask_app
39	39	0	adafruit_charlcd
85	39	0	smtp
86	39	0	current_app
145	39	0	backend_qt5agg
13	40	0	least_squares
19	40	0	vars
24	40	0	httpexceptions
28	40	0	flask_peewee
208	40	0	mpi
12	41	0	gv
12	41	0	cryptodome
13	41	0	encryption
13	41	0	module_loading
23	41	0	blender
28	41	0	internals
29	41	0	regularizers
37	41	0	backend_qt5
61	41	0	wordpress_xmlrpc
117	41	0	sgd
148	41	0	feature_selection
248	41	0	libraries
556	41	0	arcpy
12	42	0	idle
12	42	0	extra_views
12	42	0	sessionmanager
14	42	0	ajax_select
26	42	0	message_types
34	42	0	qt_compat
43	42	0	paragraph
53	42	0	ellipse
67	42	0	blocking
108	42	0	stdout
119	42	0	viewsets
146	42	0	mymodel
516	42	0	dense
10	43	0	celeryconfig
11	43	0	instances
11	43	0	cups
12	43	0	pooling
14	43	0	mosquitto
23	43	0	libsvm
29	43	0	alert
41	43	0	kernelapp
42	43	0	std
58	43	0	en
863	43	0	linear_model
10	44	0	runserver
12	44	0	netsvc
12	44	0	django_select2
16	44	0	_sparsetools
26	44	0	coo_matrix
30	44	0	imap
39	44	0	wsdl
53	44	0	walk
55	44	0	watson_developer_cloud
287	44	0	myclass
1067	44	0	in
10	45	0	appcfg
13	45	0	phantomjs
25	45	0	raw
28	45	0	square
30	45	0	get_data
34	45	0	optim
76	45	0	part
664	45	0	cross_validation
12	46	0	tweets
16	46	0	nnet
18	46	0	calls
19	46	0	freenect
19	46	0	organization
28	46	0	sense_hat
36	46	0	compute
39	46	0	tvtk
45	46	0	thumbnail
56	46	0	jpype
57	46	0	_ufuncs
65	46	0	lucene
83	46	0	firefox_binary
89	46	0	folder
182	46	0	context_processors
1960	46	0	figure
12	47	0	base_model
12	47	0	solr
12	47	0	current_process
15	47	0	coords
19	47	0	enemy
19	47	0	reviews
20	47	0	bullet
29	47	0	resize
36	47	0	testcases
49	47	0	course
93	47	0	_
129	47	0	schedulers
162	47	0	imagegrab
173	47	0	redirect
1138	47	0	spiders
10	48	0	timeouterror
12	48	0	scorer
25	48	0	digraph
10	49	0	uuidfield
11	49	0	execjs
13	49	0	communication
19	49	0	basetestcase
22	49	0	libmproxy
30	49	0	sentiment
11	50	0	ntsecuritycon
17	50	0	django_markdown
25	50	0	webelement
31	50	0	orders
37	50	0	adafruit_gpio
85	50	0	my_package
91	50	0	two
12	51	0	py3
12	51	0	relation
15	51	0	ethernet
15	51	0	csgraph
56	51	0	graph_tool
57	51	0	argparser
83	51	0	solvers
111	51	0	firebase
10	52	0	fsevents
11	52	0	neopixel
13	52	0	wnck
15	52	0	geom
16	52	0	cnn
39	52	0	clustering
44	52	0	cpu_count
74	52	0	figurecanvasagg
151	52	0	axes
250	52	0	contenttypes
10	53	0	native
11	53	0	flask_jwt
15	53	0	redmine
21	53	0	queryset
30	53	0	autocomplete_light
49	53	0	author
54	53	0	album
73	53	0	yahoo_finance
87	53	0	submodule
219	53	0	firefox
13	54	0	win32net
14	54	0	subpkg
24	54	0	database_setup
28	54	0	interaction
32	54	0	dajaxice
34	54	0	operation
11	55	0	fabfile
16	55	0	title
18	55	0	axislines
26	55	0	databases
42	55	0	axes_grid
44	55	0	art3d
45	55	0	checkbox
77	55	0	binding
1099	55	0	all
10	56	0	completer
10	56	0	nnls
14	56	0	doccer
26	56	0	lsqr
56	56	0	solve
1157	56	0	axes3d
12	57	0	sipconfig
20	57	0	_iterative
53	57	0	mean
11	58	0	base_events
14	58	0	der
24	58	0	ticket
45	58	0	probability
957	58	0	feature_extraction
996	58	0	statement
10	59	0	population
13	59	0	floppyforms
21	59	0	posixbase
24	59	0	dev
32	59	0	flask_pymongo
51	59	0	conn
53	59	0	backend_wx
60	59	0	sdl2
700	59	0	minidom
11	60	0	idl
15	60	0	hachoir_core
18	60	0	selectreactor
23	60	0	role
23	60	0	proj3d
26	60	0	producer
26	60	0	userena
27	60	0	pykeyboard
28	60	0	activations
77	60	0	subpackage
80	60	0	backend_agg
143	60	0	routers
11	61	0	metric
12	61	0	losses
12	61	0	basetest
18	61	0	testbase
27	62	0	csr
13	63	0	polymorphic
18	63	0	surface
22	63	0	iterative
58	63	0	sitemap
10	64	0	partner
12	64	0	mongodb
14	64	0	flask_marshmallow
21	64	0	private
55	64	0	section
74	64	0	vgg16
75	64	0	schedules
259	64	0	ui_mainwindow
27	65	0	formtools
34	65	0	photo
39	65	0	worksheet
44	65	0	member
46	65	0	socketio_client
64	65	0	cocos
23	66	0	_codecs
25	66	0	optimization
56	66	0	import_export
96	66	0	qgis
15	67	0	flask_sockets
20	67	0	funcs
21	67	0	sprites
27	67	0	version_info
90	67	0	student
95	67	0	gnuradio
17	68	0	courses
18	68	0	odf
21	68	0	rbf
22	68	0	pylabtools
45	68	0	homepage
11	69	0	isri
29	69	0	flask_restplus
29	69	0	internal
35	69	0	amazon
29	70	0	udp
35	70	0	django_auth_ldap
195	70	0	cbook
12	71	0	comm
17	71	0	custom_user
20	71	0	sheet
37	71	0	snowball
95	71	0	colorbar
11	72	0	spawn
11	72	0	pypy
54	72	0	cifar10
80	72	0	word
140	72	0	customer
170	72	0	compile
192	72	0	my_app
10	73	0	django_rq
11	73	0	modelcluster
12	73	0	regioninfo
13	73	0	multi
21	73	0	sharedctypes
367	73	0	pandas_datareader
11	74	0	adafruit_i2c
15	74	0	orthogonal
50	74	0	blas
66	74	0	parent
14	75	0	flask_mongoengine
16	75	0	fontforge
19	75	0	phonenumber_field
43	75	0	encode
14	76	0	_test
20	76	0	position
21	76	0	questions
39	76	0	activity
102	76	0	problem
20	77	0	aqt
30	77	0	descriptor
125	77	0	lstm
305	77	0	osv
386	77	0	input_data
10	78	0	nose_parameterized
19	78	0	scientific
63	78	0	lexers
88	78	0	pxssh
170	78	0	encoders
17	79	0	textview
18	79	0	subqueries
40	79	0	syntax
50	79	0	oauth1
11	80	0	mappings
13	80	0	openstack_dashboard
20	80	0	settings_local
25	80	0	convolve
44	80	0	problems
58	81	0	adafruit_dht
181	81	0	staticfiles
10	82	0	test_project
12	82	0	launcher
16	82	0	modname
35	82	0	printing
19	83	0	directives
20	83	0	toolbar
21	83	0	tweet
45	83	0	detail
112	83	0	hierarchy
165	84	0	formula
12	85	0	iostream
13	85	0	linear
13	86	0	preview
18	86	0	disk
19	86	0	interp
26	86	0	social_auth
70	86	0	mod2
102	86	0	v1
14	87	0	kde
31	87	0	pagination
20	88	0	stop_words
40	88	0	team
100	88	0	chart
124	88	0	opencv
23	89	0	nets
137	89	0	rand
191	89	0	pca
642	89	0	statements
17	90	0	filebrowser
21	90	0	locations
84	90	0	variables
10	91	0	cv_bridge
34	91	0	english
43	91	0	effects
26	92	0	buttons
44	92	0	number
24	93	0	hello_world
27	94	0	yacc
104	94	0	drawing
27	95	0	serialport
37	95	0	callback
47	95	0	_core
15	96	0	plotter
11	97	0	test_data
14	97	0	z3c
18	97	0	soupparser
50	97	0	forum
163	97	0	applications
15	98	0	reduction
17	98	0	flask_httpauth
15	99	0	waflib
16	99	0	tile
25	99	0	datatypes
55	99	0	elasticsearch_dsl
64	99	0	method
241	99	0	feature
12	100	0	anki
20	100	0	production
53	100	0	program
10	101	0	emb
22	101	0	load_data
25	101	0	roslib
14	102	0	drivers
15	102	0	_collections_abc
36	102	0	fake_useragent
68	102	0	django_tables2
10	103	0	qpid
14	103	0	execution
31	103	0	app2
40	103	0	libtorrent
10	104	0	visualize
47	104	0	strategy
76	104	0	ogr
77	104	0	primitives
10	105	0	locked_file
19	106	0	pykde4
38	106	0	identity
50	106	0	exporter
16	107	0	third_party
19	107	0	reddit
160	107	0	regression
179	107	0	mininet
14	108	0	main_window
95	108	0	rule
112	108	0	games
146	108	0	cursors
14	109	0	geosgeometry
18	109	0	transports
31	109	0	ckeditor
147	109	0	fftpack
20	110	0	_cffi_backend
11	111	0	analyze
14	111	0	stackless
199	111	0	optimizers
10	112	0	defs
10	113	0	flask_assets
53	113	0	eventloop
10	114	0	antlr3
14	114	0	test_app
23	114	0	pylearn2
24	114	0	flatpages
48	114	0	flask_security
56	114	0	sprite
16	115	0	sensors
32	115	0	photos
59	115	0	authorization
588	115	0	filedialog
18	116	0	requests_futures
47	116	0	articles
49	116	0	evaluation
13	117	0	_methods
150	117	0	set
28	118	0	django_countries
77	118	0	_psycopg
160	118	0	edit
319	118	0	route
107	119	0	servers
34	121	0	actor
35	121	0	machinery
12	122	0	plots
22	122	0	svg
393	122	0	aes
19	124	0	flask_bcrypt
34	124	0	_sysconfigdata
41	124	0	character
67	124	0	shapefile
39	125	0	categories
35	126	0	pushbullet
11	129	0	optimizer
76	130	0	contacts
10	131	0	ufunclike
12	131	0	matrixlib
20	131	0	rest_framework_jwt
57	131	0	updater
20	132	0	nosetester
11	133	0	decouple
24	133	0	cached_property
24	134	0	char
11	135	0	system_info
55	135	0	tix
28	136	0	cocoa
48	136	0	simpledialog
55	136	0	scene
4173	136	0	name
12	137	0	tzinfo
28	137	0	function_base
835	137	0	timezone
11	138	0	_osx_support
38	138	0	fixtk
12	139	0	numerictypes
15	139	0	login_manager
55	139	0	card
16	140	0	t1
30	140	0	_internal
275	140	0	add_newdocs
12	141	0	_heapq
21	141	0	viewer
34	141	0	aliases
47	141	0	expression
10	143	0	__config__
59	143	0	fcgi
173	143	0	multiarray
16	144	0	easy_thumbnails
158	144	0	host
36	145	0	_pickle
32	146	0	trainer
90	146	0	external
42	147	0	consumer
44	147	0	bin
13	148	0	umath
35	148	0	alsaaudio
188	148	0	font
14	149	0	flask_testing
50	149	0	memory_profiler
10	150	0	autoslug
12	150	0	flask_oauthlib
47	152	0	formatters
10	153	0	pyasn1_modules
22	153	0	pyexiv2
13	154	0	openstack
34	154	0	imageops
108	155	0	methods
15	156	0	evaluate
463	156	0	openerp
13	157	0	xx
33	158	0	wxversion
50	158	0	shop
82	158	0	renderers
100	158	0	another
33	159	0	builtin
38	159	0	requests_toolbelt
30	160	0	test_module
1503	160	0	button
31	162	0	libcloud
61	162	0	win32ui
201	162	0	bluetooth
15	163	0	imageenhance
44	163	0	apt_pkg
62	164	0	data_utils
97	164	0	flask_socketio
1267	164	0	label
25	165	0	scala
28	165	0	geometry_msgs
170	166	0	speech_recognition
36	171	0	groups
69	171	0	comments
739	172	0	date
21	175	0	tkconstants
24	177	0	sensor_msgs
53	177	0	tinymce
119	177	0	polynomial
77	179	0	contact
16	180	0	plugin_base
122	180	0	defaultfilters
1087	182	0	this
1114	182	0	repository
83	183	0	content
11	184	0	forking
56	185	0	ruamel
15	187	0	tkcolorchooser
18	188	0	flask_babel
175	188	0	category
26	190	0	abstract
28	191	0	formsets
306	194	0	imagetk
18	195	0	interpreter
35	195	0	highlight
135	198	0	scrolledtext
834	198	0	linalg
87	200	0	fixes
125	200	0	_mysql
10423	200	0	contrib
26	205	0	pdf
78	206	0	flask_admin
35	208	0	gnuplot
88	209	0	enthought
60	210	0	baz
93	210	0	frontend
28	211	0	yum
36	212	0	std_msgs
26	215	0	pysvn
35	215	0	sqlalchemy_utils
14	216	0	stemmer
16	216	0	languages
41	216	0	win32clipboard
35	217	0	inspection
83	217	0	_sqlite3
27	218	0	saxutils
31	218	0	github3
36	219	0	gdb
33	220	0	dev_appserver
67	221	0	sound
45	222	0	xmpp
94	222	0	webapp2_extras
115	222	0	mainwindow
33	224	0	providers
94	224	0	pyximport
59	225	0	local_settings
128	226	0	video
37	228	0	playhouse
40	230	0	container
84	230	0	webkit
114	230	0	dialects
212	230	0	styles
11	232	0	_multiprocessing
13	233	0	_curses
44	234	0	sorl
171	237	0	driver
39	240	0	provider
192	240	0	mysite
27	241	0	imagechops
28	243	0	win32console
28	245	0	png
28	245	0	gridfs
13	250	0	flask_debugtoolbar
121	251	0	collection
13	252	0	_md5
25	255	0	_struct
55	257	0	pygst
58	257	0	flask_mail
17	262	0	libvirt
230	263	0	osgeo
12	265	0	win32evtlog
1211	267	0	wsgi
13	268	0	oslo
118	268	0	smbus
18	269	0	_csv
144	272	0	group
41	273	0	mptt
24	274	0	repoze
65	275	0	templatetags
17	277	0	xbmcaddon
17	279	0	attributes
143	279	0	editor
36	280	0	_util
45	281	0	levenshtein
31	282	0	xbmcgui
70	286	0	rospy
270	286	0	caffe
16	287	0	_random
41	287	0	xbmc
142	291	0	_tkinter
11	296	0	prototypes
21	296	0	_functools
135	296	0	shared
13	298	0	win32evtlogutil
10	299	0	test_utils
39	299	0	loaders
89	299	0	win32serviceutil
133	299	0	authentication
90	301	0	_imaging
43	302	0	idlelib
118	305	0	enchant
766	305	0	connector
82	306	0	win32service
210	306	0	paho
18	307	0	pyobjctools
39	307	0	converters
54	309	0	_ctypes
89	309	0	allauth
714	310	0	extension
84	313	0	postgresql
251	316	0	classes
87	320	0	media
10	321	0	fuse
40	322	0	factories
20	324	0	model_utils
92	324	0	operators
219	326	0	usb
26	327	0	_weakrefset
34	327	0	imagefilter
17	330	0	_htmlparser
38	332	0	recorder
351	332	0	special
29	333	0	tokens
19	334	0	pyexpat
23	335	0	django_extensions
14	336	0	_pytest
25	339	0	linestring
53	339	0	libxml2
69	339	0	objc
13	340	0	_sha256
157	341	0	sax
19	344	0	warning
157	345	0	products
50	346	0	django_filters
60	348	0	flask_bootstrap
131	348	0	gst
79	350	0	processors
97	350	0	filter
26	352	0	gflags
60	352	0	_ssl
181	354	0	encoding
16	360	0	_locale
195	360	0	relativedelta
65	364	0	flask_cors
499	366	0	login
1962	366	0	ttk
812	372	0	management
33	376	0	sqlite
32	381	0	novaclient
56	381	0	_sre
77	382	0	_socket
150	382	0	crispy_forms
28	383	0	genericpath
14	387	0	bsddb
44	387	0	win32security
35	392	0	grammar
213	393	0	tastypie
322	393	0	examples
107	395	0	libs
70	396	0	sitemaps
89	399	0	_io
52	402	0	cgihttpserver
176	406	0	accounts
183	408	0	glib
53	410	0	_weakref
75	410	0	scikits
93	411	0	imagefont
619	411	0	cv
60	412	0	tksimpledialog
18	422	0	robotparser
236	427	0	transport
644	429	0	opengl
142	431	0	mixins
73	440	0	info
404	443	0	defaults
231	456	0	account
30	458	0	flask_migrate
120	464	0	tkfont
74	465	0	djcelery
53	466	0	vim
105	471	0	services
74	489	0	taggit
29	498	0	win32process
12	516	0	_utils
107	517	0	flaskext
695	522	0	apps
34	532	0	openid
48	541	0	flask_script
13	547	0	imagefile
136	548	0	flask_login
226	553	0	flask_restful
264	559	0	pythoncom
21	576	0	pkg1
105	580	0	requests_oauthlib
701	580	0	bio
11	584	0	win32pipe
212	586	0	managers
1597	595	0	backends
41	610	0	sre_parse
45	613	0	dj_static
44	637	0	sre_compile
13	643	0	distribute_setup
14	643	0	anydbm
89	678	0	win32event
20	688	0	filewrapper
897	722	0	application
15	733	0	zc
17	747	0	_ast
20	747	0	yourapplication
356	747	0	blog
120	759	0	imagedraw
108	780	0	__init__
19	789	0	adapter
3228	799	0	admin
22	802	0	debug_toolbar
31	812	0	inputstream
114	828	0	cairo
21	832	0	pango
157	887	0	simplexmlrpcserver
175	893	0	flask_wtf
946	905	0	handlers
13	939	0	_version
38	943	0	decoder
30	958	0	encoder
571	963	0	tkfiledialog
507	976	0	tkmessagebox
198	977	0	game
79	983	0	globals
52	997	0	win32file
10	1008	0	initialise
11	1055	0	html5parser
57	1055	0	mimetools
11	1057	0	markers
1010	1060	0	rpi
3653	1124	0	etree
555	1129	0	session
52	1134	0	win32
37	1150	0	sublime_plugin
270	1166	0	dbus
35	1168	0	copy_reg
26	1202	0	_base
127	1208	0	socks
411	1214	0	flask_sqlalchemy
16	1262	0	_compat
201	1293	0	controller
69	1298	0	pywintypes
41	1317	0	sre_constants
228	1322	0	git
11	1345	0	hotshot
2082	1361	0	error
19	1370	0	certs
10	1380	0	filepost
83	1388	0	connectionpool
64	1396	0	ntpath
36	1418	0	ntlm
112	1421	0	adapters
16	1422	0	sgmllib
15	1428	0	universaldetector
15	1440	0	ssl_match_hostname
26	1444	0	sitecustomize
70	1458	0	pysqlite2
155	1535	0	dj_database_url
568	1559	0	packages
2948	1601	0	user
1983	1624	0	mpl_toolkits
17	1646	0	userlist
26	1650	0	_collections
735	1692	0	response
186	1705	0	win32con
1463	1707	0	gi
38	1802	0	_abcoll
145	1824	0	memcache
474	1987	0	win32api
1337	2066	0	win32com
137	2082	0	simplehttpserver
1663	2209	0	rest_framework
111	2315	0	_winreg
357	2410	0	compat
167	2479	0	__version__
3233	2509	0	pylab
1493	2510	0	gtk
352	2672	0	openssl
46	2958	0	sha
66	3170	0	cookie
437	3516	0	cookielib
69	3619	0	userdict
53	3640	0	posixpath
317	3711	0	basehttpserver
4598	3939	0	forms
10	3944	0	htmlentitydefs
173	4623	0	md5
1905	5410	0	thread
696	6832	0	pkg_resources
604	6928	0	httplib
426	7999	0	cpickle
454	8636	0	cstringio
1027	11282	0	urlparse
1908	11900	0	stringio
5421	15136	0	urllib2
}{\table}

\pgfplotstableread{
x	y	type lib
11	219	1	faulthandler
136	302	1	winsound
59	329	1	chunk
26	351	1	audioop
1140	383	1	turtle
16	410	1	selectors
27	434	1	secrets
24	439	1	smtpd
21	452	1	modulefinder
20	478	1	dbm
94	554	1	sched
48	558	1	poplib
42	559	1	mailbox
10	608	1	trace
31	615	1	runpy
32	677	1	formatter
31	681	1	syslog
18	684	1	copyreg
34	749	1	pty
16	762	1	imghdr
36	768	1	posix
146	773	1	cmath
153	774	1	statistics
17	794	1	lzma
406	809	1	cgitb
230	820	1	telnetlib
290	981	1	wave
33	1080	1	filecmp
46	1105	1	winreg
84	1139	1	crypt
525	1228	1	ftplib
288	1244	1	imaplib
39	1248	1	colorsys
89	1277	1	symbol
105	1320	1	asyncore
90	1389	1	ipaddress
119	1416	1	cmd
170	1440	1	fractions
28	1481	1	rlcompleter
11	1484	1	pipes
115	1539	1	mmap
43	1561	1	tty
20	1693	1	lib2to3
16	1768	1	py_compile
13	1826	1	netrc
96	1837	1	dis
116	1845	1	sysconfig
189	1875	1	shelve
229	1875	1	typing
538	1924	1	fileinput
142	1926	1	_thread
40	1956	1	zipimport
443	1959	1	concurrent
90	1976	1	token
65	1995	1	xmlrpc
40	2001	1	plistlib
228	2037	1	profile
217	2039	1	pathlib
163	2146	1	encodings
1650	2237	1	test
477	2348	1	resource
284	2367	1	curses
203	2401	1	msvcrt
88	2435	1	numbers
17	2463	1	grp
43	2517	1	pydoc
842	2598	1	__main__
176	2647	1	linecache
968	2697	1	timeit
66	2731	1	keyword
43	2807	1	marshal
999	2870	1	asyncio
20	2898	1	pstats
17440	2927	1	tkinter
84	2942	1	bz2
568	3069	1	tokenize
290	3078	1	code
408	3145	1	enum
184	3188	1	heapq
166	3210	1	readline
258	3285	1	wsgiref
924	3366	1	parser
134	3472	1	cprofile
60	3715	1	termios
68	3765	1	pwd
2008	3819	1	html
461	3826	1	socketserver
145	3827	1	gettext
126	3870	1	builtins
554	3914	1	site
85	4090	1	pkgutil
292	4145	1	gc
107	4166	1	bisect
261	4372	1	difflib
737	4488	1	webbrowser
755	4766	1	ssl
1301	4786	1	smtplib
1013	4826	1	array
416	4928	1	ast
936	4968	1	pdb
601	4980	1	abc
125	4984	1	weakref
233	5010	1	hmac
4217	5044	1	http
375	5228	1	calendar
202	5354	1	unicodedata
214	5367	1	tarfile
908	5374	1	decimal
156	5409	1	fcntl
149	5526	1	atexit
311	5631	1	locale
145	5741	1	mimetypes
221	5917	1	zlib
175	6082	1	doctest
983	6150	1	select
132	6318	1	stat
323	6444	1	importlib
263	6488	1	fnmatch
225	6572	1	shlex
188	6624	1	getopt
464	6624	1	binascii
2458	6768	1	ctypes
1096	6830	1	cgi
390	6958	1	gzip
778	7082	1	zipfile
2431	7262	1	email
2291	7485	1	sqlite3
175	7530	1	textwrap
626	7607	1	getpass
318	7990	1	imp
526	8148	1	platform
357	8377	1	contextlib
1884	8397	1	queue
177	8657	1	errno
5492	8849	1	multiprocessing
1226	9750	1	signal
567	9901	1	uuid
364	10248	1	warnings
787	10503	1	types
715	10564	1	configparser
10014	11100	1	csv
649	11256	1	inspect
1096	11429	1	codecs
3241	11440	1	xml
1161	11610	1	struct
1635	11817	1	pickle
2426	12016	1	pprint
418	12401	1	optparse
4256	12409	1	io
1287	12484	1	base64
1385	12532	1	operator
2891	13203	1	glob
828	13329	1	traceback
623	14839	1	tempfile
1248	15103	1	functools
1192	15200	1	hashlib
1081	15238	1	copy
7225	15335	1	socket
2233	16219	1	distutils
2896	16362	1	string
5576	16673	1	threading
1405	16969	1	shutil
3799	17001	1	itertools
7098	20096	1	math
2550	20155	1	unittest
1911	20292	1	argparse
9020	21471	1	urllib
4678	22322	1	collections
7683	23049	1	subprocess
4903	23151	1	__future__
4677	23989	1	logging
15785	26089	1	random
15375	28451	1	datetime
8613	28815	1	json
12185	32818	1	re
24518	33888	1	time
33730	43591	1	sys
32364	44402	1	os
}{\tables}

\pgfplotstableread{
x	y	type lib
10	10	2	pjsua
10	10	2	emit
10	10	2	myserver
11	10	2	crowd
11	10	2	dronekit
11	10	2	zoo
11	10	2	tradingwithpython
13	10	2	zoom
15	10	2	petl
16	10	2	pydap
16	10	2	qutip
16	10	2	pymatgen
21	10	2	pwm
23	10	2	recommendation
23	10	2	pygrib
23	10	2	hybrid
23	10	2	p1
23	10	2	interop
27	10	2	openglcontext
31	10	2	openmdao
34	10	2	package_name
35	10	2	utc
36	10	2	0
83	10	2	pyomo
91	10	2	cos
135	10	2	tsa
315	10	2	print
18152	10	2	pyplot
10	11	2	ner
11	11	2	pywinusb
12	11	2	qi
13	11	2	neomodel
13	11	2	gold
13	11	2	pdfquery
14	11	2	lifelines
14	11	2	cls
14	11	2	running
14	11	2	no
16	11	2	tzwhere
16	11	2	dipy
17	11	2	usermanager
18	11	2	modbus_tk
23	11	2	ridge
25	11	2	butter
25	11	2	uber_rides
25	11	2	minimalmodbus
26	11	2	yahoofinance
26	11	2	mongoalchemy
27	11	2	googlefinance
28	11	2	aui
30	11	2	ibm_db
32	11	2	animals
35	11	2	aaa
39	11	2	sshclient
40	11	2	weka
40	11	2	tee
64	11	2	instagramapi
100	11	2	n
10	12	2	eye
10	12	2	indent
10	12	2	desc
11	12	2	spellchecker
11	12	2	unwrap
11	12	2	caesar
11	12	2	mytools
11	12	2	jws
11	12	2	pyganim
12	12	2	instabot
13	12	2	p2
13	12	2	threaded
13	12	2	optional
13	12	2	sniff
14	12	2	questionnaire
14	12	2	skyfield
16	12	2	literal
17	12	2	appjar
17	12	2	rmtree
18	12	2	gmaps
18	12	2	twittersearch
19	12	2	lasso
20	12	2	standalone
21	12	2	tabula
23	12	2	beatbox
23	12	2	httpauth
24	12	2	filepath
25	12	2	spinner
26	12	2	cosine
28	12	2	im
32	12	2	scitools
32	12	2	jaydebeapi
37	12	2	mylibrary
39	12	2	iterable
40	12	2	categorical
46	12	2	telethon
46	12	2	will
51	12	2	spyder
61	12	2	pytesser
68	12	2	loads
108	12	2	argumentparser
120	12	2	like
9862	12	2	webdriver
10	13	2	swampdragon
10	13	2	hn
11	13	2	pet
11	13	2	bioservices
11	13	2	pycorenlp
12	13	2	pytagcloud
12	13	2	bigfloat
13	13	2	factors
13	13	2	resample
13	13	2	double
13	13	2	freezer
13	13	2	expose
14	13	2	projector
14	13	2	dog
14	13	2	tesserocr
19	13	2	pdbparser
21	13	2	mlxtend
21	13	2	pyvisa
22	13	2	cplex
28	13	2	scikitlearn
30	13	2	charm
35	13	2	animal
40	13	2	projectname
40	13	2	swampy
41	13	2	cntk
47	13	2	blast
47	13	2	exscript
71	13	2	appname
108	13	2	appconfig
136	13	2	loginform
10	14	2	pickler
10	14	2	pykalman
11	14	2	shopify
11	14	2	newton
11	14	2	basicauth
12	14	2	neo4jrestclient
12	14	2	parse_rest
13	14	2	dragonfly
13	14	2	mss
14	14	2	qrtools
14	14	2	configurable
14	14	2	vadersentiment
15	14	2	playsound
16	14	2	boilerpipe
16	14	2	ortools
17	14	2	fipy
18	14	2	automation
19	14	2	rasa_nlu
19	14	2	ad
20	14	2	gutenberg
22	14	2	bittrex
31	14	2	objectlistview
31	14	2	exchangelib
32	14	2	fig
35	14	2	mainapp
35	14	2	ebaysdk
42	14	2	useragent
47	14	2	findspark
50	14	2	moment
63	14	2	pyopenssl
74	14	2	df
105	14	2	urlfetch
173	14	2	imread
176	14	2	xlwings
238	14	2	minimize
474	14	2	textinput
12	15	2	rows
12	15	2	mediainfo
14	15	2	yql
16	15	2	pylatex
18	15	2	dbconnect
18	15	2	phidgets
19	15	2	testmod
19	15	2	teradata
20	15	2	easy
22	15	2	pyhdf
22	15	2	ee
23	15	2	datatable
24	15	2	qstk
24	15	2	sock
26	15	2	polyglot
31	15	2	pyhive
31	15	2	kivymd
32	15	2	hog
64	15	2	package1
79	15	2	gof
99	15	2	letter
104	15	2	spyne
142	15	2	wtf
10	16	2	combinatorics
10	16	2	marionette
10	16	2	pypeg2
11	16	2	trading
12	16	2	earth
13	16	2	warc
14	16	2	dolfin
14	16	2	fibo
15	16	2	xy
16	16	2	l
17	16	2	measurements
18	16	2	wkhtmltopdf
18	16	2	escpos
22	16	2	require
23	16	2	clip
26	16	2	gamma
26	16	2	grouper
26	16	2	datafile
35	16	2	trial
42	16	2	1
46	16	2	can
46	16	2	pandasql
47	16	2	impala
50	16	2	bulbs
51	16	2	pyshark
51	16	2	mpl
56	16	2	ngrams
81	16	2	factorial
191	16	2	here
10	17	2	konlpy
10	17	2	sem
10	17	2	ms
11	17	2	mega
12	17	2	sasl
12	17	2	pygoogle
12	17	2	corenlp
12	17	2	me
13	17	2	project_name
14	17	2	eventdispatcher
14	17	2	rake
14	17	2	sm
15	17	2	somepackage
16	17	2	nuke
19	17	2	getdata
19	17	2	peakutils
20	17	2	bootstrapping
23	17	2	red
29	17	2	poisson
36	17	2	odo
40	17	2	pyarrow
45	17	2	gcd
47	17	2	green
52	17	2	repeat
52	17	2	spynner
10	18	2	pause
11	18	2	splunklib
11	18	2	crud
12	18	2	s3transfer
14	18	2	pifacedigitalio
15	18	2	krakenex
15	18	2	companies
16	18	2	poloniex
16	18	2	restaurant
19	18	2	python_speech_features
22	18	2	pyrebase
25	18	2	modeling
30	18	2	alpha
31	18	2	grass
33	18	2	package2
34	18	2	reload
36	18	2	use
40	18	2	garden
43	18	2	hive
47	18	2	sshtunnel
50	18	2	pillow
64	18	2	mymod
79	18	2	estimators
92	18	2	open
92	18	2	estimator
275	18	2	shuffle
423	18	2	partial
10	19	2	namedlist
10	19	2	maybe
10	19	2	pybindgen
11	19	2	textui
13	19	2	pin
15	19	2	factorization
16	19	2	mymath
16	19	2	spur
19	19	2	panda
19	19	2	shiboken
20	19	2	textstat
21	19	2	upnp
22	19	2	pydictionary
25	19	2	max
25	19	2	pyjamas
30	19	2	do
51	19	2	zipline
55	19	2	react
64	19	2	self
306	19	2	exit
10	20	2	sfml
10	20	2	nested
10	20	2	dbfpy
12	20	2	persons
12	20	2	documentation
13	20	2	padding
14	20	2	xgoogle
15	20	2	cql
15	20	2	bubble
15	20	2	slate
19	20	2	bottlenose
20	20	2	cam
22	20	2	amp
24	20	2	pow
31	20	2	netfilterqueue
32	20	2	pysal
41	20	2	pyexcel
55	20	2	lambdify
308	20	2	gis
787	20	2	integrate
12	21	2	filereader
13	21	2	pyimagesearch
14	21	2	asterisk
15	21	2	binarytree
18	21	2	callable
21	21	2	reshape
21	21	2	caches
23	21	2	ete3
29	21	2	sendkeys
41	21	2	tesseract
93	21	2	cycle
97	21	2	soaplib
261	21	2	pi
387	21	2	timedelta
10	22	2	xls
11	22	2	smart_open
11	22	2	codernitydb
11	22	2	myblog
12	22	2	concat
13	22	2	enaml
13	22	2	pykafka
15	22	2	fluidsynth
15	22	2	eggs
16	22	2	pydicom
17	22	2	minecraft
18	22	2	related
23	22	2	remove
29	22	2	contour
58	22	2	sum
59	22	2	deferred
76	22	2	restful
95	22	2	jsonify
99	22	2	cqlengine
106	22	2	ggplot
10	23	2	rds
10	23	2	forecast
11	23	2	twitterbot
11	23	2	dbmanager
12	23	2	neurolab
12	23	2	yt
13	23	2	logstash
13	23	2	libtiff
15	23	2	hcluster
17	23	2	sc
18	23	2	director
18	23	2	coinbase
20	23	2	textract
20	23	2	pytmx
23	23	2	imblearn
25	23	2	awscli
28	23	2	alchemyapi
28	23	2	delegate
30	23	2	treenode
35	23	2	requests_ntlm
37	23	2	testmodule
39	23	2	slim
41	23	2	spreadsheet
41	23	2	snippet
45	23	2	get_config
106	23	2	sha256
107	23	2	corpora
112	23	2	define
1023	23	2	mime
11	24	2	i2c
11	24	2	csvkit
12	24	2	ex
12	24	2	pynamodb
15	24	2	creator
15	24	2	int
15	24	2	sumy
17	24	2	getenv
18	24	2	qt4reactor
19	24	2	paypalrestsdk
25	24	2	mingus
27	24	2	pymacs
29	24	2	dir
29	24	2	embeddings
29	24	2	hmmlearn
46	24	2	versioning
48	24	2	obspy
58	24	2	gurobipy
124	24	2	rc
378	24	2	moves
2063	24	2	sleep
10	25	2	authenticator
10	25	2	simple_history
10	25	2	classic
10	25	2	grok
11	25	2	tickets
11	25	2	circuits
12	25	2	dnn
12	25	2	confirm
12	25	2	ffmpy
13	25	2	myhdl
14	25	2	summarize
16	25	2	sorteddict
17	25	2	ming
23	25	2	softlayer
24	25	2	importing
27	25	2	gooey
39	25	2	whatever
70	25	2	py4j
851	25	2	reverse
10	26	2	pyevolve
10	26	2	myutils
10	26	2	roster
11	26	2	pyth
11	26	2	reg
12	26	2	lcd
13	26	2	nxt
13	26	2	audit
14	26	2	lightgbm
14	26	2	user_profile
15	26	2	molecule
16	26	2	pystan
17	26	2	dajax
20	26	2	fun
38	26	2	hdfs
40	26	2	apns
42	26	2	happybase
46	26	2	pub
50	26	2	dbf
80	26	2	empty
125	26	2	scatter
147	26	2	exp
10	27	2	xadmin
10	27	2	zone
11	27	2	autopy
12	27	2	fbchat
13	27	2	handle
13	27	2	gaesessions
14	27	2	ciscoconfparse
16	27	2	blowfish
17	27	2	sketch
18	27	2	heatmap
23	27	2	ipwhois
34	27	2	coroutine
35	27	2	bag
36	27	2	p4
38	27	2	dryscrape
40	27	2	etc
42	27	2	pyocr
54	27	2	union
65	27	2	dag
79	27	2	fuzz
220	27	2	wordnet
732	27	2	dataframe
10	28	2	props
12	28	2	tab
14	28	2	sniffer
14	28	2	chef
17	28	2	say
17	28	2	rank
28	28	2	restless
30	28	2	googlesearch
46	28	2	blah
649	28	2	sequential
836	28	2	popen
10	29	2	intelhex
10	29	2	delta
12	29	2	nms
13	29	2	qtconsole
14	29	2	microbit
15	29	2	house
15	29	2	lightblue
16	29	2	scenario
18	29	2	boost
21	29	2	design
22	29	2	neuralnet
23	29	2	diagnostics
32	29	2	mm
41	29	2	h
50	29	2	bbox
81	29	2	hashtable
10	30	2	stdlib
10	30	2	mesos
10	30	2	authomatic
11	30	2	bp
14	30	2	v
14	30	2	nameko
15	30	2	sphere
15	30	2	tmp
17	30	2	livestreamer
17	30	2	nameparser
18	30	2	price
21	30	2	calculate
21	30	2	cloudinary
25	30	2	warn
25	30	2	req
29	30	2	pipelines
36	30	2	pysimplesoap
40	30	2	alphabet
48	30	2	coordinates
80	30	2	xbee
196	30	2	defer
211	30	2	mqtt
10	31	2	dumper
10	31	2	details
10	31	2	wiringpi
11	31	2	esky
13	31	2	mitmproxy
14	31	2	tld
20	31	2	zip
21	31	2	murmurhash
25	31	2	ystockquote
29	31	2	torctl
47	31	2	snowballstemmer
49	31	2	microsoft
62	31	2	share
88	31	2	xxx
100	31	2	pubnub
346	31	2	module1
10	32	2	servo
11	32	2	hex
12	32	2	llvmlite
12	32	2	w1thermsensor
14	32	2	atlas
15	32	2	disco
15	32	2	perspective
18	32	2	libarchive
18	32	2	comp
19	32	2	configmanager
20	32	2	crm
20	32	2	docstring
21	32	2	hashes
24	32	2	apple
24	32	2	story
32	32	2	blaze
43	32	2	normal
52	32	2	pynput
77	32	2	iris
88	32	2	interact
101	32	2	pymodbus
146	32	2	finance
153	32	2	y
10	33	2	goto
10	33	2	browse
10	33	2	nautilus
10	33	2	pyzabbix
11	33	2	cases
14	33	2	djangorestframework
14	33	2	broadcast
14	33	2	argument
15	33	2	kitchen
15	33	2	dao
15	33	2	boolean
17	33	2	asciitable
20	33	2	tsne
20	33	2	aplpy
21	33	2	parts
22	33	2	delorean
27	33	2	pykml
38	33	2	gearman
42	33	2	tensorboard
47	33	2	manifold
55	33	2	htmltestrunner
66	33	2	vision
126	33	2	ib
160	33	2	r
10	34	2	logfile
11	34	2	pd
12	34	2	health
12	34	2	instruments
15	34	2	face_recognition
19	34	2	basics
20	34	2	stl
31	34	2	u
34	34	2	yowsup
34	34	2	submit
36	34	2	conda
44	34	2	rdkit
48	34	2	recurrent
49	34	2	w
56	34	2	mypkg
80	34	2	pocketsphinx
137	34	2	cookiejar
181	34	2	pypyodbc
11	35	2	webview
12	35	2	qr
12	35	2	contents
12	35	2	astroml
13	35	2	traversal
13	35	2	web3
15	35	2	cfscrape
17	35	2	human
23	35	2	ftputil
24	35	2	send_email
28	35	2	pyfirmata
29	35	2	newsletter
33	35	2	timeseries
40	35	2	cvxpy
62	35	2	lru_cache
72	35	2	neural_network
84	35	2	clone
357	35	2	cmds
4227	35	2	qtgui
10	36	2	naivebayes
12	36	2	temperature
13	36	2	oursql
14	36	2	translations
15	36	2	bluepy
16	36	2	tushare
17	36	2	webservice
24	36	2	resolve
26	36	2	pythonmagick
48	36	2	pdfparser
54	36	2	initialize
111	36	2	seq
421	36	2	ensemble
489	36	2	ticker
2963	36	2	class
12	37	2	face
14	37	2	wiringpi2
14	37	2	asset
16	37	2	wit
18	37	2	stopwatch
27	37	2	vincent
29	37	2	sockets
41	37	2	thing
42	37	2	permission
50	37	2	gcloud
51	37	2	seed
69	37	2	z3
78	37	2	opt
10	38	2	cal
11	38	2	pymorphy2
11	38	2	guestbook
11	38	2	feather
12	38	2	aggregate
12	38	2	targets
12	38	2	lepl
12	38	2	parameterized
14	38	2	bind
14	38	2	front
15	38	2	glfw
15	38	2	volatility
17	38	2	nest
17	38	2	points
19	38	2	pdfrw
31	38	2	simpleitk
34	38	2	sounddevice
50	38	2	oct2py
53	38	2	my_project
56	38	2	extern
60	38	2	get_version
91	38	2	psychopy
98	38	2	pairwise
309	38	2	popup
12	39	2	braintree
13	39	2	astroquery
15	39	2	shodan
19	39	2	sieve
29	39	2	pox
32	39	2	k
33	39	2	iterator
34	39	2	http_client
35	39	2	packagename
52	39	2	barcode
52	39	2	work
57	39	2	arrays
151	39	2	test1
157	39	2	embed
692	39	2	include
10	40	2	freetype
12	40	2	ws
14	40	2	mp
15	40	2	normalizer
15	40	2	numdifftools
17	40	2	entropy
17	40	2	quiz
22	40	2	music21
24	40	2	rename
26	40	2	cat
27	40	2	ase
27	40	2	chatbot
28	40	2	quickstart
34	40	2	rasterio
34	40	2	jsonparser
42	40	2	netmiko
43	40	2	star
58	40	2	pisa
60	40	2	sha1
65	40	2	sun
166	40	2	paginator
188	40	2	pywinauto
10	41	2	addressbook
12	41	2	wikitools
14	41	2	ckan
17	41	2	goslate
24	41	2	xpath
88	41	2	pathos
383	41	2	choice
10	42	2	nipype
12	42	2	uikit
14	42	2	welcome
23	42	2	gcm
10	43	2	inception
10	43	2	payload
10	43	2	checkout
10	43	2	colormap
11	43	2	sign
13	43	2	shove
13	43	2	egg
13	43	2	mcpi
14	43	2	current
16	43	2	wrap
17	43	2	pyface
23	43	2	checkers
25	43	2	www
28	43	2	mechanicalsoup
34	43	2	observable
43	43	2	normalization
120	43	2	ec2
960	43	2	reactor
10	44	2	cement
14	44	2	alarm
15	44	2	materials
15	44	2	review
19	44	2	gp
20	44	2	uwsgidecorators
23	44	2	hand
27	44	2	ball
35	44	2	spi
58	44	2	top
67	44	2	vectors
196	44	2	userprofile
10	45	2	dnf
10	45	2	feincms
10	45	2	rabbit
11	45	2	gyp
11	45	2	nmf
14	45	2	ref
20	45	2	instrument
23	45	2	arff
23	45	2	qtpy
27	45	2	gmm
31	45	2	port
38	45	2	jenkinsapi
40	45	2	move
63	45	2	car
72	45	2	things
104	45	2	speech
1104	45	2	gpio
11	46	2	descartes
16	46	2	jellyfish
17	46	2	scripting
18	46	2	num2words
25	46	2	pymol
35	46	2	mahotas
37	46	2	kdtree
61	46	2	var
65	46	2	flatten
74	46	2	at
89	46	2	pydrive
112	46	2	neo4j
120	46	2	folium
205	46	2	one
11	47	2	dominate
13	47	2	tracer
16	47	2	switch
17	47	2	interpolation
23	47	2	texture
26	47	2	fmt
31	47	2	segmentation
33	47	2	tri
40	47	2	country
10	48	2	proxies
10	48	2	box2d
12	48	2	toolkit
12	48	2	float
13	48	2	proc
15	48	2	connexion
18	48	2	save
18	48	2	hid
19	48	2	wifi
30	48	2	pysphere
34	48	2	ship
69	48	2	configurator
79	48	2	goose
89	48	2	excel
140	48	2	geopandas
264	48	2	something
268	48	2	join
10	49	2	encrypt
12	49	2	plan
13	49	2	pandac
14	49	2	legend
23	49	2	healpy
24	49	2	googlevoice
28	49	2	soap
40	49	2	temp
61	49	2	chaco
69	49	2	activation
74	49	2	pysftp
106	49	2	pymc3
130	49	2	xlutils
10	50	2	none
10	50	2	scope
11	50	2	splitter
11	50	2	money
13	50	2	haversine
14	50	2	chalice
14	50	2	conversions
15	50	2	testpackage
15	50	2	helper_functions
15	50	2	vocabulary
15	50	2	pyv8
15	50	2	currency
16	50	2	logistic
17	50	2	deluge
17	50	2	pytube
19	50	2	ga
23	50	2	sendmail
25	50	2	spectral
27	50	2	cleaner
33	50	2	osr
36	50	2	dbapi
37	50	2	scale
79	50	2	p
81	50	2	simpy
177	50	2	airflow
616	50	2	call
11	51	2	quotes
13	51	2	riak
20	51	2	conditions
24	51	2	ijson
26	51	2	obj
33	51	2	stomp
34	51	2	arduino
35	51	2	enable
40	51	2	facepy
41	51	2	choices
46	51	2	rx
82	51	2	oscar
124	51	2	test2
153	51	2	robobrowser
155	51	2	migrations
11	52	2	vocab
11	52	2	macros
11	52	2	dicttoxml
12	52	2	rpio
13	52	2	quantities
16	52	2	bbb
17	52	2	pytumblr
20	52	2	asn1
23	52	2	tls
51	52	2	pyside2
60	52	2	talib
72	52	2	mainloop
96	52	2	measure
98	52	2	postgres
12	53	2	fn
14	53	2	fonttools
17	53	2	html5
17	53	2	datasource
20	53	2	crawl
20	53	2	workspace
24	53	2	pygeocoder
29	53	2	skype4py
30	53	2	columns
35	53	2	patsy
52	53	2	orange
64	53	2	issue
73	53	2	distributed
106	53	2	rectangle
160	53	2	cartopy
407	53	2	tensor
11	54	2	environments
12	54	2	radio
12	54	2	distro
16	54	2	viz
23	54	2	pydes
26	54	2	mapnik
27	54	2	instructions
27	54	2	inline
29	54	2	pyx
43	54	2	taskqueue
60	54	2	couchbase
96	54	2	first
160	54	2	protobuf
249	54	2	externals
10	55	2	execnet
10	55	2	oauth2_provider
11	55	2	crcmod
14	55	2	xlsx
14	55	2	topology
22	55	2	graphite
24	55	2	deps
25	55	2	pyfftw
26	55	2	rtree
32	55	2	nolearn
41	55	2	issues
83	55	2	standard
94	55	2	porterstemmer
427	55	2	some
13	56	2	frames
16	56	2	euler
16	56	2	vis
23	56	2	go
26	56	2	uncertainties
30	56	2	deprecation
33	56	2	zerorpc
35	56	2	chatterbot
44	56	2	vispy
106	56	2	avro
206	56	2	norm
10	57	2	da
10	57	2	multipoint
12	57	2	payments
16	57	2	pywapi
19	57	2	mstats
20	57	2	dataloader
23	57	2	linkedlist
24	57	2	publish
26	57	2	panda3d
27	57	2	pyes
30	57	2	str
32	57	2	validate_email
36	57	2	grab
37	57	2	material
47	57	2	vq
48	57	2	simplegui
56	57	2	lda
10	58	2	power
12	58	2	ranking
13	58	2	mido
26	58	2	write
27	58	2	tifffile
27	58	2	members
29	58	2	multiprocess
29	58	2	pigpio
34	58	2	dot
45	58	2	zbar
77	58	2	autograd
78	58	2	zeep
81	58	2	blobstore
369	58	2	streaming
508	58	2	workbook
557	58	2	cipher
10	59	2	unix
11	59	2	mocker
11	59	2	bst
12	59	2	bots
14	59	2	bank
15	59	2	gateway
30	59	2	v2
33	59	2	id
39	59	2	fib
40	59	2	plyer
51	59	2	adafruit_bbio
84	59	2	pulp
110	59	2	lmfit
189	59	2	q
2674	59	2	it
10	60	2	times
16	60	2	rdf
16	60	2	constraint
75	60	2	rpy
81	60	2	normalize
92	60	2	mode
701	60	2	cm
10	61	2	alignment
11	61	2	body
12	61	2	algo
16	61	2	tg
20	61	2	aiml
28	61	2	requirements
34	61	2	public
35	61	2	hmm
50	61	2	addons
56	61	2	pydispatch
85	61	2	proj
91	61	2	pack
148	61	2	g
12	62	2	webargs
17	62	2	colormath
19	62	2	community
22	62	2	bridge
25	62	2	option
29	62	2	countries
30	62	2	pydotplus
52	62	2	ryu
54	62	2	sparqlwrapper
106	62	2	contract
11	63	2	slimit
15	63	2	kubernetes
17	63	2	portfolio
17	63	2	ptvsd
28	63	2	des
31	63	2	exporters
33	63	2	nlp
44	63	2	timestamp
46	63	2	ipyparallel
48	63	2	lexicon
89	63	2	odoo
238	63	2	s
385	63	2	learn
11	64	2	mio
13	64	2	roles
14	64	2	envs
16	64	2	steps
19	64	2	pyowm
20	64	2	detect
35	64	2	netcdf
227	64	2	layouts
242	64	2	my_module
11	65	2	uri
16	65	2	characters
16	65	2	voice
19	65	2	snappy
20	65	2	flash
31	65	2	weasyprint
32	65	2	pymunk
48	65	2	weave
182	65	2	chrome
264	65	2	ml
14	66	2	edge
16	66	2	pyvmomi
16	66	2	simplekml
18	66	2	logutils
19	66	2	bibtexparser
22	66	2	fdb
78	66	2	foobar
166	66	2	row
12	67	2	datadog
15	67	2	ngram
20	67	2	images2gif
41	67	2	clean
41	67	2	colored
61	67	2	pdfkit
78	67	2	ode
86	67	2	word2vec
184	67	2	hash
11	68	2	codes
11	68	2	geoalchemy2
13	68	2	pyqrcode
13	68	2	langdetect
13	68	2	testlib
14	68	2	filechunkio
18	68	2	pyvim
23	68	2	nose2
39	68	2	classifiers
47	68	2	pyscreenshot
54	68	2	gps
128	68	2	msg
182	68	2	blueprint
11	69	2	scp
15	69	2	cairosvg
15	69	2	rst
15	69	2	chess
19	69	2	de
22	69	2	multidict
25	69	2	xvfbwrapper
34	69	2	dict
39	69	2	serve
40	69	2	vm
69	69	2	portal
74	69	2	circle
94	69	2	vcs
122	69	2	xyz
189	69	2	easygui
196	69	2	quandl
273	69	2	ndb
11	70	2	expr
17	70	2	ffmpeg
20	70	2	kernels
21	70	2	uploader
23	70	2	recipe
30	70	2	axis
31	70	2	pg
36	70	2	loop
52	70	2	builders
56	70	2	mpld3
60	70	2	bundle
74	70	2	blob
91	70	2	rnn
111	70	2	ghost
10	71	2	vk
12	71	2	ec
17	71	2	pusher
19	71	2	err
25	71	2	gmail
36	71	2	aa
37	71	2	align
41	71	2	pymouse
108	71	2	stuff
166	71	2	charts
11	72	2	restkit
14	72	2	transfer
17	72	2	forward
20	72	2	billiard
24	72	2	us
15	73	2	wolframalpha
19	73	2	klein
21	73	2	pygraph
36	73	2	spline
116	73	2	simplecv
170	73	2	other
15	74	2	preprocessor
21	74	2	survey
27	74	2	dash
28	74	2	pywt
33	74	2	value
38	74	2	arch
44	74	2	ascii
157	74	2	receiver
13	75	2	res
16	75	2	area
16	75	2	knn
16	75	2	decode
23	75	2	appscript
28	75	2	rl
30	75	2	size
36	75	2	collector
37	75	2	room
38	75	2	consumers
40	75	2	evernote
41	75	2	deap
55	75	2	ip
10	76	2	deployment
14	76	2	sort
16	76	2	pluginmanager
16	76	2	maze
16	76	2	bob
18	76	2	usage
19	76	2	song
20	76	2	pycallgraph
33	76	2	detector
39	76	2	clang
150	76	2	slider
492	76	2	maya
12	77	2	ca
24	77	2	payment
33	77	2	whois
53	77	2	physics
74	77	2	symbols
545	77	2	translation
11	78	2	journal
12	78	2	wallet
13	78	2	feedback
18	78	2	glob2
23	78	2	movies
23	78	2	w3lib
43	78	2	lookup
47	78	2	pyinstaller
47	78	2	op
118	78	2	mod1
10	79	2	apsw
16	79	2	djangoappengine
17	79	2	locust
28	79	2	pony
31	79	2	aws
53	79	2	geocoder
62	79	2	fpdf
62	79	2	publisher
80	79	2	xarray
85	79	2	morphology
97	79	2	fiona
100	79	2	variable
109	79	2	show
137	79	2	mixer
16	80	2	spotify
20	80	2	pint
45	80	2	mesh
632	80	2	sparse
13	81	2	glue
14	81	2	parsley
20	81	2	pafy
31	81	2	property
42	81	2	split
49	81	2	commons
53	81	2	storm
74	81	2	pyftpdlib
17	82	2	analytics
18	82	2	hub
54	82	2	ldap3
65	82	2	newspaper
71	82	2	cart
190	82	2	classify
237	82	2	mrjob
13	83	2	lazy
14	83	2	modeltranslation
20	83	2	compress
23	83	2	region
24	83	2	perceptron
28	83	2	failure
28	83	2	imapclient
29	83	2	values
59	83	2	quartz
93	83	2	traitsui
10	84	2	parameter
12	84	2	horizon
13	84	2	segment
15	84	2	brain
46	84	2	paypal
57	84	2	gpiozero
10	85	2	compose
25	85	2	pcapy
47	85	2	impl
70	85	2	rpyc
135	85	2	py2neo
168	85	2	question
13	86	2	matcher
26	86	2	pmw
37	86	2	directory
52	86	2	cloudstorage
13	87	2	es
35	87	2	setting
133	87	2	sites
331	87	2	multipart
28	88	2	calculator
34	88	2	capture
60	88	2	start
115	88	2	gtts
11	89	2	filelist
12	89	2	emcee
15	89	2	conversion
15	89	2	impacket
19	89	2	password
24	89	2	hyperopt
27	89	2	ns
57	89	2	conv
83	89	2	proto
589	89	2	dates
913	89	2	i
10	90	2	etcd
12	90	2	predict
27	90	2	exchange
52	90	2	twitterapi
62	90	2	find
17	91	2	access
30	91	2	inputs
33	91	2	executor
53	91	2	luigi
55	91	2	shape
68	91	2	dicom
27	92	2	term
39	92	2	xhtml2pdf
47	92	2	cursor
56	92	2	range
100	92	2	pyttsx
145	92	2	get
664	92	2	my
10	93	2	sensor
11	93	2	yapsy
19	93	2	mongokit
26	93	2	rect
35	93	2	deck
57	93	2	jnius
123	93	2	rango
11	94	2	titlecase
12	94	2	triangle
13	94	2	pywikibot
14	94	2	documents
19	94	2	updates
22	94	2	filer
24	94	2	bookmarks
24	94	2	notes
53	94	2	training
164	94	2	pymc
15	95	2	target
16	95	2	supervisor
43	95	2	tester
59	95	2	cycler
100	95	2	t
177	95	2	master
931	95	2	corpus
25	96	2	nacl
50	96	2	panel
93	96	2	eve
13	97	2	definitions
26	97	2	dump
30	97	2	toolbox
31	97	2	dal
113	97	2	matlab
13	98	2	transitions
14	98	2	args
34	98	2	charset
35	98	2	five
36	98	2	gmpy2
51	98	2	jenkins
53	98	2	porter
92	98	2	vlc
11	99	2	caching
12	99	2	flow
23	99	2	lettuce
27	99	2	startup
37	99	2	calc
38	99	2	pympler
61	99	2	getch
85	99	2	escape
91	99	2	dlib
121	99	2	pubsub
135	99	2	pyhook
147	99	2	comtypes
1677	99	2	selector
17	100	2	auto
19	100	2	flags
31	100	2	scraperwiki
53	100	2	uno
83	100	2	m
248	100	2	chain
283	100	2	gdal
556	100	2	cloud
10	101	2	collective
11	101	2	amqp
13	101	2	emails
14	101	2	notifier
15	101	2	javascript
57	101	2	address
66	101	2	classification
79	101	2	company
135	101	2	create
13	102	2	sortedcontainers
19	102	2	summary
34	102	2	dependencies
41	102	2	note
109	102	2	execute
132	102	2	artist
10	103	2	colour
15	103	2	dice
16	103	2	zsi
21	103	2	astroid
28	103	2	distribution
33	103	2	mongo
47	103	2	spidev
112	103	2	httpclient
13	104	2	bottleneck
15	104	2	histogram
24	104	2	play
28	104	2	leap
41	104	2	deploy
73	104	2	type
727	104	2	svm
14	105	2	ping
17	105	2	constant
17	105	2	moto
20	105	2	py3compat
23	105	2	elixir
39	105	2	lmdb
111	105	2	observer
158	105	2	wsgiserver
63	106	2	words
101	106	2	entry
197	106	2	visual
239	106	2	igraph
12	107	2	timeline
16	107	2	tablib
24	107	2	mapreduce
31	107	2	codec
56	107	2	routing
76	107	2	app1
164	107	2	lines
228	107	2	pyautogui
889	107	2	animation
10	108	2	reflect
35	108	2	ws4py
19	109	2	scandir
48	109	2	imports
61	109	2	nmap
14	110	2	newrelic
17	110	2	profiler
18	111	2	gallery
36	111	2	threads
45	111	2	movie
51	111	2	dictionary
342	111	2	stopwords
12	112	2	sqlobject
14	112	2	pyamf
16	112	2	pyro
16	112	2	desktop
36	112	2	extra
39	112	2	jupyter_client
39	112	2	readability
42	112	2	sage
47	112	2	imagekit
74	112	2	dpkt
170	112	2	func
22	113	2	css
83	113	2	pyro4
93	113	2	geos
108	113	2	required
13	114	2	dateparser
13	114	2	boot
15	114	2	flaskr
17	114	2	keystoneauth1
20	114	2	prompt
21	114	2	simulator
27	114	2	reference
28	114	2	interval
52	114	2	pp
70	114	2	spec
577	114	2	mlab
10	115	2	pefile
10	115	2	ddt
16	115	2	sentry
20	115	2	readers
21	115	2	djangotoolbox
22	115	2	nibabel
22	115	2	policy
22	115	2	tldextract
37	115	2	midi
40	115	2	visualization
46	115	2	mouse
136	115	2	pipe
257	115	2	pysnmp
17	116	2	natsort
21	116	2	analyzer
27	116	2	cairocffi
96	116	2	binary
17	117	2	graphs
33	117	2	pyudev
77	117	2	background
87	117	2	read
20	118	2	messaging
31	118	2	devices
40	118	2	https
103	118	2	wordcloud
197	118	2	kmeans
713	118	2	interpolate
16	119	2	cc
50	119	2	shapes
13	120	2	fetcher
16	120	2	cd
17	120	2	autocomplete
27	120	2	extract
29	120	2	algorithm
36	120	2	merge
68	120	2	instagram
139	120	2	crontab
216	120	2	gen
21	121	2	np
32	121	2	listener
33	121	2	keywords
43	121	2	pydevd
1356	121	2	patterns
20	122	2	instance
21	122	2	hosts
31	122	2	maps
41	122	2	pb
364	122	2	docx
381	122	2	tflearn
651	122	2	discovery
25	123	2	transactions
39	123	2	jedi
55	123	2	bb
316	123	2	dask
408	123	2	csrf
882	123	2	spider
11	124	2	recaptcha
24	124	2	pcap
25	124	2	pycassa
51	124	2	bucket
57	124	2	twill
11	125	2	terminaltables
21	125	2	music
34	125	2	rethinkdb
58	125	2	mapping
77	125	2	step
156	125	2	traits
18	126	2	unit
22	126	2	schemas
37	126	2	notify
542	126	2	testcase
12	127	2	benchmark
16	127	2	pt
22	127	2	experiment
43	127	2	translator
249	127	2	add
43	128	2	mxnet
48	128	2	spotipy
137	128	2	datastore
169	128	2	ipywidgets
13	129	2	networks
15	129	2	diff
29	129	2	configs
46	129	2	soundcloud
51	129	2	kernel
103	129	2	youtube
11	130	2	ai
21	130	2	links
21	130	2	cards
26	130	2	behave
21	131	2	track
43	131	2	match
140	131	2	structure
37	132	2	graphene
57	132	2	vendor
190	132	2	azure
14	133	2	pgdb
21	133	2	queries
33	133	2	apis
108	133	2	which
14	134	2	osc
33	134	2	people
48	134	2	clients
62	134	2	librosa
98	134	2	helloworld
121	134	2	wagtail
12	135	2	geojson
13	135	2	svgwrite
142	135	2	mylib
10	136	2	texttable
22	136	2	scimath
37	136	2	stock
43	136	2	ipykernel
113	136	2	pyopencl
29	137	2	salt
47	137	2	inventory
75	137	2	docs
126	137	2	protorpc
134	137	2	book
46	138	2	sendgrid
66	138	2	object
302	138	2	series
15	139	2	heap
16	139	2	rrdtool
19	139	2	tf
29	139	2	reports
105	139	2	cvxopt
127	139	2	pkg
21	140	2	kazoo
30	140	2	cssselect
61	140	2	grpc
77	140	2	sandbox
104	140	2	cmdline
10	142	2	pysolr
11	142	2	logs
16	142	2	fetch
24	142	2	img
44	142	2	telebot
55	142	2	sanic
101	142	2	poll
16	143	2	theme
17	143	2	exifread
21	143	2	level
28	143	2	expressions
85	143	2	nn
220	143	2	pytesseract
395	143	2	stem
15	144	2	strings
17	144	2	workflow
37	144	2	todo
43	144	2	wizard
52	144	2	telepot
94	144	2	mezzanine
14	145	2	visitor
29	145	2	extractor
173	145	2	kafka
254	145	2	patch
339	145	2	netcdf4
411	145	2	futures
12	146	2	pudb
13	146	2	bitcoin
22	146	2	clipboard
65	146	2	rauth
88	146	2	cell
195	146	2	jira
62	147	2	assets
159	147	2	ma
237	147	2	cassandra
388	147	2	pybrain
27	148	2	box
37	148	2	trac
67	148	2	pygal
282	148	2	connect
16	149	2	construct
22	149	2	deprecated
42	149	2	simulation
48	149	2	jobs
53	149	2	z
57	149	2	tinydb
71	149	2	spark
12	150	2	agents
16	150	2	subprocess32
677	150	2	lang
13	151	2	xmlrunner
17	151	2	clientform
43	151	2	timeout
64	151	2	ephem
12	152	2	retrying
14	152	2	doc
37	152	2	manage
466	152	2	patches
476	152	2	mnist
15	153	2	nbformat
24	153	2	menus
27	153	2	mlp
32	153	2	chainer
40	153	2	plone
87	153	2	snippets
15	154	2	trollius
19	154	2	configurations
23	154	2	dulwich
787	154	2	webapp
77	155	2	distributions
182	155	2	imutils
204	155	2	element
357	155	2	dispatch
28	156	2	pycountry
31	156	2	ctypeslib
32	156	2	batch
33	156	2	importer
26	157	2	localflavor
1398	157	2	optimize
45	158	2	stripe
135	158	2	controls
19	159	2	mathutils
33	159	2	serialization
33	159	2	interactive
29	160	2	singleton
89	160	2	header
29	161	2	printer
36	161	2	unicode
251	161	2	pycuda
24	162	2	tracker
66	162	2	runtime
80	162	2	poster
93	162	2	posts
104	162	2	linkedin
130	162	2	googlemaps
134	162	2	tutorial
24	163	2	parsedatetime
38	163	2	eyed3
48	163	2	geventwebsocket
50	163	2	x509
85	164	2	pydub
122	164	2	threadpool
12	165	2	pycparser
29	165	2	buildbot
37	165	2	evdev
43	165	2	weather
10	166	2	testdata
10	166	2	timing
25	166	2	generators
16	167	2	bencode
60	167	2	cfg
78	167	2	news
17	168	2	pysam
18	168	2	toolz
18	168	2	pyshell
20	168	2	router
25	168	2	record
86	168	2	namespace
15	169	2	pbkdf2
17	169	2	secret
18	169	2	states
29	169	2	jsonrpc
34	169	2	debugger
41	169	2	catalog
57	169	2	win
140	169	2	callbacks
15	170	2	pyreadline
16	170	2	slacker
17	170	2	guardian
57	170	2	imageio
232	170	2	export
416	170	2	pyqtgraph
16	171	2	pygeoip
239	171	2	line
24	172	2	containers
180	172	2	frame
657	172	2	ioloop
24	173	2	svn
60	173	2	general
78	173	2	keyboard
158	173	2	wmi
247	173	2	vtk
335	173	2	mayavi
26	174	2	swiftclient
71	174	2	servicemanager
1100	174	2	counter
50	175	2	names
68	175	2	imdb
80	175	2	layer
83	176	2	gluon
10	177	2	statsd
11	177	2	sdist
12	177	2	solver
39	177	2	chat
71	177	2	dashboard
114	177	2	fileutils
327	177	2	function
24	178	2	influxdb
44	178	2	echo
106	178	2	connections
64	179	2	feed
70	179	2	component
80	179	2	link
709	179	2	init
36	180	2	numexpr
38	180	2	packet
55	180	2	pyproj
170	180	2	moviepy
219	180	2	autobahn
410	180	2	post
31	181	2	gio
65	181	2	direct
12	182	2	preferences
30	182	2	sleekxmpp
45	182	2	domain
52	182	2	social
112	182	2	d
30	183	2	notifications
209	183	2	pyvirtualdisplay
38	184	2	language
57	184	2	sources
1877	184	2	url
55	185	2	projects
99	185	2	website
801	185	2	plotting
14	186	2	dml
15	186	2	formencode
17	186	2	integration
30	186	2	release
66	186	2	repo
102	186	2	lock
19	187	2	pygit2
20	188	2	validator
35	188	2	bitstring
108	188	2	mpmath
16	189	2	ecdsa
66	189	2	resolver
153	189	2	splinter
36	191	2	where
54	191	2	processor
290	191	2	parallel
33	192	2	block
64	192	2	hook
76	192	2	testapp
78	192	2	robot
246	192	2	protocols
150	194	2	gspread
29	195	2	results
48	195	2	async
128	195	2	signature
1182	195	2	root
39	196	2	wxpython
50	197	2	grequests
98	197	2	e
473	197	2	product
18	198	2	logbook
30	198	2	params
593	198	2	clock
19	199	2	unipath
58	199	2	jsonpickle
262	199	2	ops
12	200	2	latex
33	200	2	piston
138	200	2	transforms
15	201	2	myapplication
179	201	2	fft
315	201	2	distance
25	202	2	mailer
59	202	2	processing
158	202	2	wand
381	202	2	gl
112	203	2	download
177	203	2	units
179	204	2	spacy
67	205	2	notebook
11	206	2	records
194	206	2	person
19	207	2	reversion
149	207	2	f
154	207	2	screen
181	207	2	pymssql
818	208	2	declarative
14	209	2	codegen
76	209	2	bitarray
158	209	2	endpoints
62	210	2	classifier
25	211	2	geo
11	212	2	streams
12	212	2	objgraph
13	212	2	cast
35	212	2	custom
56	212	2	apt
205	212	2	article
318	213	2	style
558	213	2	android
21	214	2	reflection
32	214	2	prompt_toolkit
93	214	2	pyfits
29	215	2	traitlets
48	215	2	flickrapi
72	215	2	translate
14	216	2	renderer
17	216	2	about
140	216	2	order
87	217	2	soappy
123	217	2	sub
424	217	2	pdfminer
17	218	2	uwsgi
17	218	2	jieba
75	218	2	algorithms
83	218	2	location
169	218	2	xgboost
44	219	2	notification
89	219	2	wait
10	220	2	py2app
13	221	2	stevedore
24	221	2	downloader
18	222	2	pelican
30	222	2	memory
255	222	2	javax
12	223	2	tzlocal
42	223	2	fs
100	223	2	audio
21	225	2	preprocess
275	225	2	bootstrap
35	226	2	blocks
46	226	2	wget
70	226	2	draw
767	226	2	item
33	227	2	motor
353	228	2	xlsxwriter
15	229	2	scons
96	230	2	pattern
13	231	2	suite
56	231	2	parsing
133	231	2	vector
392	231	2	httpserver
57	232	2	demo
25	233	2	help
38	234	2	formats
16	235	2	geoip2
34	235	2	scan
37	235	2	machine
72	235	2	websockets
73	235	2	integer
683	236	2	cluster
29	238	2	generate
10	239	2	uvloop
12	239	2	pystache
39	239	2	gnupg
121	239	2	pypdf
158	239	2	camera
196	240	2	job
13	241	2	kerberos
111	241	2	matrix
130	241	2	grid
13	242	2	dynamic
255	242	2	mypackage
27	243	2	checker
1011	244	2	widget
14	245	2	rpm
14	248	2	annotation
621	249	2	qt
1389	249	2	preprocessing
74	250	2	logic
279	250	2	s3
794	250	2	static
77	251	2	dispatcher
219	251	2	home
85	252	2	bindings
90	252	2	comment
229	252	2	canvas
17	253	2	humanize
18	253	2	oracle
224	253	2	pypdf2
99	254	2	stack
52	255	2	marshmallow
144	255	2	load
21	256	2	checks
25	256	2	case
57	256	2	profiles
23	257	2	wiki
772	257	2	x
56	258	2	graphviz
149	259	2	pages
10	260	2	clint
14	260	2	constraints
76	261	2	wikipedia
35	262	2	somewhere
279	262	2	stream
46	263	2	writer
631	263	2	scapy
158	264	2	thrift
178	264	2	rdflib
74	266	2	world
35	267	2	action
35	268	2	captcha
87	268	2	terminal
325	268	2	security
46	269	2	queues
152	270	2	transform
210	270	2	rest
218	270	2	list
17	271	2	dependency
42	271	2	scraper
52	271	2	falcon
83	271	2	enums
107	271	2	extras
122	271	2	pyperclip
323	271	2	textblob
86	272	2	books
124	272	2	unicodecsv
136	272	2	converter
154	272	2	simple
43	273	2	zodb
120	273	2	channels
84	275	2	rq
12	276	2	invoke
112	276	2	lasagne
96	277	2	credentials
124	277	2	bpy
71	278	2	ssh
27	279	2	update
26	280	2	dialogs
39	281	2	headers
50	282	2	format
101	282	2	dill
43	283	2	storages
108	284	2	remote
163	284	2	windows
40	285	2	field
49	286	2	qrcode
54	286	2	pylons
89	286	2	components
14	287	2	persistent
53	287	2	youtube_dl
53	287	2	functional
94	287	2	agent
40	288	2	distlib
274	288	2	script
535	288	2	column
104	289	2	mod
722	290	2	graphics
160	291	2	page
2400	291	2	pyspark
30	292	2	convert
73	292	2	jsonfield
91	292	2	map
26	293	2	datastructures
67	293	2	rpc
176	293	2	entity
41	294	2	nodes
131	299	2	sample
15	301	2	strategies
261	302	2	tag
45	304	2	train
128	304	2	fuzzywuzzy
212	304	2	discord
234	305	2	env
23	306	2	hypothesis
954	307	2	rpy2
212	309	2	report
2296	310	2	render
329	311	2	picamera
37	312	2	guppy
1736	313	2	layers
30	314	2	lists
164	314	2	cms
351	315	2	crawler
76	317	2	default
68	318	2	mapper
424	318	2	clr
2236	318	2	python
43	319	2	braces
51	319	2	couchdb
356	319	2	color
270	320	2	layout
290	322	2	framework
16	325	2	persistence
197	325	2	images
201	325	2	scheduler
366	325	2	c
75	326	2	utility
111	326	2	permissions
85	327	2	analysis
990	327	2	plotly
147	329	2	xlib
311	329	2	socketio
812	329	2	build
19	331	2	monitor
1149	331	2	display
49	332	2	urwid
1424	332	2	items
34	333	2	dialog
71	333	2	geoip
214	333	2	mpi4py
314	333	2	net
34	337	2	parameters
56	338	2	oauthlib
31	339	2	pylint
213	339	2	input
32	341	2	adodbapi
287	342	2	plot
20	344	2	dammit
132	344	2	proxy
142	344	2	status
240	344	2	apscheduler
15	345	2	dotenv
16	345	2	bleach
294	345	2	key
66	346	2	channel
69	346	2	atom
31	347	2	elements
662	347	2	pipeline
2312	347	2	bokeh
674	348	2	properties
83	349	2	slackclient
134	349	2	templates
349	351	2	browser
386	354	2	numba
931	354	2	window
31	356	2	keystoneclient
46	356	2	output
126	357	2	device
219	357	2	googleapiclient
228	357	2	monkey
263	358	2	twython
21	359	2	history
253	362	2	rsa
211	363	2	shell
1220	363	2	metrics
35	364	2	irc
12	366	2	migration
40	366	2	loading
62	366	2	lex
77	366	2	source
24	367	2	icalendar
71	367	2	tz
149	368	2	whoosh
41	370	2	entities
318	370	2	timer
330	374	2	astropy
276	375	2	hello
22	376	2	lockfile
42	376	2	board
115	378	2	local
196	378	2	haystack
1163	381	2	orm
146	384	2	slugify
123	385	2	dropbox
128	385	2	schedule
26	386	2	torchvision
31	387	2	check
62	387	2	generator
1183	387	2	datasets
435	389	2	document
104	390	2	sequence
541	390	2	suds
684	390	2	mail
13	391	2	responses
88	392	2	exception
36	393	2	rules
62	393	2	spam
4327	394	2	ext
10	399	2	markdown2
193	400	2	dummy
95	404	2	objects
22	405	2	carbon
93	406	2	foundation
84	407	2	faker
149	407	2	environ
130	408	2	worker
17	410	2	setproctitle
147	410	2	control
577	410	2	mymodule
22	411	2	waitress
781	413	2	b
368	414	2	shapely
70	417	2	store
76	417	2	console
54	420	2	jwt
76	420	2	pydot
92	421	2	exc
324	422	2	run
92	423	2	kombu
2625	424	2	keys
66	425	2	appkit
444	426	2	files
46	430	2	beaker
136	431	2	parsers
3390	432	2	shortcuts
3675	436	2	file
56	439	2	pygraphviz
159	442	2	polygon
1375	443	2	sql
1351	444	2	generic
115	446	2	menu
31	447	2	validate
465	448	2	task
74	450	2	html2text
27	451	2	tagging
30	454	2	operations
105	454	2	configuration
226	454	2	basic
196	455	2	reader
125	456	2	controllers
138	456	2	register
280	457	2	library
459	459	2	xlwt
65	461	2	meta
65	461	2	scanner
100	461	2	i18n
68	462	2	validation
95	462	2	telegram
151	463	2	ansible
655	464	2	bar
288	467	2	index
308	468	2	mongoengine
157	469	2	scripts
101	470	2	regex
89	471	2	gym
1488	471	2	pool
211	472	2	peewee
61	473	2	sync
68	474	2	oauth
37	484	2	netifaces
322	486	2	joblib
102	490	2	sip
339	492	2	pyaudio
3730	492	2	a
228	495	2	polls
82	496	2	bot
863	499	2	statsmodels
1559	500	2	openpyxl
218	501	2	geopy
41	502	2	pynotify
482	504	2	form
824	504	2	pyodbc
16	505	2	filesystem
2154	507	2	support
155	512	2	extensions
42	515	2	msgpack
287	516	2	numeric
68	517	2	sh
534	517	2	colors
92	518	2	passlib
38	520	2	mercurial
53	521	2	packaging
170	521	2	registration
200	522	2	pika
239	522	2	messages
16	526	2	sqlparse
47	527	2	tags
41	530	2	tool
111	531	2	middleware
89	533	2	result
112	538	2	pyquery
181	538	2	point
56	541	2	features
168	541	2	migrate
20	544	2	nt
203	548	2	example
203	550	2	environment
473	550	2	elasticsearch
28	551	2	attr
141	551	2	mutagen
83	552	2	xmltodict
70	559	2	netaddr
582	561	2	package
62	562	2	actions
102	562	2	botocore
481	565	2	tree
1536	567	2	process
9478	567	2	kivy
212	568	2	interface
55	570	2	cffi
4854	572	2	urls
25	573	2	virtualenv
190	575	2	node
457	579	2	view
207	583	2	win32gui
35	584	2	feeds
316	588	2	geometry
63	590	2	configobj
639	590	2	options
34	591	2	mod_python
81	593	2	markup
432	593	2	command
42	596	2	arrow
254	596	2	twilio
307	596	2	manager
410	596	2	service
194	597	2	facebook
313	598	2	dataset
54	599	2	github
37	600	2	const
391	600	2	websocket
57	603	2	docker
169	603	2	praw
516	603	2	backend
193	618	2	ldap
414	618	2	pyglet
634	620	2	project
65	628	2	whitenoise
1667	629	2	stats
709	632	2	cx_freeze
596	634	2	protocol
150	636	2	install
44	637	2	raven
121	637	2	bcrypt
394	638	2	gdata
136	645	2	handler
540	645	2	pymysql
221	653	2	utilities
77	656	2	ply
323	657	2	message
1812	659	2	sympy
33	660	2	ujson
205	660	2	tables
748	660	2	builder
1016	666	2	mysql
365	667	2	pyramid
35	669	2	upload
150	669	2	m2crypto
470	669	2	gensim
22	671	2	blueprints
13	672	2	pypandoc
236	672	2	ipdb
88	684	2	prettytable
560	690	2	storage
278	693	2	torch
1016	694	2	reportlab
92	698	2	interfaces
3145	699	2	setup
83	702	2	dist
122	702	2	tabulate
129	704	2	future
23	707	2	archive
300	710	2	graph
206	711	2	player
355	711	2	table
2046	719	2	template
970	722	2	modules
1114	751	2	xlrd
20	757	2	fixtures
106	759	2	unidecode
221	759	2	query
170	760	2	helper
24	762	2	gunicorn
3192	762	2	path
94	769	2	magic
150	770	2	dns
76	773	2	paths
38	777	2	webtest
122	777	2	network
536	785	2	users
106	794	2	wrappers
303	794	2	celementtree
346	797	2	pexpect
212	798	2	plugins
124	802	2	routes
655	803	2	functions
111	806	2	progressbar
36	809	2	itsdangerous
241	819	2	testing
1147	821	2	parse
98	832	2	plugin
1623	832	2	pyside
34	833	2	serialize
110	835	2	transaction
137	842	2	daemon
1648	848	2	text
78	849	2	registry
16	850	2	cheetah
39	861	2	webob
6262	861	2	conf
253	866	2	oauth2
26	868	2	easy_install
153	868	2	debug
61	874	2	wrapper
247	877	2	event
937	879	2	mechanize
337	880	2	validators
873	881	2	system
13	882	2	ipaddr
61	883	2	state
1054	887	2	seaborn
89	888	2	babel
1437	893	2	foo
1449	899	2	ui
200	900	2	watchdog
622	900	2	misc
653	906	2	loader
98	909	2	decorator
137	922	2	cli
206	952	2	pycurl
212	953	2	events
523	953	2	zmq
30	955	2	jsonschema
14	966	2	markupsafe
20	971	2	psyco
818	972	2	webapp2
45	977	2	termcolor
733	979	2	apiclient
234	983	2	search
27	986	2	progress
1308	988	2	theano
370	989	2	engine
248	994	2	myproject
307	999	2	gui
374	1003	2	context
376	1005	2	com
102	1013	2	runner
447	1020	2	src
59	1027	2	retry
248	1027	2	schema
31	1039	2	pyinotify
32	1043	2	keyring
65	1048	2	tokenizer
261	1050	2	cryptography
367	1059	2	h5py
4058	1060	2	pyqt5
63	1061	2	alembic
746	1095	2	dom
312	1113	2	pyparsing
52	1116	2	greenlet
428	1116	2	cx_oracle
1277	1123	2	module
384	1131	2	tasks
1025	1134	2	skimage
92	1162	2	sublime
904	1181	2	serializers
24	1188	2	coverage
12	1192	2	mx
44	1200	2	certifi
41	1207	2	serializer
215	1228	2	tqdm
21	1247	2	pylibmc
394	1255	2	twitter
742	1262	2	boto3
226	1265	2	bson
1302	1274	2	oauth2client
314	1290	2	factory
185	1293	2	eventlet
928	1330	2	paramiko
61	1338	2	flup
1573	1338	2	tweepy
80	1341	2	mako
875	1342	2	myapp
284	1353	2	logger
8126	1357	2	scrapy
203	1361	2	aiohttp
679	1376	2	py2exe
1547	1378	2	web
290	1379	2	pygtk
82	1381	2	poolmanager
83	1383	2	metadata
710	1387	2	bottle
47	1397	2	wheel
125	1402	2	docopt
4113	1405	2	keras
24	1430	2	cookies
1005	1438	2	decorators
417	1442	2	cherrypy
119	1452	2	colorama
2085	1467	2	elementtree
90	1473	2	paste
14	1486	2	structures
3055	1501	2	wx
708	1521	2	widgets
812	1543	2	log
44	1549	2	pyasn1
1240	1585	2	tools
32	1590	2	backports
297	1596	2	psutil
91	1597	2	hooks
437	1602	2	cache
942	1621	2	networkx
93	1671	2	html5lib
161	1673	2	zope
121	1700	2	sessions
590	1732	2	org
546	1739	2	errors
291	1773	2	resources
576	1783	2	signals
557	1795	2	wtforms
961	1801	2	server
309	1806	2	feedparser
5079	1811	2	request
564	1818	2	ordereddict
1147	1819	2	py
33	1822	2	ez_setup
84	1828	2	new
55	1877	2	sphinx
228	1884	2	filters
2083	1915	2	serial
1178	1955	2	java
314	1966	2	gobject
428	1982	2	fabric
867	1991	2	connection
529	2024	2	urllib3
2044	2053	2	celery
61	2100	2	unittest2
5857	2145	2	tensorflow
905	2154	2	boto
480	2162	2	database
1860	2169	2	data
114	2232	2	compiler
7772	2268	2	db
655	2295	2	cython
5974	2331	2	pyqt4
122	2332	2	markdown
812	2344	2	fields
5093	2374	2	auth
4340	2377	2	nltk
4577	2377	2	client
2226	2391	2	lib
626	2447	2	httplib2
63	2454	2	south
5354	2511	2	cv2
992	2659	2	model
181	2673	2	pygments
702	2722	2	gevent
1307	2756	2	pymongo
6998	2773	2	pygame
89	2775	2	docutils
10785	2783	2	selenium
1408	2792	2	crypto
1341	2804	2	main
117	2879	2	click
871	2900	2	pip
8998	2900	2	beautifulsoup
2091	2942	2	tornado
88	2944	2	chardet
302	2954	2	helpers
337	2994	2	version
1003	3088	2	psycopg2
284	3126	2	tests
2296	3152	2	mysqldb
3958	3184	2	common
519	3237	2	commands
3365	3292	2	twisted
442	3389	2	constants
475	3547	2	werkzeug
1299	3549	2	ipython
171	3570	2	xmlrpclib
5857	3596	2	image
8743	3649	2	core
496	3663	2	pytz
438	3736	2	htmlparser
418	3837	2	redis
3346	3950	2	api
8074	4040	2	app
250	4168	2	nose
10395	4254	2	sklearn
3519	4546	2	google
669	4558	2	dateutil
1809	4709	2	base
87	4773	2	sets
21624	5350	2	pandas
6151	5400	2	sqlalchemy
1632	5487	2	exceptions
551	5550	2	jinja2
819	5669	2	mock
6258	5893	2	views
3380	5923	2	settings
714	6134	2	six
1959	6379	2	util
468	6577	2	pytest
3250	6795	2	lxml
524	6995	2	simplejson
9960	7017	2	scipy
7777	7019	2	bs4
408	7499	2	yaml
4662	7671	2	pil
1963	7826	2	config
30561	9699	2	matplotlib
20446	10423	2	models
12160	10469	2	flask
5172	11433	2	utils
39595	13222	2	django
43939	15598	2	numpy
11190	17694	2	requests
4080	21862	2	setuptools
}{\tabless}

\pgfplotstableread{
x	y	type lib
8613	28815	1	json
32364   44402   1   os
21624   5350    2   pandas
8613    28815   1   json
8126    1357    2   scrapy
418 12401   1   optparse
10423   200 0   contrib
55   1877 0   sphinx
%
}{\tablebig}

\begin{tikzpicture}


\begin{axis}[
    align =center,
    title = {Stack Overflow and GitHub Popularity},
    width=3.250in,
    clip=false,
    height=2.75in,
    ymin = 10,
    xmin = 10,
    ymode=log,
    xmode=log,
    xlabel = {Num. Stackoverflow Posts about $\ell$ (log)},
    ylabel = {Num. Users of $\ell$ (log)},
    y label style={at={(axis description cs:0.08,.5)}},
    ]
    \addplot [only marks, mark=*, red, mark size=0.8pt, fill opacity=0.6, draw opacity=0.4]  table [x=x, y=y]   {\tables};
    
    \addplot [only marks, mark=*, green, mark size=0.6pt, fill opacity=0.5, draw opacity=0.0]  table [x=x, y=y]   {\tabless};
    
    \addplot [only marks, mark=*, blue, mark size=0.6pt, fill opacity=0.5, draw opacity=0.0]  table [x=x, y=y]   {\table};
    
    \addplot+ [only marks, mark=o, black, thick, mark size=2.0pt, draw opacity=1]  table [x=x, y=y]   {\tablebig};  
    \node[anchor=south west] at (rel axis cs:.80,.88) {\scriptsize{\textsf{os}}}; 
    \node[anchor=south west] at (rel axis cs:.60, 0.60)  {\scriptsize{\textsf{pandas}}}; 
    \node[anchor=south west] at (rel axis cs:.68, 0.80)  {\scriptsize{\textsf{json}}}; 
    \node[anchor=south west] at (rel axis cs:.52, 0.50)  {\scriptsize{\textsf{scrapy}}}; 
    \node[anchor=south west] at (rel axis cs:.20, 0.70)  {\scriptsize{\textsf{optparse}}}; 
    \node[anchor=south west] at (rel axis cs:.70, 0.30)  {\scriptsize{\textsf{contrib}}}; 
    \node[anchor=south west] at (rel axis cs:.06, 0.55)  {\scriptsize{\textsf{sphinx}}}; 
    
    \addplot[red,domain=10:100000] {194.05*x^0.5088};
    \addplot[green,domain=10:100000] {4.37*x^0.7032};
    \addplot[blue,domain=10:100000] {2.70*x^0.5135};
    \end{axis}
    
    \begin{customlegend}[legend columns=-1,
      legend style={
        at={(5, 5.25)},
        draw=none,
        column sep=1ex,
      },
    legend entries={Builtin, PyPI, Local}]
    \addlegendimage{red,only marks, mark=*}
    \addlegendimage{green,only marks,mark=*}
    \addlegendimage{blue,only marks, mark=*}
    \end{customlegend}
    
\end{tikzpicture}
    \caption{Number of users of $\ell$ in GitHub dataset as a function of the number of Stack Overflow posts about $\ell$} 
    \label{fig:StackOverflow_users}
\end{figure}

\begin{figure*}[t]
    \centering
\pgfplotstableread{
x	zero	one	hund	thou
1	1	1	1.015151515	1.121428571
2	1.033333333	1	1.052188552	1.254761905
3	1.1	1.007575758	1.107744108	1.383333333
4	1.195238095	1.049242424	1.17688378	1.466666667
5	1.273015873	1.10479798	1.233180077	1.5
6	1.33968254	1.171296296	1.288735632	1.555555556
7	1.395959596	1.240740741	1.355959596	1.638888889
8	1.451515152	1.296296296	1.412457912	1.75
9	1.484848485	1.355555556	1.468013468	1.796527778
10	1.5	1.355555556	1.483164983	1.835218254
11	1.555555556	1.384990253	1.523809524	1.82172619
12	1.611111111	1.418323587	1.534920635	1.886309524
13	1.694444444	1.507212476	1.568253968	1.930952381
14	1.70751634	1.533333333	1.594738562	2
15	1.771008403	1.533333333	1.605222135	2
16	1.780267663	1.5	1.613555469	2.094298246
17	1.830687831	1.511965812	1.632731397	2.094298246
18	1.878306878	1.55958486	1.686894289	2.039853801
19	1.952380952	1.567594013	1.678560955	2.012222222
20	2	1.638961534	1.775757576	2.061672772
21	2	1.62467582	1.866666667	2.257531358
22	2	1.709259259	2	2.524198024
23	2	1.744973545	2	2.641414141
24	2	1.767195767	2	2.580952381
25	2	1.707936508	2	2.362698413
26	2	1.677256619	2	2.196031746
27	2	1.705034396	1.951948052	2.281746032
28	2	1.796701063	1.921645022	2.435763889
29	2.033333333	1.875	1.921645022	2.593658626
30	2.033333333	1.875	1.945887446	2.645245927
31	2.033333333	1.884042553	1.976190476	2.393741298
32	2.043693292	1.80070922	1.976190476	2.485846561
33	2.110359958	1.884042553	2	2.434259259
34	2.169883768	1.916666667	2	2.609360692
35	2.226190476	2	2.025641026	2.692694025
36	2.215079365	2	2.025641026	2.792694025
37	2.252194211	2	2.092307692	2.785185185
38	2.173721989	1.958333333	2.066666667	2.701851852
39	2.1598331	1.88539886	2.066666667	2.763616558
40	2.174305556	1.88539886	2	2.807598039
41	2.232777778	1.927065527	2	2.724264706
42	2.324444444	2	2	2.729166667
43	2.308571429	2	2	2.666666667
44	2.287129987	2	2.025641026	2.611111111
45	2.241515952	1.948717949	2.044017094	2.563123495
46	2.312944523	1.948717949	2.096648673	2.729790162
47	2.407885618	1.948717949	2.114969	3.007567939
48	2.463023463	2	2.116044037	2.917415138
49	2.407467907	2.051956815	2.125140854	2.717791786
50	2.370634921	2.185290148	2.163386708	2.751125119
51	2.394444444	2.185290148	2.143935602	2.555932203
52	2.416666667	2.133333333	2.142813268	2.448888889
53	2.430434783	2.064724919	2.060606061	2.066896552
54	2.430434783	2.138798993	2.060606061	2.733563218
55	2.463768116	2.138798993	2.029661985	3.414614122
56	2.533333333	2.169312169	2.103468866	3.52021757
57	2.566666667	2.169312169	2.12727839	3.078550903
58	2.633002833	2.180585618	2.180949738	2.579166667
59	2.6496695	2.168680856	2.125661376	2.655555556
60	2.624577924	2.104047595	2.188900684	3.097222222
61	2.53972324	2.140393194	2.155335869	3.148148148
62	2.552108223	2.168170972	2.226103065	2.830367044
63	2.668866465	2.296825397	2.232969577	2.49703371
64	2.770718317	2.315873016	2.329136803	2.878800092
65	2.652777778	2.363492063	2.406517756	3.307692308
66	2.514957265	2.225396825	2.418157967	3.498168498
67	2.578282595	2.277393753	2.585147392	3.139194139
68	2.689393706	2.237711214	2.532369615	3.139194139
69	2.778684807	2.317174578	2.593480726	3.51233802
70	2.59869281	2.274271754	2.54371345	3.769969278
71	2.662184874	2.279033658	2.622575239	3.651219278
72	2.699398496	2.366236961	2.69400381	2.92093254
73	2.804160401	2.457142857	2.555845916	2.539980159
74	2.601779449	2.4	2.681277056	2.603174603
75	2.654761905	2.423148148	2.6746633	3.094627595
76	2.583333333	2.423148148	2.666534779	3.142865421
77	2.833333333	2.467592593	2.52721475	3.376992405
78	2.833333333	2.35902067	2.295733268	2.804936777
79	3	2.35902067	2.241185235	2.620191014
80	2.916666667	2.35902067	2.228990113	3.084476729
81	2.916666667	2.38034188	2.367879002	3.373412698
82	2.916666667	2.38034188	2.543494152	4.223771121
83	3	2.394230769	2.710160819	3.997580645
84	3	2.295008032	2.704605263	4.266258806
85	3	2.211674699	2.633333333	3.375159642
86	3	2.126357238	2.451103426	2.625159642
87	3	2.273700805	2.589275469	2.759259259
88	3	2.273700805	2.660255861	3.383680556
89	3	2.567351598	2.803270082	3.595801768
90	3	2.662280702	2.865098039	3.579928752
91	3.053333333	2.838206628	2.803671787	3.140692641
92	3.053333333	2.679476469	2.74406178	3.078571429
93	3.053333333	2.705048875	2.559422192	2.503107345
94	2.965789474	2.668011838	2.716534719	2.586440678
95	2.965789474	2.671186441	2.675121279	2.503107345
96	3.099122807	2.490909091	2.720700819	2.953921569
97	3.133333333	2.496906566	2.621667003	3.703921569
98	3.169208494	2.504811115	2.774726648	3.984477124
99	3.053812741	2.620853035	2.870680045	4.020833333
}{\tablenotfound}

\pgfplotstableread{
x	zero	one	hund	thou
1	1.058894961	1	1	1.116666667
2	1.125561627	1	1.005747126	1.227777778
3	1.171857923	1	1.03908046	1.323476703
4	1.261111111	1.006726457	1.113154534	1.406810036
5	1.361111111	1.038472489	1.190740741	1.462365591
6	1.444444444	1.071805822	1.252645503	1.5
7	1.444444444	1.097156085	1.293877027	1.533333333
8	1.444444444	1.176521164	1.362058845	1.588888889
9	1.444444444	1.254298942	1.433487417	1.644444444
10	1.526386404	1.333333333	1.484848485	1.694444444
11	1.559719738	1.31965812	1.5	1.738888889
12	1.615275293	1.303785104	1.5	1.794444444
13	1.597660819	1.303785104	1.5	1.877777778
14	1.683375104	1.373015873	1.5	1.944444444
15	1.683375104	1.444444444	1.5	2
16	1.714667277	1.5	1.512820513	2
17	1.762286325	1.445048309	1.512820513	2
18	1.855853576	1.460199824	1.529551779	2
19	1.98245614	1.460199824	1.572286822	2
20	1.98245614	1.570707071	1.627842377	2
21	1.948717949	1.555555556	1.682539683	2
22	1.948717949	1.611111111	1.71031746	2
23	1.948717949	1.648148148	1.784391534	2
24	2	1.703703704	1.784391534	2
25	2	1.703703704	1.819223986	2.050949914
26	2	1.777777778	1.856261023	2.102800617
27	2	1.888888889	1.932435937	2.169467283
28	2	1.833333333	1.980936819	2.229628481
29	2	1.833333333	1.980936819	2.22793501
30	2	1.745614035	1.968253968	2.225370908
31	2	1.912280702	1.968253968	2.225370908
32	2.038126362	1.745614035	1.931216931	2.286324786
33	2.038126362	1.833333333	1.925925926	2.333333333
34	2.038126362	1.833333333	1.925925926	2.355555556
35	2	2	1.962962963	2.411111111
36	2.01754386	2	2	2.466666667
37	2.036062378	1.964240102	1.960784314	2.444444444
38	2.064444021	1.964240102	1.960784314	2.444444444
39	2.046900161	1.964240102	1.960784314	2.444444444
40	2.083937198	2	2	2.466666667
41	2.055555556	2	2	2.466666667
42	2.065569991	2	2	2.466666667
43	2.010014435	2	2	2.5
44	2.010014435	2	2	2.5
45	2	1.833333333	2	2.5
46	2	1.833333333	2	2.555555556
47	2.111111111	1.833333333	2	2.611111111
48	2.277777778	2	2	2.644444444
49	2.277777778	2	2	2.644444444
50	2.166666667	2	2	2.688888889
51	2.083333333	2	2	2.732323232
52	2.173423423	2	2	2.76010101
53	2.284534535	2	2	2.743434343
54	2.201201201	2	2	2.766666667
55	2.271604938	2.111111111	2.111111111	2.806521739
56	2.261458014	2.349206349	2.194444444	2.778743961
57	2.37684263	2.349206349	2.277777778	2.821601104
58	2.383015469	2.349206349	2.214285714	2.865079365
59	2.282051282	2.277777778	2.146103896	2.976190476
60	2.207535121	2.277777778	2.062770563	2.975274725
61	2.151979566	2.166666667	2.060606061	2.975274725
62	2.318646232	2	2.128787879	2.975274725
63	2.545811187	2.222222222	2.185887506	3
64	2.47914452	2.222222222	2.223766294	2.9
65	2.375222952	2.247863248	2.251544072	2.9
66	2.156830568	2.092307692	2.311111111	2.9
67	2.223497235	2.258974359	2.338888889	3
68	2.271863248	2.316666667	2.394444444	3
69	2.288888889	2.25	2.325396825	3
70	2.177777778	2.083333333	2.325396825	3
71	2.233333333	2.111111111	2.158730159	3
72	2.166666667	2.111111111	2.222222222	3
73	2.333333333	2.111111111	2.277777778	3
74	2.333333333	2.107638889	2.440568475	3
75	2.533760684	2.107638889	2.329457364	3
76	2.503205128	2.186491935	2.230437756	3.033333333
77	2.598008887	2.301075269	2.134313725	3.2
78	2.564248203	2.467741935	2.277170868	3.2
79	2.523375188	2.555555556	2.352380952	3.203703704
80	2.428571429	2.5	2.369047619	3.12037037
81	2.428571429	2.444444444	2.301636292	3.12037037
82	2.555555556	2.388888889	2.199612483	3.133333333
83	2.5	2.222222222	2.249612483	3.162698413
84	2.481481481	2.277777778	2.340833333	3.162698413
85	2.444444444	2.261904762	2.466666667	3.255555556
86	2.503333333	2.317460317	2.4	3.226190476
87	2.530294629	2.150793651	2.288888889	3.35952381
88	2.678442777	2.182137834	2.288888889	3.35
89	2.84177611	2.459915612	2.444444444	3.418302387
90	2.933333333	2.543248945	2.555555556	3.451635721
91	2.85	2.666666667	2.555555556	3.451635721
92	2.85	2.555555556	2.533333333	3.466666667
93	2.805555556	2.638888889	2.533333333	3.466666667
94	2.875901876	2.574074074	2.551375684	3.472222222
95	2.84018759	2.518518519	2.494232827	3.472222222
96	2.945766198	2.518518519	2.549788383	3.472222222
97	2.958753211	2.555555556	2.531746032	3.5
98	3.045749548	2.588888889	2.555555556	3.444444444
99	3.076923077	2.633333333	2.5	3.416666667
}{\tablepypi}

\pgfplotstableread{
x	zero	one	hund	thou
1	1.087387387	1	1	1.118920973
2	1.170720721	1	1.012345679	1.230032084
3	1.227777778	1	1.054012346	1.327777778
4	1.31523569	1.022222222	1.137345679	1.411111111
5	1.398569024	1.06621087	1.208333333	1.466666667
6	1.454124579	1.110655315	1.277777778	1.5
7	1.466666667	1.127648779	1.325396825	1.533333333
8	1.466666667	1.194771242	1.408730159	1.588888889
9	1.5	1.261437908	1.464285714	1.644444444
10	1.585057471	1.355555556	1.5	1.694444444
11	1.640613027	1.355555556	1.5	1.738888889
12	1.662835249	1.355555556	1.5	1.794444444
13	1.666666667	1.333333333	1.518518519	1.877777778
14	1.751461988	1.388888889	1.518518519	1.944444444
15	1.773684211	1.444444444	1.518518519	2
16	1.773684211	1.5	1.529198636	2
17	1.8	1.458333333	1.529198636	2
18	1.888888889	1.505952381	1.560448636	2
19	2	1.505952381	1.586805556	2
20	2	1.603174603	1.642361111	2
21	1.984496124	1.591555556	1.694444444	2
22	1.984496124	1.647111111	1.738888889	2
23	1.984496124	1.688046784	1.85	2
24	2	1.713058688	1.85	2
25	2	1.728931704	1.916666667	2.055555556
26	2	1.799107143	1.916666667	2.111111111
27	2.03030303	1.904761905	2	2.177777778
28	2.03030303	1.833333333	2	2.233333333
29	2.03030303	1.833333333	2	2.233333333
30	2	1.766666667	2	2.233333333
31	2	1.933333333	2	2.233333333
32	2.066666667	1.8	2	2.288888889
33	2.122222222	1.866666667	2	2.333333333
34	2.122222222	1.866666667	2	2.355598733
35	2.103174603	2	2	2.411154289
36	2.095238095	2	2	2.466709845
37	2.150793651	2	1.972222222	2.444444444
38	2.186507937	2	1.972222222	2.444444444
39	2.138888889	2	1.972222222	2.444444444
40	2.194444444	2	2	2.466666667
41	2.111111111	2	2	2.466666667
42	2.198830409	2	2	2.466666667
43	2.087719298	2	2	2.5
44	2.118374689	2	2	2.5
45	2.077167019	1.833333333	2	2.5
46	2.094261036	1.833333333	2	2.555555556
47	2.230272312	1.833333333	2	2.611111111
48	2.392094017	2	2	2.666666667
49	2.422619048	2	2	2.666666667
50	2.255952381	2	2	2.722222222
51	2.138015151	2.027777778	2	2.75
52	2.201507214	2.075396825	2	2.787037037
53	2.368173881	2.075396825	2	2.759259259
54	2.277777778	2.047619048	2	2.781986532
55	2.351851852	2.133333333	2.111111111	2.823802542
56	2.318518519	2.466666667	2.194444444	2.796024764
57	2.485185185	2.466666667	2.28968254	2.856630824
58	2.471994025	2.456439394	2.234126984	2.888888889
59	2.388111242	2.289772727	2.206349206	3
60	2.296421991	2.289772727	2.131944444	3
61	2.276167676	2.166666667	2.127327534	3
62	2.393383793	2.077493242	2.182883089	3
63	2.618406377	2.410826576	2.244970548	3
64	2.590277778	2.410826576	2.305143014	2.916666667
65	2.534722222	2.4	2.309695225	2.916666667
66	2.298644639	2.15	2.369631576	2.916666667
67	2.286144639	2.316666667	2.401377608	3
68	2.298139588	2.361111111	2.452380952	3
69	2.324116162	2.277777778	2.365079365	3
70	2.213005051	2.111111111	2.333333333	3
71	2.256565657	2.166666667	2.166666667	3
72	2.211411411	2.239130435	2.25462963	3
73	2.378078078	2.239130435	2.310185185	3
74	2.385653836	2.167987007	2.476851852	3
75	2.549242424	2.095523239	2.333333333	3
76	2.549242424	2.209412128	2.277777778	3.033333333
77	2.708333333	2.447222222	2.194444444	3.2
78	2.710526316	2.669444444	2.361111111	3.2
79	2.710526316	2.722222222	2.416666667	3.214285714
80	2.613859649	2.555555556	2.428571429	3.14021164
81	2.614444444	2.5	2.373015873	3.14021164
82	2.781111111	2.444444444	2.273015873	3.145224172
83	2.711111111	2.361111111	2.344444444	3.185422907
84	2.633333333	2.361111111	2.455555556	3.185422907
85	2.522222222	2.416666667	2.588888889	3.286637482
86	2.577777778	2.416666667	2.501010101	3.25188537
87	2.694444444	2.25	2.362121212	3.389816404
88	2.805555556	2.25	2.328787879	3.378827393
89	2.916666667	2.5	2.5	3.447454844
90	2.933333333	2.611111111	2.583333333	3.476190476
91	2.871780303	2.733134921	2.607142857	3.466666667
92	2.871780303	2.646596459	2.579365079	3.466666667
93	2.847537879	2.702152015	2.579365079	3.466666667
94	2.909090909	2.635683761	2.611111111	3.476190476
95	2.909090909	2.555555556	2.555555556	3.476190476
96	3	2.555555556	2.680555556	3.476190476
97	3.024539877	2.59454191	2.625	3.5
98	3.157873211	2.653021442	2.625	3.444444444
99	3.236809816	2.729532164	2.5	3.416666667
}{\tablebuiltin}

\pgfplotstableread{
x	zero	one	hund	thou
0	1	1	1	1
1	0.985294118	0.372834373	0.786785334	1
2	1	0.649985584	0.933333333	1
3	1	0.810254944	1	1
4	1	0.937420572	1	1
5	1	0.993602694	1	1
6	1	1	1	1
7	1	1	1	1
8	1	1	1	1
9	1	1	1	1
10	1	1	1	1
11	1	1	1	1
12	1	1	1	1
13	1	1	1	1
14	1	1	1	1
15	1	1	1	1.016429942
16	1	1	1	1.033202094
17	1	1	1	1.060979872
18	1	1	1	1.081586967
19	1.001096491	1	1	1.085033898
20	1.001096491	1	1	1.090589453
21	1.001644737	1	1	1.08774045
22	1.013563862	1	1	1.114032995
23	1.02431655	1	1	1.12831871
24	1.023768304	1	1	1.149686231
25	1.010752688	1	1	1.16232493
26	1.022932436	1	1	1.175311943
27	1.026404658	1	1	1.186423054
28	1.035875262	1	1	1.198701299
29	1.01354016	1	1	1.179761905
30	1.010067938	1	1	1.187169312
31	1.02623836	0.833333333	1	1.199074074
32	1.062678063	0.833333333	1	1.230324074
33	1.062678063	0.833333333	1	1.233173077
34	1.052910053	1	1	1.245077839
35	1.043805977	1	1	1.255494505
36	1.080843014	1	1	1.220238095
37	1.094521628	1	1	1.220238095
38	1.092229693	1	1	1.231499356
39	1.072736516	1	1	1.23851483
40	1.046765527	1	1	1.232301125
41	1.029738554	1	1	1.22504444
42	1.012194694	1	1	1.259695634
43	1.008614052	1	1	1.261580334
44	1.027777778	1	1	1.257575758
45	1.027777778	1	1	1.285353535
46	1.062865497	1	1	1.305555556
47	1.100577188	1	1	1.317460317
48	1.118121048	1	1	1.317460317
49	1.083033328	1	1.003703704	1.317460317
50	1.045980517	1	1.003703704	1.333333333
51	1.092539222	1	1.003703704	1.377777778
52	1.092539222	1	1	1.392162698
53	1.064102564	1	1.003679176	1.393603912
54	1.039673913	1	1.01706276	1.363048356
55	1.067451691	1	1.029547676	1.33279042
56	1.118733742	1	1.025868501	1.331349206
57	1.162393162	1	1.020388234	1.338225233
58	1.134615385	1.031017214	1.007903318	1.38584428
59	1.112745098	1.031017214	1.007903318	1.38584428
60	1.058049916	1.031017214	1.006944444	1.420634921
61	1.085600007	1	1.007715194	1.420634921
62	1.14234943	1	1.007715194	1.420634921
63	1.124552104	1	1.015757574	1.420634921
64	1.115520531	1	1.014986825	1.444444444
65	1.029359343	1	1.042764603	1.5
66	1.05496811	1	1.041111111	1.5
67	1.072723244	1	1.07771412	1.474747475
68	1.111817894	1	1.063059702	1.474747475
69	1.075368302	1	1.049726369	1.461926962
70	1.105761317	1	1.01312336	1.453846154
71	1.086131387	1	1.039243225	1.438694639
72	1.135514103	1	1.057265554	1.427705628
73	1.148291881	1	1.075203134	1.460102631
74	1.195493827	1.047619048	1.035959909	1.391920813
75	1.180306513	1.047619048	1.01793758	1.415730337
76	1.181270232	1.047619048	1	1.416666667
77	1.169743286	1.018518519	1.019607843	1.5
78	1.157314401	1.018518519	1.039759445	1.5
79	1.139969301	1.018518519	1.039759445	1.5
80	1.14605407	1	1.020151602	1.481481481
81	1.197084488	1.000533484	1	1.481481481
82	1.134021425	1.000533484	1.005493736	1.481481481
83	1.139463602	1.001976485	1.025101579	1.488781014
84	1.141672881	1.001443001	1.030657135	1.488781014
85	1.205836098	1.001443001	1.025163399	1.488781014
86	1.168005693	1	1.005555556	1.5
87	1.176332812	1	1	1.5
88	1.221801378	1.035087719	1.05688607	1.5
89	1.248520672	1.035087719	1.080803819	1.524107143
90	1.25796238	1.073549258	1.148195123	1.524107143
91	1.268765717	1.054528128	1.114306469	1.538439657
92	1.324321273	1.054528128	1.100900883	1.514332514
93	1.342657343	1.019625308	1.070546615	1.514332514
94	1.298778999	1.012105727	1.050327041	1.5
95	1.31610367	1.012105727	1.120888626	1.5
96	1.327805192	1.041880342	1.083851589	1.5
97	1.393156628	1.044205415	1.096160478	1.5
98	1.397081269	1.066308123	1.022630097	1.5
99	1.425724638	1.032616245	1.045260194	1.5

}{\tablebuiltinlq}

\pgfplotstableread{
x	zero	one	hund	thou
0	1	1	1	1
1	1	0.876319759	0.958333333	1
2	1	0.974358974	1	1.000107939
3	1	1	1	1.009345979
4	1	1	1	1.029547999
5	1	1	1	1.057609074
6	1	1	1	1.087586721
7	1	1	1	1.112429745
8	1.003777148	1	1	1.139816287
9	1.006312039	1	1	1.160947113
10	1.020180382	1	1	1.182568735
11	1.029223747	1	1	1.200542591
12	1.048764879	1	1	1.223298572
13	1.062674314	1	1	1.237977885
14	1.071534018	1	1	1.257041066
15	1.072093773	1	1.012000768	1.26998377
16	1.068961964	1	1.012429257	1.284649676
17	1.076267255	1	1.034651479	1.303168195
18	1.098075921	1	1.0309251	1.318234108
19	1.115096618	1	1.056137637	1.342136752
20	1.127777778	1	1.04906693	1.342136752
21	1.131206548	1	1.077829578	1.358971769
22	1.145095437	1.001289851	1.089225589	1.372390572
23	1.148362372	1.001289851	1.109376694	1.401600803
24	1.145943703	1.001289851	1.100117435	1.416511818
25	1.145943703	1.002038206	1.100919727	1.434839324
26	1.167676768	1.003957898	1.10343732	1.432199141
27	1.188888889	1.005598047	1.119137803	1.442495152
28	1.202777778	1.011795282	1.128917521	1.442349464
29	1.202777778	1.009875589	1.132763975	1.471334971
30	1.219444444	1.01035877	1.122619048	1.484848485
31	1.222052402	1.00212333	1.112037037	1.5
32	1.248908867	1.009567499	1.121553166	1.5
33	1.253294832	1.025197216	1.154886499	1.5
34	1.287195324	1.05283327	1.173405018	1.515339733
35	1.296411918	1.045389102	1.168552668	1.517687151
36	1.294141654	1.037159864	1.168552668	1.525040092
37	1.271522093	1.024061853	1.170968126	1.517546679
38	1.274521248	1.024993845	1.190571747	1.540752412
39	1.272405548	1.03145252	1.172389929	1.538449976
40	1.272405548	1.031407231	1.195857987	1.572270322
41	1.277777778	1.053672926	1.174368366	1.588383838
42	1.293543544	1.071596055	1.208262652	1.626190476
43	1.321321321	1.077865288	1.18742964	1.632142857
44	1.304327857	1.066922725	1.186566166	1.632142857
45	1.291457844	1.054214285	1.187520365	1.620659722
46	1.26399335	1.053431268	1.193580971	1.611502213
47	1.280986814	1.06732999	1.219033531	1.615290092
48	1.285666819	1.085222049	1.222499817	1.629370629
49	1.315067852	1.111314827	1.239166484	1.646464646
50	1.292758824	1.122198018	1.242596722	1.669969471
51	1.312960844	1.120145822	1.255797101	1.572227344
52	1.321848251	1.104574329	1.270614392	1.572227344
53	1.344157279	1.104574329	1.309261735	1.589533822
54	1.338785048	1.095461269	1.315946717	1.697376959
55	1.316320892	1.109868821	1.313938347	1.727954473
56	1.331163566	1.10730674	1.290707981	1.687143171
57	1.336535796	1.097086113	1.290927761	1.694887666
58	1.337064897	1.09903627	1.264861264	1.664310152
59	1.322222222	1.082105173	1.229165977	1.674411162
60	1.322222222	1.092775446	1.199267099	1.683015873
61	1.334821429	1.101473038	1.207495391	1.683015873
62	1.390376984	1.141395949	1.238643266	1.727460317
63	1.390376984	1.168577882	1.286637382	1.774603175
64	1.38252147	1.144907489	1.282407407	1.798738335
65	1.326965915	1.135059766	1.282407407	1.782071668
66	1.349188137	1.109024907	1.247150997	1.738292867
67	1.342929293	1.128991597	1.256410256	1.77764977
68	1.365151515	1.122826631	1.284188034	1.783205325
69	1.334382284	1.164734505	1.278705087	1.785765254
70	1.357109557	1.146215986	1.297935856	1.772578879
71	1.370258559	1.163585434	1.297935856	1.812261419
72	1.388495367	1.142856163	1.298388604	1.800740916
73	1.39255955	1.135163856	1.306935612	1.816611698
74	1.373950778	1.127622378	1.318631519	1.792206937
75	1.381631196	1.143589744	1.337815066	1.832083492
76	1.414876195	1.167594108	1.337815066	1.627815228
77	1.443324471	1.186487495	1.32611916	1.629709167
78	1.430785171	1.171777644	1.333333333	1.632220021
79	1.420412016	1.18879892	1.323207128	1.830216415
80	1.430756844	1.204549977	1.320180445	1.840097838
81	1.481481481	1.219259828	1.294112444	1.84333838
82	1.479166667	1.193881583	1.282527721	1.859822667
83	1.479166667	1.148543903	1.252748693	1.886905454
84	1.435820071	1.148351004	1.270464593	1.89802333
85	1.456653404	1.148729248	1.281064409	1.859729679
86	1.423320071	1.1748071	1.330964137	1.820240217
87	1.466666667	1.151344086	1.339316239	1.82132755
88	1.452146465	1.204915515	1.390831391	1.861416594
89	1.470328283	1.204915515	1.425369659	1.887235448
90	1.467980865	1.240453782	1.432538117	1.901982934
91	1.53727415	1.240857333	1.408317223	1.910775564
92	1.58341579	1.240857333	1.38843097	1.948088023
93	1.540332636	1.245641646	1.393513223	1.975865801
94	1.489640129	1.205492852	1.417734118	1.969347319
95	1.470554766	1.222159518	1.380558793	1.917053122
96	1.515985338	1.196297449	1.392864707	1.917053122
97	1.466450216	1.188423645	1.366349555	1.897547575
98	1.431818182	1.157635468	1.391001605	1.924762658
99	1.363636364	1.142857143	1.375	1.849525316
}{\tablepypilq}

\pgfplotstableread{
x	zero	one	hund	thou
0	1	1	1	1
1	0.689393939	0.277777778	0.72962963	0.914980326
2	0.850695015	0.5	0.876802452	0.998313659
3	0.934028348	0.642309797	0.943469119	1
4	0.994634409	0.753420909	0.980506156	1
5	1	0.827494983	1	1
6	1	0.907407407	1	1
7	1	0.935185185	1	1
8	1	0.972222222	1	1.000331741
9	1	0.956313131	1	1.004636178
10	1	0.984090909	1	1.004636178
11	1	0.984090909	1	1.008264336
12	1	1	1	1.022234754
13	1	1	1	1.022234754
14	1	1	1	1.049829111
15	1	1	1	1.048213194
16	1	1	1	1.071010129
17	1	1	1	1.039455871
18	1	1	1	1.037247802
19	1	1	1.013729495	1.046531881
20	1	1	1.013729495	1.077396079
21	1	1	1.025599734	1.207638412
22	1	1	1.016213484	1.262390643
23	1	1	1.018882136	1.311547199
24	1	1	1.007011897	1.190047759
25	1	1	1.018577744	1.117206474
26	1	1	1.015909091	1.070409525
27	1	1	1.015909091	1.096818939
28	1	1	1.018020095	1.19393809
29	1	1	1.018020095	1.160714286
30	1	1	1.018020095	1.114876307
31	1	1	1.002788907	1.15362514
32	1.007874016	1	1.005278084	1.216312827
33	1.007874016	1	1.012941592	1.291714298
34	1.007874016	1	1.065141347	1.237104523
35	1	1	1.06752645	1.235527947
36	1	1	1.069317541	1.213096587
37	1	1	1.037135896	1.228957529
38	1	1	1.042573801	1.226753539
39	1	1	1.041249284	1.253351566
40	1.002845839	1	1.036960785	1.192410523
41	1.002845839	1	1.0266486	1.146003401
42	1.002845839	1	1.027678441	1.115848915
43	1	1	1.018049663	1.065678847
44	1	1	1.018920744	1.128575673
45	1	1	1.026108509	1.095935546
46	1	1	1.019302101	1.169588745
47	1	1	1.018431021	1.20530303
48	1	1	1.021691176	1.240708366
49	1	1.029495761	1.045388831	1.230596834
50	1	1.029495761	1.045388831	1.119485723
51	1	1.029495761	1.025780988	1.063541667
52	1.005238737	1	1.010566302	1.035774411
53	1.012485114	1.003623188	1.010566302	1.245410694
54	1.017750003	1.003623188	1.040114325	1.313087461
55	1.031448429	1.022470728	1.047648926	1.339693613
56	1.03195399	1.01884754	1.052699432	1.16036036
57	1.060877135	1.01884754	1.070770457	1.192683593
58	1.055273306	1	1.060190042	1.13030303
59	1.047521368	1	1.055139537	1.261458333
60	1.013333333	1	1.049187156	1.214089912
61	1.005325674	1	1.041666667	1.191537002
62	1.033103452	1.006848358	1.05322262	1.092342171
63	1.040072177	1.006848358	1.046780425	1.215636518
64	1.034746503	1.006848358	1.062284301	1.215535706
65	1.006968726	1	1.078808058	1.380544932
66	1.017195969	1.018997446	1.131285921	1.368681506
67	1.018464537	1.034101351	1.115782045	1.391335227
68	1.02373179	1.034101351	1.09041519	1.367518464
69	1.006535821	1.015103905	1.017452085	1.250281361
70	1.022497207	1.007609014	1.039327085	1.250281361
71	1.022502059	1.028885609	1.133480326	1.046825397
72	1.025027994	1.042914932	1.118741097	0.988095238
73	1.00779804	1.035305918	1.191551339	1.038095238
74	1.04648919	1.065511694	1.136351909	1.095258539
75	1.043963255	1.065975125	1.139700123	1.140983647
76	1.092923082	1.065975125	1.045014881	1.106121919
77	1.090626495	1.014592197	0.936681548	1.042477136
78	1.116820059	1.003159457	0.933333333	1.076910758
79	1.084940347	1.003159457	0.933333333	1.172883598
80	1.098829236	1.008072545	1.033333333	1.310919386
81	1.120254718	1.012678337	1.091205936	1.430165418
82	1.12356588	1.012678337	1.129861124	1.42803889
83	1.107226011	1.007665805	1.124776378	1.421484583
84	1.130627169	1.006666667	1.066903775	1.240604504
85	1.132180109	1.006666667	1.031563354	1.191847194
86	1.13000146	1.02598219	1.004764042	1.33272021
87	1.10064792	1.019315523	1.050473613	1.555568888
88	1.139190461	1.08598219	1.084195883	1.603581315
89	1.173581995	1.114285714	1.189413275	1.571260931
90	1.170505933	1.154349817	1.143703704	1.465371877
91	1.147056212	1.08768315	1.1152218	1.460667531
92	1.119253986	1.079676779	1.015910983	1.234361276
93	1.080663382	1.054140521	1.015910983	1.083560726
94	1.053882755	1.080341394	1.00735585	0.715794393
95	1.116154508	1.040728718	1	0.80256511
96	1.142607007	1.049071287	1.029433681	0.940897225
97	1.165375288	1.081693944	1.050591967	1.427350427
98	1.089215785	1.122540915	1.07588795	1.5
99	1.099074074	1.176470588	1.063474858	1.666666667
}{\tablenotfoundlq}

\pgfplotstableread{
x	zero	one	hund	thou
0	1	1	1	1
1	2.166666667	1.888888889	1.888888889	2
2	2.277777778	2	2	2.111111111
3	2.5	2	2	2.277777778
4	2.722222222	2	2	2.5
5	2.888888889	2	2.047619048	2.697293447
6	2.916666667	2	2.1995671	2.863960114
7	2.916666667	2.166666667	2.366233766	2.975071225
8	3.053508772	2.333333333	2.531156806	3
9	3.303508772	2.52173913	2.573653199	3.041666667
10	3.470175439	2.52173913	2.629208754	3.139318885
11	3.444444444	2.52173913	2.75	3.282176028
12	3.611111111	2.555555556	2.888888889	3.407176028
13	3.777777778	2.722222222	3	3.531746032
14	3.857142857	2.888888889	3	3.638888889
15	3.69047619	3	3	3.805555556
16	3.69047619	2.952380952	3	3.916666667
17	3.833333333	3.015343915	3	4
18	4.066666667	3.015343915	3	4
19	4.066666667	3.146296296	3.056944444	4
20	4.066666667	3.194444444	3.123611111	4
21	4.208333333	3.336309524	3.234722222	4.111111111
22	4.208333333	3.443452381	3.344444444	4.277777778
23	4.408333333	3.499007937	3.444444444	4.444444444
24	4.311111111	3.468253968	3.5	4.555555556
25	4.480982906	3.611111111	3.666666667	4.611111111
26	4.614316239	3.777777778	3.813265306	4.666666667
27	4.669871795	3.833333333	3.979931973	4.779513889
28	4.833333333	3.833333333	3.979931973	4.890625
29	4.666666667	3.722222222	3.919590643	5.001736111
30	4.738756614	3.888888889	3.919590643	5
31	4.754137272	3.888888889	3.919590643	5.011904762
32	4.920803939	4	4	5.011904762
33	5.015380658	4.027777778	4.066666667	5.123015873
34	5.043025362	4.361111111	4.095652174	5.444444444
35	5.125634058	4.639619883	4.198429952	5.777777778
36	5.181189614	4.611842105	4.131763285	5.777777778
37	5.471497585	4.278508772	4.171296296	5.611111111
38	5.388888889	3.920833333	4.151851852	5.611111111
39	5.666666667	4.254166667	4.285185185	5.666666667
40	5.333333333	4.320833333	4.438888889	5.833333333
41	5.333333333	4.483333333	4.442676768	5.804761905
42	5.111111111	4.483333333	4.476010101	5.971428571
43	5.111111111	4.583333333	4.533549784	5.942443064
44	5.186026936	4.5	4.613095238	5.859903382
45	5.074915825	4.5	4.743055556	5.877958937
46	5.408249158	4.666666667	4.743055556	5.906944444
47	5.784722222	5	4.788510101	6.018055556
48	5.784722222	4.877840909	4.714105339	6.083333333
49	5.451388889	4.877840909	4.767676768	6.191919192
50	5.286675821	4.977350713	4.888888889	6.303030303
51	5.397786932	5.432843137	5	6.553030303
52	5.729683484	5.349509804	4.866666667	6.611111111
53	5.276340996	5.10952381	4.811111111	6.633333333
54	5.498563218	5.183597884	4.899252137	6.466666667
55	5.333333333	5.266931217	5.03258547	6.3
56	5.833333333	5.407407407	5.088141026	6.388888889
57	5.8625	5.333333333	5	6.888888889
58	5.778174603	5.666666667	5	7.111111111
59	5.490818281	5.666666667	5	7.222222222
60	5.350540504	5.555555556	5	6.680555556
61	5.601532567	5.405555556	5.01510832	6.791666667
62	6.114686469	5.399382716	5.01510832	6.791666667
63	6.128575358	5.377160494	5.126219431	6.92745098
64	6.075780326	5.527160494	5.194444444	6.829890005
65	5.738871636	5.616666667	5.305555556	6.77054657
66	5.836093858	5.861111111	5.361111111	6.84309559
67	5.872751323	5.861111111	5.611111111	7.107323232
68	5.983862434	5.666666667	5.5	7.277777778
69	5.650529101	5.604938272	5.555555556	7.444444444
70	5.833333333	5.604938272	5.388888889	7.5
71	5.5	5.781200898	5.588888889	7.555555556
72	5.722222222	5.731818182	5.7	7.555555556
73	5.555555556	5.641706539	5.866666667	7.425
74	6.093376068	5.826555024	5.666666667	7.425
75	6.204487179	5.826555024	5.5	7.508333333
76	6.311630037	5.583333333	5.418803419	8.305555556
77	6.218253968	6.166666667	5.810470085	8.145178197
78	6.218253968	6.388888889	5.977136752	8.22851153
79	6.388888889	6.888888889	5.876205451	7.685776488
80	6.388888889	6.555555556	5.817872117	8.012820513
81	6.611111111	6.395299145	5.817872117	8.012820513
82	6.5	6.061965812	6.074894382	8.113715278
83	6.555555556	6.228632479	6.408227716	8.126535791
84	6.666666667	6.5	6.574894382	8.348758013
85	6.909722222	6.739583333	6.5	8.235042735
86	7.076388889	6.295138889	6.055555556	8.351851852
87	7.409722222	6.42590812	6.051767677	8.425925926
88	7.469047619	7.307014441	6.218434343	8.592592593
89	7.469047619	7.529236664	6.396212121	8.796296296
90	7.433630952	7.509578544	6.280952381	8.833333333
91	7.708442982	6.914906988	6.280952381	9
92	7.541776316	6.803795877	6.214285714	8.916666667
93	6.910526316	6.854965468	6.333333333	8.916666667
94	6.894230769	6.828947368	6.166666667	8.830357143
95	7.227564103	7.145614035	6.366666667	8.913690476
96	7.672008547	7.030952381	6.2	8.913690476
97	7.777777778	7.030952381	6.311111111	8.916666667
98	7.916666667	7.071428571	6.166666667	8.875
99	8.5	7	6.333333333	8.75
}{\tablebuiltinhq}

\pgfplotstableread{
x	zero	one	hund	thou
0	1	1	1	1
1	2.083333333	1.888888889	1.888888889	2.192307692
2	2.25	2	2.02797619	2.458974359
3	2.472222222	2.083333333	2.139087302	2.766666667
4	2.722222222	2.229166667	2.305753968	3.035897436
5	2.888888889	2.395833333	2.5	3.294871795
6	3.111111111	2.5625	2.722222222	3.569381599
7	3.289316239	2.75	2.888888889	3.80015083
8	3.601816239	2.916666667	3	4.036414566
9	3.824038462	3	3.047619048	4.261904762
10	3.979166667	3.031385281	3.121693122	4.484126984
11	4.111111111	3.114718615	3.288359788	4.722222222
12	4.29968254	3.248051948	3.407407407	4.888888889
13	4.521904762	3.35952381	3.535443723	5
14	4.744126984	3.456843157	3.590999278	5.166666667
15	4.888888889	3.556843157	3.757665945	5.333333333
16	5.041904762	3.636208236	3.888888889	5.527777778
17	5.184761905	3.788888889	4.043859649	5.694444444
18	5.406984127	3.888888889	4.077192982	5.861111111
19	5.698412698	4	4.097471582	6.130555556
20	5.888888889	4.066666667	4.2202786	6.263888889
21	6.05771729	4.161904762	4.353611933	6.412037037
22	6.224383956	4.295238095	4.5	6.473941799
23	6.35771729	4.33968254	4.555555556	6.593277802
24	6.357971014	4.411111111	4.611111111	6.722907431
25	6.324637681	4.527777778	4.777777778	6.863780447
26	6.514221014	4.638888889	4.888888889	6.944444444
27	6.789583333	4.805555556	5	7.104166667
28	7.15625	4.888888889	5	7.305448718
29	7.5	4.961012312	5.066666667	7.51756993
30	7.714646465	4.961012312	5.066666667	7.746736597
31	7.985159285	4.961012312	5.177777778	7.878787879
32	7.930249724	5	5.222222222	8
33	8.213351008	5.133333333	5.412698413	8
34	8.220615965	5.244444444	5.523809524	8.119642857
35	8.608858859	5.494444444	5.654553049	8.286309524
36	8.738970588	5.583333333	5.797410192	8.597420635
37	8.96119281	5.638888889	5.797410192	8.738888889
38	8.84338978	5.588888889	5.888888889	8.863888889
39	8.849217562	5.7	5.888888889	8.886111111
40	8.726717562	5.806060606	6	9.093055556
41	9.066742814	5.939393939	6	9.326231656
42	9.433055556	6.022727273	6	9.469088799
43	9.756280193	6.083333333	6.111111111	9.57729587
44	9.67294686	6.058809524	6.135912698	9.656619769
45	9.67294686	5.97547619	6.278769841	9.815113978
46	9.805555556	6.058809524	6.334325397	10.07218468
47	10.14479167	6.083333333	6.491341991	10.13468468
48	10.04128372	6.191919192	6.681818182	10.23333333
49	10.04128372	6.241919192	6.726957071	10.24962121
50	10.18857539	6.491919192	6.795138889	10.39247835
51	10.49222428	6.716666667	6.795138889	10.25914502
52	10.82555761	6.75	6.983333333	10.24285714
53	10.64458539	6.801515152	7.066666667	10.325
54	10.87301587	6.81112499	7.204761905	10.68333333
55	10.9265873	6.977791657	7.304761905	10.62777778
56	11.14115646	7.009609838	7.630030722	10.59325397
57	11.3042517	7	7.491935484	10.5953373
58	11.34591837	7	7.325268817	10.6902868
59	11.38134921	7	7	10.86092172
60	11.12301587	7	7	10.95407648
61	11.19965278	7.166666667	7.166666667	11.08134921
62	11.83854167	7.277777778	7.480728336	11.3202381
63	12.171875	7.501736111	7.814061669	11.6
64	12.22256944	7.335069444	7.758506114	12.0010181
65	12.01145833	7.484540344	7.777777778	12.07130795
66	12.01145833	7.295341369	7.611111111	12.07130795
67	12.12222222	7.545341369	7.722222222	11.85547504
68	12.01515152	7.479835116	7.722222222	11.87037037
69	11.71515152	7.806583694	7.888888889	11.96263228
70	11.78266385	7.826280664	7.866666667	12.27744709
71	12.34270173	7.96453824	7.961904762	12.61950549
72	13.08714617	7.936363636	7.961904762	12.52724359
73	13.37104994	7.916666667	8.171448921	12.24946581
74	13.24030499	7.833333333	8.177660102	12.30732861
75	12.96252721	7.907407407	8.110993435	12.64066194
76	13.11111111	8.021043771	8.034782609	12.41843972
77	13.16666667	8.400673401	7.933333333	12
78	13.38492063	8.326599327	8.26171875	12.10144928
79	13.4765873	8.642532855	8.413462249	12.79546637
80	13.70912698	8.679569892	8.635684471	13.33713304
81	14.24503968	9.012903226	8.519373884	13.45790598
82	14.5422619	9	8.700963719	13.72685185
83	14.14305556	8.867592593	8.706519274	13.96990741
84	13.61919192	8.800925926	9.075925926	14.22536376
85	13.73030303	8.707852333	9.075925926	14.28621032
86	13.84141414	8.748268398	9.153703704	14.08482143
87	14.19444444	8.814935065	8.972222222	14.0515873
88	13.98703023	9.278379028	9.194444444	14.02777778
89	14.37591912	9.37037037	9.4625	14.27777778
90	14.53068102	9.537037037	9.729166667	14.58333333
91	15.1547619	9.373502467	9.729166667	15.08333333
92	15.47142857	9.540169133	9.743518519	15.18333333
93	15.48346547	9.373502467	9.858101852	15.14761905
94	15.22751309	9.5	10.0275463	14.81719269
95	15.11672877	9.5	10.27291667	14.81302602
96	15.4251552	9.333333333	10.55833333	14.23872569
97	15.08495373	9.268181818	10.66666667	14.42916667
98	15.06860707	9.152272727	10.75	14.25
99	14.46153846	9.804545455	10.5	15.08004386
}{\tablepypihq}

\pgfplotstableread{
x	zero	one	hund	thou
0	1	1	1	1
1	1.916666667	1.7	1.815705128	2.083333333
2	2	1.866666667	1.949038462	2.25
3	2.111111111	2	2.038751987	2.44047619
4	2.277777778	2	2.11718336	2.607142857
5	2.5	2.043154762	2.260040503	2.773809524
6	2.655555556	2.186011905	2.430812325	3.007575758
7	2.822222222	2.352678571	2.584325397	3.090909091
8	2.933333333	2.531746032	2.719246032	3.178409091
9	3.029260686	2.611111111	2.843055556	3.194642857
10	3.195927353	2.694444444	2.944444444	3.240695489
11	3.362594019	2.805555556	3	3.469862155
12	3.611111111	2.916666667	3.034259259	3.627401838
13	3.694444444	3	3.116203704	3.783923732
14	3.861111111	3	3.199537037	3.800590399
15	3.916666667	3.025897436	3.272916667	4.049796748
16	4	3.081452991	3.435416667	4.086404915
17	4.066666667	3.248119658	3.560416667	3.919738248
18	4.233333333	3.333333333	3.552777778	4.072516026
19	4.4	3.444444444	3.641666667	4.166666667
20	4.5	3.441570881	3.766666667	4.666666667
21	4.606060606	3.552681992	4.140229885	4.736111111
22	4.689393939	3.719348659	4.290336723	4.958333333
23	4.758669007	3.777777778	4.531408151	4.86875
24	4.819275068	3.732026144	4.557844933	5.021527778
25	4.958163957	3.636788049	4.564347291	4.799305556
26	5.055555556	3.74789916	4.323275862	4.777777778
27	5.103174603	3.904761905	4.267720307	4.555555556
28	5.214285714	4	4.329812207	4.955555556
29	5.380952381	4	4.502034429	5.308333333
30	5.5	4.069230769	4.557589984	5.308333333
31	5.63986014	4.069230769	4.672222222	5.345833333
32	5.806526807	4.095641026	4.754166667	5.336706349
33	5.973193473	4.312124542	4.920833333	5.424503968
34	6.066666667	4.52740232	4.831547619	5.752876984
35	6.066666667	4.66765873	4.910714286	5.65922619
36	6.233333333	4.715277778	4.956827593	5.880952381
37	6.166666667	4.698039216	5.046113307	5.68452381
38	6.194444444	4.769467787	5.097491753	5.878968254
39	6.194444444	4.769467787	4.974455369	5.902777778
40	6.394444444	4.904761905	4.974455369	5.873363095
41	6.626587302	5	4.810331825	6.227529762
42	6.571031746	5	5.093315508	5.910863095
43	6.461358086	5	5.174197861	5.810416667
44	6.534770785	4.833333333	5.420276292	5.914583333
45	6.756993007	4.833333333	5.51004902	6.255132114
46	7	4.935897436	5.475897563	6.68546748
47	6.961328976	5.110169923	5.467564229	6.032689702
48	6.948508463	5.310169923	5.338397563	6.258807588
49	6.948508463	5.540939153	5.361904762	6.047222222
50	7.011881868	5.553835979	5.170238095	5.838888889
51	7.13367674	5.456684982	5.253571429	5.088888889
52	7.276533883	5.441116119	5.627777778	4.652777778
53	7.418498168	5.587280141	5.916666667	5.138888889
54	7.547619048	5.817764471	5.718849991	5.895833333
55	7.619667208	5.875886525	5.450421806	5.916326729
56	7.786333874	6.042553191	5.478199584	5.516326729
57	7.88157197	5.799497636	5.759349593	4.965264637
58	7.952380952	5.756944444	5.643809524	5.22254902
59	7.841269841	5.450231481	5.636898955	5.669771242
60	7.591269841	5.572074916	5.443717137	5.880555556
61	7.638888889	5.710963805	5.523412258	5.746085859
62	7.75	6.128787879	5.613656161	5.605726381
63	8.222222222	6.513888889	6.199695122	6.370877897
64	7.972222222	6.808333333	6.392857143	6.67201426
65	7.972222222	6.697222222	6.476190476	6.808568882
66	8.035714286	6.6	6.383333333	5.837451458
67	8.396825397	6.237962963	6.433333333	6.118701458
68	8.73015873	6.237962963	6.419444444	7.325284091
69	8.267973856	6.237962963	6.302777778	8.233630952
70	8.490196078	6.5	6.330555556	8.235714286
71	8.323529412	6.833333333	6.427777778	6.652380952
72	8.833333333	7	6.444444444	5.745454545
73	8.477777778	6.86969697	6.7	5.692513369
74	8.644444444	6.86969697	7.2	6.992513369
75	8.311111111	6.758585859	7.222222222	7.317429194
76	8.919642857	6.775757576	6.888888889	7.282706972
77	8.94206292	6.580445076	5.833333333	6.774373638
78	9.289868674	6.691556187	5.544850949	6.629382056
79	8.870225817	6.8046875	5.694850949	7.267045455
80	8.958916865	6.9375	6.417073171	6.935795455
81	9.277777778	7.346861472	6.816666667	8.960416667
82	9.543345543	7.346861472	6.846428571	8.888988095
83	9.717948718	7.199802648	6.690873016	8.916666667
84	9.717948718	7.12377451	6.702777778	6.4
85	9.73015873	7.091254185	6.68968254	5.335675883
86	10.00326797	7.308943089	7.100793651	6.722580645
87	10.00326797	6.714952334	7.116666667	8.211974585
88	10.39215686	7.080805993	7.838888889	8.902922078
89	10.16666667	7.239342578	8.041666667	7.861255411
90	10.48787879	7.68438914	7.583333333	7.638528139
91	10.48787879	7.351055807	6.852136752	7.479166667
92	10.48787879	7.476055807	6.060470085	7.5625
93	10.02083333	7.625	6.605840456	7.139583333
94	9.854166667	7.958333333	6.596957672	7.360416667
95	10.1875	7.666666667	6.891829467	8.235416667
96	10.5	7.732017544	6.443681319	9.991666667
97	10.61111111	8.232017544	6.961538462	10.20833333
98	10.41666667	8.848026316	7	10
99	10.33333333	9.5	8	8
}{\tablenotfoundhq}

\begin{tikzpicture}

 \begin{groupplot}[
    group style={%
        group size=4 by 1,%
        x descriptions at=edge bottom,%
        y descriptions at=edge left,%
        horizontal sep=10pt,%
        vertical sep=4pt,%
    },
    y label style={at={(axis description cs:0.20,.5)}, anchor=south},
    clip=true,
    clip mode=individual,
    width=2.0in,
    height=1.75in, 
    ymin = 0,
    ymax = 16,
    ymode = log,
    xmin=1,
    ymin = 0.9,
    xtick={1,20,40,60,80,100},
    yticklabels={1,5x,10x,15x},
    ytick={1,5,10,15},
    xmax=100,
    legend columns=3,
    legend style={draw=none},
    ]

            \nextgroupplot[
            ylabel style={align=center},
            ylabel = {Growth from Adoption (LOC$_\ell$)},
            title = {SO Posts = 0},
            xlabel = {Commit \# After Adoption},
        ]
    \addplot [blue!80!white, thick, dashed, ]  table [x=x, y=zero]   {\tablepypihq};
    \addplot [blue, thick, mark=square,  mark repeat=25]  table [x=x, y=zero]   {\tablepypi};
    \addplot [blue!80!white, thick, dashed, ]  table [x=x, y=zero]   {\tablepypilq};
    
            \nextgroupplot[
            title = {{SO Posts $\in$ [1,100)}},
            xlabel = {Commit \# After Adoption},
        ]
    \addplot [blue!80!white, thick, dashed, ]  table [x=x, y=one]   {\tablepypihq};
    \addplot [blue, thick, mark=square,  mark repeat=25]  table [x=x, y=one]   {\tablepypi};
    \addplot [blue!80!white, thick, dashed, ]  table [x=x, y=one]   {\tablepypilq};
    
        \nextgroupplot[
        title = {{SO Posts $\in$ [100,1000)}},
        xlabel = {Commit \# After Adoption},
        ]
    \addplot [blue!80!white, thick, dashed, ]  table [x=x, y=hund]   {\tablepypihq};
    \addplot [blue, thick, mark=square,  mark repeat=25]  table [x=x, y=hund]   {\tablepypi};
    \addplot [blue!80!white, thick, dashed, ]  table [x=x, y=hund]   {\tablepypilq};
    
            \nextgroupplot[
            title = {{SO Posts $\in$ [1000,$\infty$)}},
            xlabel = {Commit \# After Adoption},
        ]
    \addplot [blue!80!white, thick, dashed, ]  table [x=x, y=thou]   {\tablepypihq};
    \addplot [blue, thick, mark=square,  mark repeat=25]  table [x=x, y=thou]   {\tablepypi};
    \addplot [blue!80!white, thick, dashed, ]  table [x=x, y=thou]   {\tablepypilq};
  
    \end{groupplot}

\end{tikzpicture}    
    \caption{Growth of library usage after adoption grouped by Stack Overflow usage. Q1, Median, and Q3 growth in lines of code referencing $\ell$ are represented from bottom to top respectively.}
    \label{fig:StackOverflow}
\end{figure*}
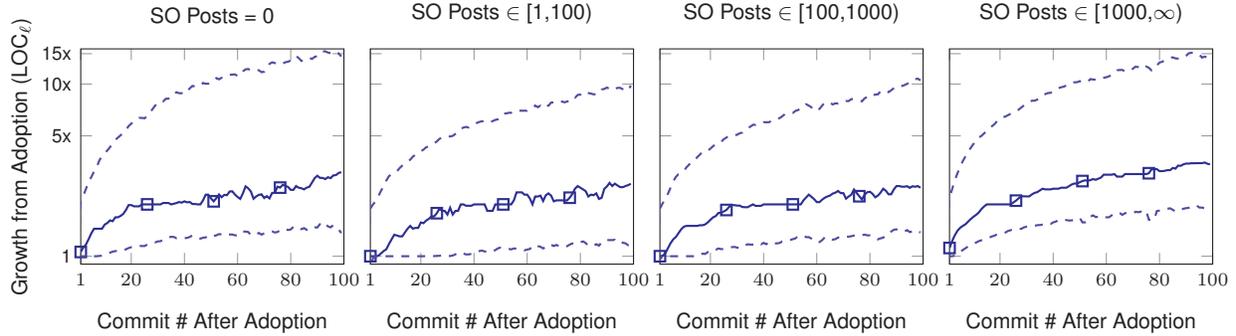

\smallskip
\noindent\textbf{Stack Overflow.}  \qquad
To find the impact of Stack Overflow on the speed of library adoptions, we plot the number of users of $\ell$ by the mean average number of Stack Overflow posts (across all adoption times) in Fig.~\ref{fig:StackOverflow_users} that existed when $\ell$ was referenced. We observe a strong positive correlation between the number of library users and the number of Stack Overflow mentions for standard libraries ($R^2$=0.625, $p<$0.001) and PyPi libraries ($R^2$=0.410, $p<$0.001). There is a small positive correlation between usage of unknown libraries and Stack Overflow posts ($R^2$=0.08, $p<$0.001).

We also investigate the impact of Stack Overflow popularity on library growth. Formally, we compute the growth of a library $\ell$ within a project as follows. If $x$=$0$, then let $y_{x}$=$1$; otherwise
$y_{x} = y_{x-1} \left(\frac{\sum\limits_{i=0}^{x-1}{(n_i)} + n_x}{\sum\limits_{i=0}^{x-1}{(n_i)}}\right),$
where $n_i$ is the number of changed lines of code in commit $i$ that contain $\ell$.

We plot the median growth (in LOC referencing $\ell$) as a function of the number of commits after the adoption in Fig.~\ref{fig:StackOverflow}. The primary distinction is in the growth rates for libraries with more than 1000 Stack Overflow posts. 100 commits after the adoption, the adoption of a highly mentioned library on Stack Overflow will have approximately 350\% growth (on average) compared to only 250\% growth for less mentioned libraries. 


When a library does not appear on Stack Overflow, the growth rate is similar to libraries that have over 1000 posts. Libraries that do not appear at all in Stack Overflow mostly consist of libraries that were written by developers who are also the authors committing the library to the repository. This may explain why growth is large in unknown libraries -- the adopters know how to use the library because they wrote it. 


\smallskip
\noindent\textbf{Project Team Size.}  \qquad Next, we investigate differences in library adoptions as a function of team size. Git and GitHub directly store how team members collaborate and the types of activities that they perform~\cite{middleton2018contributions}. For example, researchers have found that diversity and team makeup have a significant impact on the productivity of GitHub teams~\cite{vasilescu2015data}, and larger teams tend to process more pull requests on average~\cite{vasilescu2015quality}.

\begin{figure}[t]

\pgfplotstableread{
x	one	two	three	six	ten
1	1.03030303	1.018518519	1.007673132	1.006259919	1
2	1.096969697	1.066137566	1.041006465	1.019080432	1.011111111
3	1.18030303	1.149470899	1.107673132	1.05754197	1.025603865
4	1.264942529	1.242063492	1.183333333	1.122710623	1.050603865
5	1.344109195	1.324461785	1.254166667	1.193223443	1.08578905
6	1.427442529	1.407795118	1.298611111	1.25	1.140079365
7	1.479166667	1.463350674	1.356691919	1.278106222	1.181746032
8	1.5	1.5	1.419191919	1.361439555	1.233010711
9	1.538461538	1.5	1.474747475	1.432868127	1.275338753
10	1.594017094	1.533333333	1.5	1.5	1.29200542
11	1.64957265	1.588888889	1.515151515	1.5	1.305555556
12	1.686868687	1.644444444	1.559011164	1.5	1.305555556
13	1.75036075	1.738506705	1.592344498	1.5	1.333333333
14	1.819805195	1.7515786	1.577192982	1.5	1.388888889
15	1.910714286	1.796023044	1.588888889	1.5	1.444444444
16	1.958333333	1.795179175	1.632791328	1.533333333	1.405797101
17	2	1.893218391	1.688346883	1.557142857	1.305797101
18	2	1.959885057	1.726287263	1.612698413	1.298852657
19	2	2	1.749051491	1.634920635	1.393055556
20	2	2	1.860162602	1.675925926	1.493055556
21	2	2	1.85	1.703703704	1.5
22	2	2	1.916666667	1.814814815	1.523809524
23	2	2	1.916666667	1.903846154	1.54985119
24	2.023809524	2	1.986111111	1.977179487	1.54985119
25	2.023809524	2	1.986111111	1.977179487	1.559375
26	2.09047619	2.007246377	1.986111111	1.99	1.575
27	2.09047619	2.007246377	2	2	1.630555556
28	2.157142857	2.007246377	2	2	1.669025747
29	2.201587302	2.062622549	2	2	1.682914636
30	2.288888889	2.094368581	2	2	1.62735908
31	2.340073529	2.161035247	2	2	1.638888889
32	2.362295752	2.199861974	2	2	1.527777778
33	2.391782614	2.245039019	2	2	1.5638322
34	2.415782414	2.311705686	2	2	1.588013075
35	2.449115747	2.364102564	2	2	1.810235297
36	2.47518444	2.398290598	2	2.03030303	1.903810505
37	2.5	2.442552893	2.077350427	2.03030303	1.796296296
38	2.49122807	2.455373406	2.077350427	2.03030303	1.701058201
39	2.435672515	2.455373406	2.188461538	2	1.738095238
40	2.48112706	2.444444444	2.194444444	2	1.793650794
41	2.545454545	2.444444444	2.254662005	2	1.7858827
42	2.621843434	2.526067123	2.143550894	2	1.738263652
43	2.631944444	2.50153025	2.121756022	2	1.804839662
44	2.576388889	2.50153025	2.144871795	2	1.907845851
45	2.626984127	2.475463127	2.172649573	2	1.788798232
46	2.571428571	2.55	2.245543346	2.066666667	1.80877193
47	2.708769473	2.55	2.273321123	2.066666667	1.7784689
48	2.752364375	2.583333333	2.356654457	2.177777778	1.94114355
49	2.88649136	2.588888889	2.361111111	2.111111111	1.910149398
50	2.915817125	2.672222222	2.413265306	2.177777778	1.940452428
51	2.967460317	2.694444444	2.419801254	2.133333333	1.944444444
52	3	2.638888889	2.414245698	2.133333333	2.006535948
53	3	2.583333333	2.341889483	2.15	2.006535948
54	3	2.611111111	2.390909091	2.183333333	2.006535948
55	2.985915493	2.777777778	2.457575758	2.266666667	2
56	2.985915493	2.861111111	2.533333333	2.35	2
57	2.985915493	2.87037037	2.588888889	2.341954023	2
58	3	2.850762527	2.566666667	2.341954023	2.060606061
59	3	2.906318083	2.6	2.234791714	2.108225108
60	3.025641026	2.847058824	2.522705314	2.253948802	2.074891775
61	2.858974359	2.866666667	2.554451346	2.337282135	2.014285714
62	2.883699634	2.866666667	2.558034799	2.425925926	1.966666667
63	2.913614164	3	2.603583454	2.505291005	2
64	3.08028083	3	2.627392977	2.495487084	2
65	3.055555556	3	2.673809524	2.484593838	2
66	3	3	2.733333333	2.460784314	2
67	3.03030303	3	2.785714286	2.563180828	2
68	3.03030303	3.066666667	2.702380952	2.592592593	2.0625
69	3.113636364	3.066666667	2.719047619	2.592592593	2.0625
70	3.157407407	3.064804469	2.6	2.5	2.0625
71	3.296296296	3.023778828	2.683333333	2.523809524	2
72	3.37962963	3.023778828	2.75	2.49047619	2
73	3.362466125	3.06027306	2.833333333	2.546031746	2
74	3.318815331	3.034632035	2.848232057	2.51965812	2
75	3.235481998	3.034632035	2.755639465	2.552991453	1.962962963
76	3.305892383	3.052380952	2.705639465	2.543359046	1.962962963
77	3.321765399	3.052380952	2.774074074	2.535465633	2.014591011
78	3.371765399	3.076326272	2.825	2.574681319	2.134961381
79	3.411111111	3.066701081	2.891666667	2.614616756	2.151552633
80	3.411111111	3.150034415	2.780555556	2.625074272	2.211035696
81	3.462962963	3.192755762	2.822222222	2.585858586	2.127702363
82	3.376756066	3.261111111	2.888888889	2.5	2.135802469
83	3.265644955	3.288888889	3.014015844	2.666666667	2.024691358
84	3.173793103	3.388888889	3.014015844	2.7	2.024691358
85	3.226666667	3.28989066	3.014015844	2.866666667	2
86	3.44630491	3.257677775	2.857142857	2.81359953	2.035087719
87	3.600834824	3.142162338	2.869047619	2.724710641	2.035087719
88	3.634168157	3.163382789	2.888655462	2.724710641	2.035087719
89	3.552625153	3.160824889	3.114845938	2.777777778	2.166666667
90	3.46031746	3.232887945	3.102941176	3	2.269230769
91	3.543650794	3.282887945	3.083333333	2.962962963	2.435897436
92	3.6498315	3.35589051	3	2.962962963	2.462213225
93	3.760942611	3.375533367	3	2.962962963	2.382871233
94	3.716070816	3.458866701	2.960784314	3	2.327315677
95	3.802041785	3.476190476	2.960784314	2.990196078	2.194939282
96	3.841469448	3.533333333	3.047496025	2.990196078	2.260668762
97	3.969674577	3.584242424	3.253378378	2.990196078	2.205113207
98	3.833872107	3.584242424	3.364489489	2.939632546	2.232226444
99	3.791666667	3.576363636	3.416666667	2.909448819	2.214912281
100	3.583333333	3.5	3.333333333	2.818897638	2.263157895

}{\table}
\pgfplotstableread{
Size    Number
1	0.590837131
2	0.241474811
3	0.072450147
4	0.029347529
5	0.01516488
6	0.009294604
7	0.005989685
8	0.004329521
9	0.003004472
10	0.002438245
11	0.001952907
12	0.001629348
13	0.001413643
14	0.001194085
15	0.001009194
16	0.000855119
17	0.000697192
18	0.000620154
19	0.000624006
20	0.000516153
21	0.000543116
22	0.000473782
23	0.000516153
24	0.000377485
25	0.000323559
26	0.000365929
27	0.000342818
28	0.00034667
29	0.000238817
30	0.000246521
31	0.000342818
32	0.000215706
33	0.000223409
34	0.000227261
35	0.000227261
36	0.000215706
37	0.000234965
38	0.000154075
39	0.000157927
40	0.000165631
41	0.000157927
42	0.000146372
43	0.00014252
44	9.24453E-05
45	0.000130964
46	0.00014252
47	0.000119409
48	0.000154075
49	0.000119409
50	0.000130964
51	0.000146372
52	0.000169483
53	0.000100149
54	0.000157927
55	0.000177187
56	0.000130964
57	0.00014252
58	0.000104001
59	7.70377E-05
60	0.000154075
61	0.00012326
62	0.000130964
63	0.000146372
64	0.00012326
65	0.000154075
66	8.08896E-05
67	0.000119409
68	8.47415E-05
69	0.00014252
70	0.000107853
71	4.23708E-05
72	7.31859E-05
73	7.31859E-05
74	0.000115557
75	0.000104001
76	0.000104001
77	7.31859E-05
78	7.70377E-05
79	7.70377E-05
80	6.54821E-05
81	4.23708E-05
82	4.23708E-05
83	6.54821E-05
84	0.000173335
85	9.62972E-05
86	6.9334E-05
87	5.00745E-05
88	8.08896E-05
89	5.39264E-05
90	4.23708E-05
91	7.31859E-05
92	2.31113E-05
93	2.31113E-05
94	4.62226E-05
95	3.08151E-05
96	4.62226E-05
97	5.39264E-05
98	7.70377E-05
99	6.54821E-05
100	8.08896E-05
101	3.08151E-05
102	3.08151E-05
103	5.39264E-05
104	2.69632E-05
105	2.69632E-05
106	4.23708E-05
107	7.70377E-06
108	1.92594E-05
109	3.85189E-05
110	2.31113E-05
111	3.4667E-05
112	1.92594E-05
113	3.4667E-05
114	2.31113E-05
115	1.92594E-05
116	1.54075E-05
117	2.31113E-05
118	3.08151E-05
119	1.54075E-05
120	7.70377E-06
121	1.92594E-05
122	1.92594E-05
123	3.08151E-05
124	7.70377E-06
125	7.70377E-06
126	3.08151E-05
127	2.31113E-05
128	7.70377E-06
129	3.4667E-05
130	2.69632E-05
131	1.54075E-05
132	2.69632E-05
133	3.85189E-05
134	1.54075E-05
135	1.54075E-05
136	7.70377E-06
137	1.15557E-05
138	1.54075E-05
139	2.31113E-05
140	1.54075E-05
141	3.4667E-05
142	7.70377E-06
143	1.15557E-05
144	7.70377E-06
145	2.31113E-05
146	1.54075E-05
147	1.15557E-05
148	2.69632E-05
149	1.15557E-05
150	1.92594E-05
151	1.54075E-05
152	1.92594E-05
153	1.54075E-05
154	1.15557E-05
155	7.70377E-06
156	1.54075E-05
157	1.92594E-05
158	1.15557E-05
159	3.08151E-05
160	1.54075E-05
161	3.85189E-06
162	1.54075E-05
163	1.15557E-05
164	7.70377E-06
165	1.92594E-05
166	1.15557E-05
167	2.31113E-05
169	1.54075E-05
171	2.69632E-05
172	7.70377E-06
173	7.70377E-06
174	1.15557E-05
175	1.15557E-05
176	1.54075E-05
177	1.92594E-05
179	1.15557E-05
180	1.92594E-05
181	7.70377E-06
182	1.15557E-05
183	3.85189E-06
184	1.15557E-05
186	7.70377E-06
187	7.70377E-06
188	1.15557E-05
189	2.31113E-05
190	1.54075E-05
191	2.31113E-05
192	1.15557E-05
193	1.15557E-05
194	3.85189E-06
195	1.92594E-05
196	3.85189E-06
198	1.15557E-05
199	7.70377E-06
200	7.70377E-06
201	1.54075E-05
202	7.70377E-06
203	3.85189E-06
204	1.15557E-05
205	3.85189E-06
206	3.85189E-06
209	3.85189E-06
210	1.54075E-05
211	1.15557E-05
212	7.70377E-06
213	3.85189E-06
214	7.70377E-06
215	1.54075E-05
216	1.15557E-05
220	7.70377E-06
221	7.70377E-06
222	3.85189E-06
223	3.85189E-06
224	7.70377E-06
225	1.15557E-05
226	7.70377E-06
227	3.85189E-06
228	1.15557E-05
229	1.54075E-05
230	1.15557E-05
231	1.54075E-05
232	1.54075E-05
233	1.15557E-05
234	1.15557E-05
235	3.85189E-06
236	7.70377E-06
237	7.70377E-06
238	3.85189E-06
240	1.15557E-05
242	3.85189E-06
243	1.15557E-05
244	7.70377E-06
245	3.85189E-06
246	3.85189E-06
247	3.85189E-06
248	3.85189E-06
249	3.85189E-06
250	1.15557E-05
251	7.70377E-06
252	3.85189E-06
253	1.15557E-05
254	3.85189E-06
255	3.85189E-06
256	3.85189E-06
258	3.85189E-06
259	7.70377E-06
260	3.85189E-06
261	7.70377E-06
262	7.70377E-06
263	3.85189E-06
265	3.85189E-06
266	3.85189E-06
267	3.85189E-06
268	3.85189E-06
269	1.15557E-05
270	7.70377E-06
271	7.70377E-06
272	3.85189E-06
273	1.15557E-05
276	7.70377E-06
277	1.54075E-05
283	3.85189E-06
285	1.15557E-05
286	3.85189E-06
287	1.15557E-05
288	3.85189E-06
289	3.85189E-06
290	3.85189E-06
291	3.85189E-06
292	3.85189E-06
294	3.85189E-06
295	3.85189E-06
297	3.85189E-06
298	3.85189E-06
300	3.85189E-06
301	7.70377E-06
302	3.85189E-06
303	7.70377E-06
304	1.15557E-05
305	3.85189E-06
309	3.85189E-06
310	3.85189E-06
311	7.70377E-06
314	1.15557E-05
315	7.70377E-06
316	1.15557E-05
317	3.85189E-06
318	3.85189E-06
320	3.85189E-06
322	3.85189E-06
323	3.85189E-06
327	1.54075E-05
329	3.85189E-06
330	3.85189E-06
331	3.85189E-06
332	7.70377E-06
333	3.85189E-06
335	1.15557E-05
338	7.70377E-06
339	7.70377E-06
341	3.85189E-06
342	3.85189E-06
344	7.70377E-06
346	3.85189E-06
350	3.85189E-06
351	3.85189E-06
353	3.85189E-06
354	7.70377E-06
355	3.85189E-06
358	7.70377E-06
360	3.85189E-06
361	3.85189E-06
365	3.85189E-06
367	7.70377E-06
368	3.85189E-06
370	3.85189E-06
372	7.70377E-06
373	3.85189E-06
375	3.85189E-06
376	7.70377E-06
378	3.85189E-06
381	7.70377E-06
382	3.85189E-06
383	3.85189E-06
386	1.15557E-05
388	7.70377E-06
391	7.70377E-06
395	3.85189E-06
397	1.54075E-05
399	3.85189E-06
400	7.70377E-06
402	3.85189E-06
403	3.85189E-06
405	3.85189E-06
408	3.85189E-06
409	3.85189E-06
413	3.85189E-06
414	3.85189E-06
415	1.54075E-05
416	3.85189E-06
417	3.85189E-06
419	7.70377E-06
421	3.85189E-06
425	3.85189E-06
433	3.85189E-06
435	3.85189E-06
437	7.70377E-06
439	3.85189E-06
445	3.85189E-06
446	3.85189E-06
447	3.85189E-06
448	3.85189E-06
449	3.85189E-06
450	3.85189E-06
452	3.85189E-06
458	3.85189E-06
459	3.85189E-06
461	3.85189E-06
464	3.85189E-06
467	3.85189E-06
468	7.70377E-06
472	1.15557E-05
474	3.85189E-06
475	7.70377E-06
476	3.85189E-06
477	3.85189E-06
482	3.85189E-06
483	7.70377E-06
487	3.85189E-06
490	3.85189E-06
492	3.85189E-06
496	3.85189E-06
497	3.85189E-06
500	3.85189E-06
501	3.85189E-06
509	7.70377E-06
520	3.85189E-06
529	3.85189E-06
533	3.85189E-06
538	3.85189E-06
542	3.85189E-06
545	3.85189E-06
547	3.85189E-06
550	3.85189E-06
557	3.85189E-06
558	3.85189E-06
560	3.85189E-06
2613	3.85189E-06
567	3.85189E-06
574	3.85189E-06
575	3.85189E-06
583	3.85189E-06
584	3.85189E-06
585	3.85189E-06
586	3.85189E-06
593	3.85189E-06
594	3.85189E-06
598	3.85189E-06
601	3.85189E-06
603	3.85189E-06
605	3.85189E-06
610	3.85189E-06
613	3.85189E-06
621	3.85189E-06
622	3.85189E-06
623	3.85189E-06
624	3.85189E-06
625	3.85189E-06
638	3.85189E-06
644	3.85189E-06
645	3.85189E-06
655	7.70377E-06
656	3.85189E-06
658	3.85189E-06
660	3.85189E-06
671	3.85189E-06
2723	3.85189E-06
676	3.85189E-06
680	3.85189E-06
681	3.85189E-06
684	3.85189E-06
687	3.85189E-06
689	3.85189E-06
693	3.85189E-06
699	3.85189E-06
704	3.85189E-06
707	3.85189E-06
710	3.85189E-06
715	3.85189E-06
717	3.85189E-06
720	3.85189E-06
723	3.85189E-06
730	3.85189E-06
732	3.85189E-06
742	3.85189E-06
747	3.85189E-06
758	3.85189E-06
768	3.85189E-06
771	3.85189E-06
784	3.85189E-06
790	3.85189E-06
793	3.85189E-06
799	3.85189E-06
802	7.70377E-06
804	7.70377E-06
805	3.85189E-06
806	3.85189E-06
807	3.85189E-06
818	3.85189E-06
828	7.70377E-06
837	3.85189E-06
839	3.85189E-06
855	3.85189E-06
857	3.85189E-06
869	3.85189E-06
886	3.85189E-06
896	3.85189E-06
898	3.85189E-06
902	3.85189E-06
943	3.85189E-06
948	3.85189E-06
962	3.85189E-06
993	3.85189E-06
1056	3.85189E-06
1092	3.85189E-06
1094	3.85189E-06
1146	3.85189E-06
1152	3.85189E-06
1157	3.85189E-06
1160	3.85189E-06
1185	7.70377E-06
1186	3.85189E-06
1229	3.85189E-06
3307	3.85189E-06
1300	3.85189E-06
1319	3.85189E-06
1320	3.85189E-06
3450	3.85189E-06
1451	3.85189E-06
3532	3.85189E-06
3543	3.85189E-06
1561	3.85189E-06
1595	3.85189E-06
1769	3.85189E-06
3905	3.85189E-06
675	3.85189E-06

}{\teamsize}

{
\begin{tikzpicture}
\begin{groupplot}
[
    group style={
        group name=my plots,
        group size=2 by 1,
        xlabels at=edge bottom,
        xticklabels at=edge bottom,
        vertical sep=15pt,
        horizontal sep=1cm,
    },
    legend style={draw=none, fill=none, nodes={scale=0.75, transform shape}},
    width=1.7in,
    height=1.5in, 
    ]
    \nextgroupplot[
    xlabel={Commits \# After Adoption}, 
    ylabel={median \% change (LOC$_\ell$)}, 
    title = Adoptions by Team Size,
    clip=false,
    xtick = {1,20,40,60,80,100},
    yticklabels={0,100\%,200\%,300\%,400\%,500\%},
    ytick={0,1,2,3,4,5},
    ymax=5,
    xmin = -1,
    legend columns=2, 
    legend pos=north west,
    ]
    
    \addplot [black, thick]  table [x=x, y=one]   {\table}; 
    \addlegendentry{1}
    
    \addplot [blue, thick]  table [x=x, y=two]   {\table};
    \addlegendentry{2}
    
    \addplot [blue!50!green, thick]  table [x=x, y=three]   {\table};
    \addlegendentry{3--5}
    
    \addplot [green, thick]  table [x=x, y=six]   {\table};
    \addlegendentry{6--9}
    
    \addplot [orange, thick]  table [x=x, y=ten]   {\table};
    \addlegendentry{10+}
    
    \nextgroupplot[
    title=Team Size Distribution,
    clip=false,
    ymode=log,
    xmode=log,
    xlabel = {Team Size},
    ylabel = {$p(x)$},
    y label style={at={(axis description cs:0.15,.5)}},
    ]
    \addplot [only marks, mark=*, blue, mark size=1pt, fill opacity=0.75, draw opacity=0.2]  table [x=Size, y=Number]   {\teamsize};
    \end{groupplot}
    
\node at (-0.5, -0.75) {{\textbf{(a)}}};
\node at (4.0, -0.75) {{\textbf{(b)}}};
\end{tikzpicture}
}

    \caption{(a) Median percentage change in lines of code referencing $\ell$ after adoption. (b) Like the commit and adoption distributions illustrated in Figs. \ref{fig:commit_dist} and \ref{fig:adopt_dist}, the team size distribution follows a power law.}
    \label{fig:change}
\end{figure}
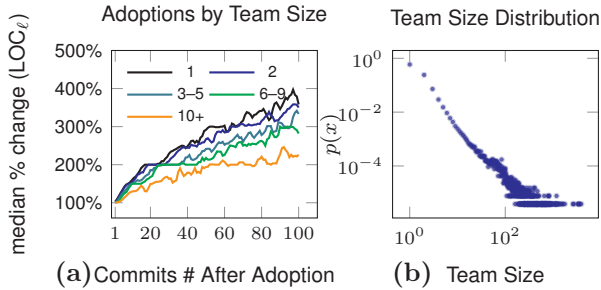


We observe in  Fig.~\ref{fig:change} that larger teams exhibit slower adoption growth and appear to be less agile than larger teams. A possible explanation could be due to perspective differences between two or more committers to a project. Users might feel more comfortable making more commits or experimenting with newly adopted libraries in smaller teams, or if they are working alone. Also, we can theorize that larger teams might do more work while using fewer commits, which results in there being less opportunity for growth later.

Additionally, Fig~\ref{fig:change} shows us that the distribution of team sizes has a power law-like heavy tail wherein 59\% of projects have only a single committer; 24\% and 7\% of projects have two and three distinct committers respectively.

\section{Code Fights}

In the context of library adoptions in collaborative projects, we informally define a code fight as a series of commits that include back-and-forth additions and deletions of the same code containing a newly adopted $\ell$. A \textit{round} is a series of commits by one user that is uninterrupted by the other user. Formally, let $n^{(r)}$ represent the net change in lines of code referencing $\ell$ in round $r$; $r=0$ indicates the round of the adoption event. Also let $n^{\le r}$ be the sum of all lines of code referencing $\ell$ up to and including $r$, \ie, the running total. 

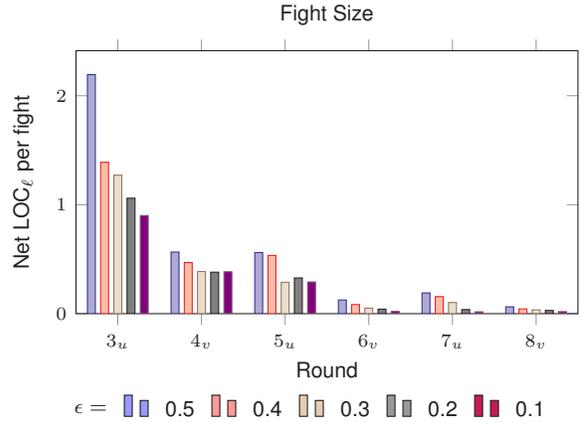
\begin{figure}[t]
    \centering
    \begin{tikzpicture}
\begin{axis}[       
        ybar, 
        bar width=3pt, 
        xmax=7.5,
        xmin=1.5,
        ymin=0,
        height=2.0in,
        title=Fight Size,
        width=3.25in,
        y label style={at={(axis description cs:0.08,.5)}},
        xlabel = {Round},
        xtick = {2,3,4,5,6,7},
        xticklabels = {$3_u$,$4_v$,$5_u$,$6_v$,$7_u$,$8_v$},
        ylabel = {Net LOC$_\ell$ per fight},
]

\addplot+ coordinates
	{
(0, 24.58701664403882)
(1, -20.84974480816614)
(2, 2.193977221300347)
(3, 0.5654824759893398)
(4, 0.5608689093377583)
(5, 0.12530799014431537)
(6, 0.19011791622668076)
(7, 0.06275456328254639)
(8, 0.05572735958163624)
(9, 0.04858701664403882)
};

\addplot+ coordinates
	{
(0, 28.49826319774567)
(1, -25.323942892380707)
(2, 1.38937935664161)
(3, 0.4690771140200923)
(4, 0.5344121021942252)
(5, 0.0831557501814860)
(6, 0.156663466267007)
(7, 0.04325223052244573)
(8, 0.03600839909764341)
(9, 0.03197277318533436)
};

\addplot+ coordinates
	{
(0, 27.221312110131212)
(1, -25.13420520542052)
(2, 1.2710174231017424)
(3, 0.3865691546569155)
(4, 0.2876575607657561)
(5, 0.05107334910733491)
(6, 0.10212949021294902)
(7, 0.034699935469993545)
(8, 0.02906431490643149)
(9, 0.025837814583781458)
};

\addplot+ coordinates
	{
(0, 24.095782216914557)
(1, -23.079352585676546)
(2, 1.0595574132377954)
(3, 0.3807404190641624)
(4, 0.3277240979987696)
(5, 0.04141786993811747)
(6, 0.03759092389534253)
(7, 0.029258495277385735)
(8, 0.024481598089241125)
(9, 0.01049469836789346)
};

\addplot+ coordinates
	{
(0, 18.120241398049366)
(1, -17.92301385273956)
(2, 0.9004090038040214)
(3, 0.3845207791093442)
(4, 0.29087892915367675)
(5, 0.020593198524154106)
(6, 0.016922651564987748)
(7, 0.01855294644814995)
(8, 0.00824681329786727)
(9, 0.0021546587344716795)
};

\end{axis}
\end{tikzpicture}
\begin{tikzpicture}
    \begin{customlegend}[legend columns=-1,
      legend style={
        draw=none,
        column sep=1ex,
      },
    legend entries={$\epsilon=$, 0.5, 0.4, 0.3, 0.2, 0.1}]
    \addlegendimage{ybar,ybar legend, fill opacity=0.6, draw opacity=0.0}
    \addlegendimage{ybar,ybar legend, fill=blue, fill opacity=0.40, draw opacity=0.9}
    \addlegendimage{ybar,ybar legend, fill=red, fill opacity=0.40, draw opacity=0.9}
    \addlegendimage{ybar,ybar legend, fill=brown, fill opacity=0.40, draw opacity=0.9}
    \addlegendimage{ybar,ybar legend, fill=black, fill opacity=0.40, draw opacity=0.9}
    \addlegendimage{ybar,ybar legend, fill=purple, fill opacity=0.90, draw opacity=0.9}
    \end{customlegend}
\end{tikzpicture}
    \caption{When a two-person fight occurs, the adopter $u$ (indicated by odd-numbered $x$) tends to commit more code than $v$ on average.}
    \label{fig:fights}
\end{figure}

A \textit{code fight} occurs if there exists any $r$ such that $(1-\epsilon)n^{\le (r-1)} \le n^{\le r}$, where we set $\epsilon$ to represent the percent reduction that must occur, with $\epsilon \in \{.10, .20, .30, .40, .50\}$.

We observe that fights are relatively rare, occurring between 1 and 3 times for every 100,000 commits on average. Also, choice for $\epsilon$ has a limited effect on the probability of a fight. The probability of a fight increases with team size, but with diminishing returns that resemble a Zipfian Distribution.


Next, we analyze what happens during a two-person fight. Technically speaking, the first round of a fight is the adoption event and the second round of the fight is the removal of at least 100(1 - $\epsilon$) percent lines of the adopter's code. After this point, the two fighters may continue with more rounds of back and forth commits. 

Despite the dropoff in number of fights, the adopter tends to fight back with more lines of code. We observe that odd-numbered rounds, corresponding to the adopter, have more net LOC  referencing $\ell$ per round, than the deleter's round that comes afterwards. This is shown in Fig~\ref{fig:fights}. We see that the larger the original deletion of the code was, the less likely the adoptor is to fight back with lines of code.

\nop{\begin{figure}
    \centering
    \include{./figs/Fight_win_distribution}
    \caption{We can see that in the majority of fights, the fighters, who are the people who originally delete code from a repository, win the fight as the last committer. Non-adopters, who are parties who were not involved as either the deleter of code or the adopter, win fights as someone who enters a fight later slightly more often than the adopter themselves.}
    \label{fig:fights_win_distribution}
\end{figure}}

We define a fight's winner as the user who was the last commiter referencing $\ell$. By our definition of rounds, the deleter wins approximately 90\% of the fights because the adopter only fights back 10\% of the time. 

What role does experience play in winning a fight? The current work maintains the standard set by prior studies~\cite{rotabi2017competition} and therefore defines experience as the time since the user's first commit (in any project). The more experienced committer wins the fights between 70\% and 80\% of the time. Results from alternative $\epsilon$ values were nearly identical to $\epsilon = 0.1$. Interestingly, the more experienced users have about a 75\% win probability even when the experience differences are less than a week or even a day.

We observe that common debugging libraries \texttt{pdb}, \texttt{pprint}, and \texttt{syslog} comprise three of the top four most common causes of fights. It is not surprising to see the \texttt{distutils} library counted among the top fight starters. This particular library is used to generate source code distributions \ie, code releases, but it is strongly encouraged that users use the \texttt{setuptools} library instead. So most cases importing \texttt{disutils} is likely an error.

\nop{
\begin{table}[t]
    \centering
    \begin{tabular}{c|c}
        Fight Threshold  & \% Median Net Commit Size \\ \hline
0.5 & 24.58 \\
0.4 & 28.49 \\
0.3 & 27.22 \\
0.2 & 24.09 \\
0.1 & 18.12 \\
    \end{tabular}
    \caption{Various values for median net commit lines before a fight occurs, in which the next commit results in net lines of code that are below the set fight threshold.}
    \label{tab:fight_before}
\end{table}
}

\section{Discussion}

\smallskip
\noindent\textbf{RQ1: What does it look like when a team adopts a library for the first time?}  \qquad In Fig.~\ref{fig:adopt_per_commit} we observe that library adoptions tend to happen early in a project's history. We can expect that it is difficult to adopt a new library once a project has matured. Perhaps this is because new libraries may introduce instability into a repository, or because the primary innovation within a project occurs early on in its lifespan.

\smallskip
\noindent\textbf{RQ2: Are commits containing new libraries more likely to have deletions than other types of commits?}  \qquad Once an adoption has occurred, we track how long it takes the library to become stable within the project by examining how many additions and deletions occur in the commits after a library is adopted. In Fig.~\ref{fig:afteradoptionteamsize}, we observe that activity involving a newly adopted library is relatively high after a commit occurs.  Over time, the number of lines of code referencing the adopted library stabilizes. We can safely conclude that users tend to write most lines of code that involve a newly adopted library relatively soon (within 10-15 commits) after library adoption.

\smallskip
\noindent\textbf{RQ3: Do the answers to these questions vary by library type, team size, or the amount of information available on Stack Overflow?}  \qquad When team sizes are larger, the lines of library code do not grow as quickly relative to the first adoption commit as they do when team sizes are smaller. This may be because larger team projects require more communication and planning and are therefore less agile than small teams or individual projects. Additionally, we showed that the number of times a library appears in Stack Overflow is highly correlated with the number of adoptions. 

\smallskip
\noindent\textbf{RQ4: What does it look like when team members fight over new library usage?}  \qquad When working on a team, there is bound to be conflict. Different team members have various opinions about which library is best to use in a repository. The probability of these fights occurring increases with team size. The winner of these fights tends to be more experienced.

\bigskip
\noindent\textbf{Implications.} \qquad  We see that the number of commits and adoptions per project, along with team size, follow a power law distribution. We found positive correlations between the number of times libraries appear in Stack Overflow and GitHub. We discovered that popular libraries on Stack Overflow have faster rates of adoption for projects in Git. Additionally, smaller teams are more agile and they can adopt to using new libraries more quickly than larger teams, when productivity is measured as a function of median percentage growth. We also find that code fights are rare, but when they occur, they tend to be won by more experienced coders, and involve libraries which are used for debugging purposes.


\section*{Acknowledgements}
This work is sponsored by grant W911NF-17-1-0448 from the US Army Research Office and grant W911NF-17-C-0094 from DARPA.


\end{document}